%% file: designExp.tex
\documentclass[final,5p,twocolumn,10pt]{elsarticle} 

\include{preamble}

\journal{Computer-Aided Design}

\begin{document}

\begin{frontmatter}

\title{Exploring Feasible Design Spaces for Heterogeneous Constraints}

\author{Amir M. Mirzendehdel, Morad Behandish, and Saigopal Nelaturi}

\address{\rm
	Palo Alto Research Center (PARC),
	3333 Coyote Hill Road, Palo Alto, California 94304
	\vspace{-15.0pt}
}

\input{abstract}

\end{frontmatter}

\input{introduction}
\input{contributions}
\input{relatedWork}

\input{definitions}
\input{pruning}
\input{exploration}
\input{conclusion}

\bibliography{designExp}

\end{document}

%% file: preamble.tex
\usepackage{lineno}
\usepackage{hyperref}
\usepackage{enumitem}
\modulolinenumbers[5]

\usepackage{subcaption}
\usepackage{mathtools}
\usepackage{amsmath}
\usepackage{amssymb}
\usepackage{stmaryrd}
\usepackage{amsthm}
\usepackage{amsfonts}
\usepackage{dsfont}
\usepackage{graphicx}
\usepackage{upgreek}
\usepackage{bm}
\usepackage{color}
\usepackage{float}
\usepackage{tikz-cd}
\usepackage{scrextend}

\usepackage{todonotes}

\biboptions{numbers,sort&compress}

\DeclareFontFamily{OT1}{pzc}{}
\DeclareFontShape{OT1}{pzc}{m}{it}{<-> s * [1.200] pzcmi7t}{}
\DeclareMathAlphabet{\mathpzc}{OT1}{pzc}{m}{it}

\newcommand{\bx}{{\mathbf{x}}}

\newcommand{\bt}{{\mathbf{t}}}
\newcommand{\bu}{{\mathbf{u}}}

\newcommand{\bff}{{\mathbf{f}}}

\newcommand{\sign}{\mathsf{sign}}

\newcommand{\indic}{{\mathbf{1}}}

\newcommand{\dimm}{\mathrm{d}}
\newcommand{\vol}{{\mathsf{vol}}}

\newcommand{\sweep}{\mathsf{sweep}}
\newcommand{\unsweep}{\mathsf{unsweep}}
\newcommand{\obs}{\mathsf{obs}}

\newcommand{\powerset}{\mathcal{P}}
\newcommand{\Fourier}{\mathcal{F}}

\newcommand{\R}{{\mathds{R}}}

\newcommand{\Z}{{\mathds{Z}}}

\newcommand{\SO}[1]{{\mathrm{SO}(#1)}}
\newcommand{\SE}[1]{{\mathrm{SE}(#1)}}

\newcommand{\conf}{{\mathsf{C}}}

\renewcommand{\th}{$^\text{th}$ }

\theoremstyle{definition}

\newtheorem{prop}{Proposition}

\newcommand{\dspace}{\mathsf{D}}
\newcommand{\pspace}{\mathsf{P}}
\newcommand{\fspace}{\mathsf{F}}
\newcommand{\feaspace}{{\dspace^\ast}}
\newcommand{\halfspace}{\mathsf{H}}
\newcommand{\range}{\mathbb{Q}}

\newcommand{\analysis}{\mathcal{A}}
\newcommand{\test}{\mathpzc{c}}
\newcommand{\gglob}{g} 
\newcommand{\gloc}{g} 
\newcommand{\gsloc}{g^\ast} 
\newcommand{\tsf}{{\mathcal{T}}}

\newcommand{\eq}[1]{(\ref{#1})} 
\newcommand{\com}[1]{} 

%% file: abstract.tex
\begin{abstract}

We demonstrate an approach of exploring design spaces to simultaneously satisfy
kinematics- and physics-based requirements. We present a classification of
constraints and solvers to enable postponing optimization as far down the design
workflow as possible. The solvers are organized into two broad classes of design
space `pruning' and `exploration' by considering the types of constraints they
can satisfy. We show that pointwise constraints define feasible design subspaces
that can be represented and computed as {\it first-class entities} by their
maximal feasible elements. The design space is pruned upfront by intersecting
maximal elements, without premature optimization. To solve for other
constraints, we apply topology optimization (TO), starting from the pruned
feasible space. The optimization is steered by a topological sensitivity field
(TSF) that measures the global changes in violation of constraints with respect
to local topological punctures. The TSF for global objective functions is
augmented with TSF for global constraints, and penalized/filtered to incorporate
local constraints, including set constraints converted to differentiable
(in)equality constraints. We demonstrate application of the proposed workflow to
nontrivial examples in design and manufacturing. Among other examples, we show
how to explore pruned design spaces via TO to simultaneously satisfy
physics-based constraints (e.g., minimize compliance and mass) as well as
kinematics-based constraints (e.g., maximize accessibility for machining).

\end{abstract}


%% file: introduction.tex
\section{Introduction} \label{sec_intro}

Mechanical design problems require reasoning about diverse, multiple, and often
conflicting objectives and constraints arising from requirements across a
product's lifecycle. The key engineering design challenge lies in traversing the
trade space of these requirements to synthesize feasible designs. This challenge
has recently been amplified by rapid advances in manufacturing processes.
Light-weight, high-performance, and multi-material composite structures with
complex geometry and material distribution can now be fabricated using various
additive manufacturing (AM) processes. Yet, existing computer-aided design (CAD)
systems are far behind in their representations and algorithms to navigate the
high-dimensional trade spaces that grow exponentially in the number of available
decisions per spatial elements. Additional functional constraints such as
manufacturability, ease of assembly, motion in presence of obstacles, and
aesthetics dramatically increase the trade space complexity.

Specialized domain-specific computational tools are used to generate designs
that satisfy specific types of functional requirements. For example, to maximize
a part's performance with as little cost or material as possible, one may employ
topology optimization (TO) tools
\cite{Sigmund2013topology,Dijk2013level,Rozvany2009critical}. In most TO
approaches, an objective function is defined over a design domain in terms of
the physical performance (e.g., strength and/or stiffness) with additional {\it
	constraints} on total mass or volume as well as boundary conditions (e.g.,
loading and/or restraints) that often account for interfaces with other parts.
TO produces valid designs with nonintuitive shapes, topologies, and material
distributions that meet physical requirements, but is rarely aware of other
important design criteria such as kinematic constraints. On the other hand, to
ensure collision-free motion of a part in an assembly, one may need to examine
its free configuration space
\cite{Latombe2012robot,Lozano-Perez1983spatial,Nelaturi2012rapid} to guarantee
collision avoidance. Similarly, for subtractive manufacturing (SM), the
machinability of a designed part is predicated on whether the volume to be
removed from a raw stock is accessible within the cutting tool assembly's
non-colliding configurations \cite{Nelaturi2015automatic,Behandish2018turning}.
For AM, one may need to consider the part's morphology, minimum feature size,
and skeleton
\cite{Nelaturi2015manufacturability,Nelaturi2015representation,
	Behandish2018characterizing,Behandish2019detection}
Hybrid manufacturing (combined AM and SM) requires more complicated logical
reasoning \cite{Behandish2018automated}. These problems require nontrivial
interference analysis of shapes in relative motion that rely on different tools
of reasoning than physics-driven design tools such as TO. The latter often
ignore motion related constraints by considering them out-of-scope.

\subsection{Kinematic, Physical, \& Manufacturing Constraints}

Generating practical designs requires {\it simultaneous} reasoning about shape,
motion, materials, physics, manufacturing, and assembly, among other factors.
For example, a machine part that moves relative to other parts in a mechanical
device has to avoid collisions with both stationary and moving obstacles
\cite{Ilies2000shaping}. These requirements are imposed as {\it kinematic}
constraints, expressed in terms of pointset containment or non-interference
(Section \ref{sec_setcons}). The same part has to sustain mechanical loads at
its joints or points of contact with other parts. These requirements are imposed
as {\it physical} constraints, expressed in terms of (in)equalities of
mathematical functions that represent physical response
\cite{Lai2009introduction} (e.g., bounds on deflection or stress). Moreover, the
part has to be manufacturable with one or more AM or SM capabilities
\cite{Behandish2018automated}. {\it Manufacturability} constraints can be of
both kinematic and physical types; for instance, accessibility in SM
\cite{Nelaturi2015automatic,Behandish2018turning} and post-processing of AM
(e.g., support removal \cite{Nelaturi2019automatic}) are of predominantly
kinematic nature, whereas achieving desired material properties in AM requires
in situ physical analysis \cite{Michopoulos2018multiphysics}. With few
exceptions (e.g., TO for AM with minimized support
\cite{Mirzendehdel2016support}) TO algorithms are not developed with
manufacturability provisions built into their objective functions.

\subsection{Generating Feasible Designs with Multiple Solvers}
\label{sec_introExam}

Different computational services, herein called {\it solvers}, are used to
assist the design process with heterogeneous constraints by providing
either {\it analysis} tools to evaluate the performance of one or more given
designs, or {\it synthesis} tools to generate one or more designs that satisfy a
given set of performance criteria analysis. We distinguish between these two
types of solvers as {\it forward} and {\it inverse} problem solvers
(`FP/IP-solvers'), respectively (Section \ref{sec_classify}). Specifically, {\it
	generative design} tools are IP-solvers which solve the inverse problem by
systematically generating candidate designs and evaluating their performance
(using FP-solvers) to guide refinement of designs until the criteria are met.

It is unlikely that a single IP-solver is capable of simultaneous reasoning
about all design criteria (e.g., objective functions and constraints).
Therefore, a typical computational design workflow requires carefully organizing
and reasoning about several multidisciplinary solvers to address heterogeneous
design criteria. While every IP-solver reasoning about a {\it subset} of these
criteria may produce designs that are feasible (and ideally, optimal) only with
respect to the specific criteria it considers, it provides no guarantees about
the rest of the criteria. Different IP-solvers are thus likely to generate
designs distinct from one another, while none of them simultaneously satisfies
all criteria. Except for extremely simple criteria, it appears impossible to
combine these solutions in any obvious way that preserves the constraints
satisfied separately by each solver, or at least provides the best compromise.
Even if such a solution exists, there may not exist any particular ordering of
these solvers to find it, simply because each solver performs {\it premature
	optimization} with regard to the subset of criteria it cares about.

\begin{figure} [ht!]
	\centering
	\includegraphics[width=0.32\textwidth]{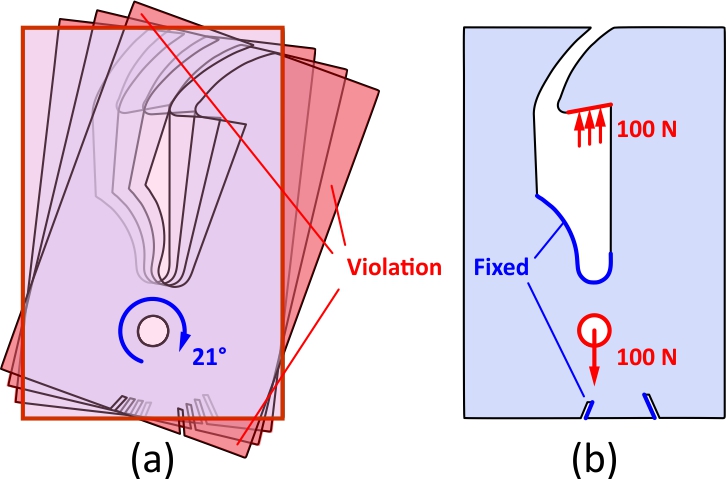}
	\caption{Functional requirements for designing a car hood latch, including (a)
	containment under given motion after assembly; and (b) boundary conditions on
	interfaces for assembly.} \label{fig_latchBCs}
\end{figure}

Consider the problem of designing a car hood latch (adopted from
\cite{Ilies2000shaping}). An initial design domain with boundary conditions is
provided, and the goal is to find a design that is as stiff as possible with the
least mass. Moreover, the latch has to be free to rotate clockwise by 20 degrees
around a revolute joint, without exiting the initial design domain, so that it
would not collide with other parts that are possibly outside that envelope.
Functional load-bearing surfaces where the latch mates with other parts are
shown in Fig. \ref{fig_latchBCs}. 
All feasible designs must retain these surfaces as specified. The said
requirements immediately suggest using two IP-solvers that are well-positioned
to deal with them (each solver satisfying a subset of them):
\begin{itemize}
	\item Let \textsf{Unsweep} \cite{Ilies1999dual} be a solver that generates a
	design that remains within a given region of space while moving according to a
	prescribed motion (in this case, a clockwise rotation of 21 degrees).
	\item Let \textsf{PareTO} \cite{Suresh2013efficient} be a solver that, starting
	from an initial design, generates a design on the Pareto front of the two
	objectives (compliance and mass), i.e., one that satisfies the stiffness
	requirement and the given boundary conditions with minimal mass.
\end{itemize}

\begin{figure} [ht!]
	\centering
	\includegraphics[width=0.46\textwidth]{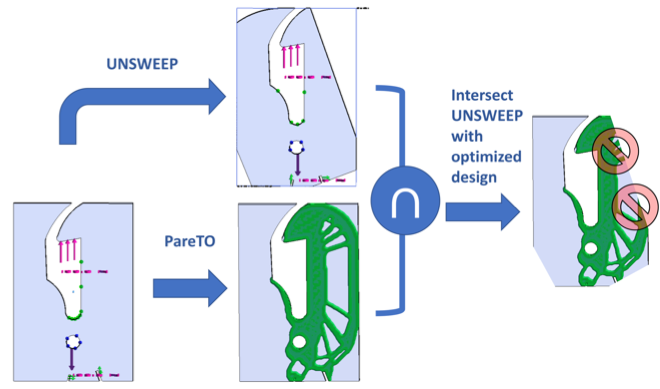}
	\caption{Two partially feasible designs are generated separately by two
	solvers. Each solver optimizes for a subset of constraints. Arbitrarily
	combining them (e.g., via set intersection in this example) may result in an
	infeasible design with respect to all constraints.} \label{fig_latchIntersect}
\end{figure}

Using these solvers separately, one may generate two distinct designs, as
illustrated in Fig. \ref{fig_latchIntersect}. However, there is no clear
operation with which to combine these two, in order to generate a design that
satisfies both kinematics- and physics-based constraints. For example, the
intersection of the two designs, shown in Fig. \ref{fig_latchIntersect}
generates a design that does not violate the constraints satisfied by
\textsf{Unsweep}, because every subset of the unswept volume also satisfies the
containment constraints. However, the constraints satisfied by \textsf{PareTO}
are no longer satisfied, because the load paths are changed due to the changed
topology and the compliance target is no longer met.

In this case, \textsf{Unsweep} has a property that its solution can be
interpreted not as a single design, but as a representation of all designs that
satisfy the containment constraints. This family of designs is {\it closed under
	set intersection}, i.e., intersecting any of the feasible designs with another
set leads to another feasible design. Similarly, \textsf{PareTO} can generate a
family of designs that satisfy compliance requirements for different mass
budgets. However, this family is not closed under set intersection. The goal of
this paper is to show how we can exploit such information to organize the design
workflows such that they produce solutions that satisfy all criteria.

\begin{figure} [ht!]
	\centering
	\includegraphics[width=0.46\textwidth]{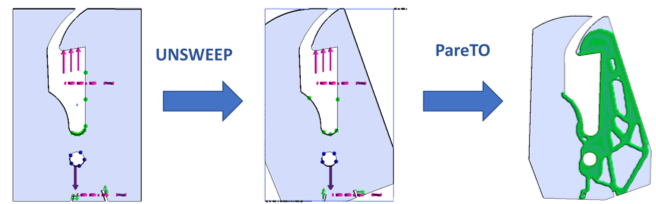}
	\caption{Both constraints can be satisfied by first finding a shape that
	satisfies the containment constraint and then using this shape as the input to
	TO (or any other material reducing IP-solver).} \label{fig_latchProperSeq}
\end{figure}

It is possible to obtain a feasible solution to the latch design problem by
using the same set of IP-solvers, if the workflow is organized differently, as
shown in Fig. \ref{fig_latchProperSeq}. Suppose we first solve for the
containment constraint using \textsf{Unsweep} and generate a valid design that
does not exit a given envelope while moving. We can use this design as the
initial design input to \textsf{PareTO} and optimize its shape and topology to
achieve the compliance target with minimal mass. This approach works because
\textsf{PareTO} is a material reducing solver, i.e., its solutions remain
strictly contained within the initial design. Hence any design generated
downstream will be faithful to the containment constraint that was satisfied
upstream. The same argument is not true if the order of applying the solvers is
swapped. There is no reason to believe that applying \textsf{Unsweep} to a
topologically optimized solution of \textsf{PareTO} will remain on the Pareto
front. The fundamental differences between the two IP-solvers should be taken
into account when deciding on their arrangement in a workflow---in this case,
choosing between parallel or sequential execution and the proper order for the
latter. We elaborate on these differences in the classification of solvers in
Section \ref{sec_classify}.

\begin{figure} [ht!] 
	\centering
	\includegraphics[width=0.46\textwidth]{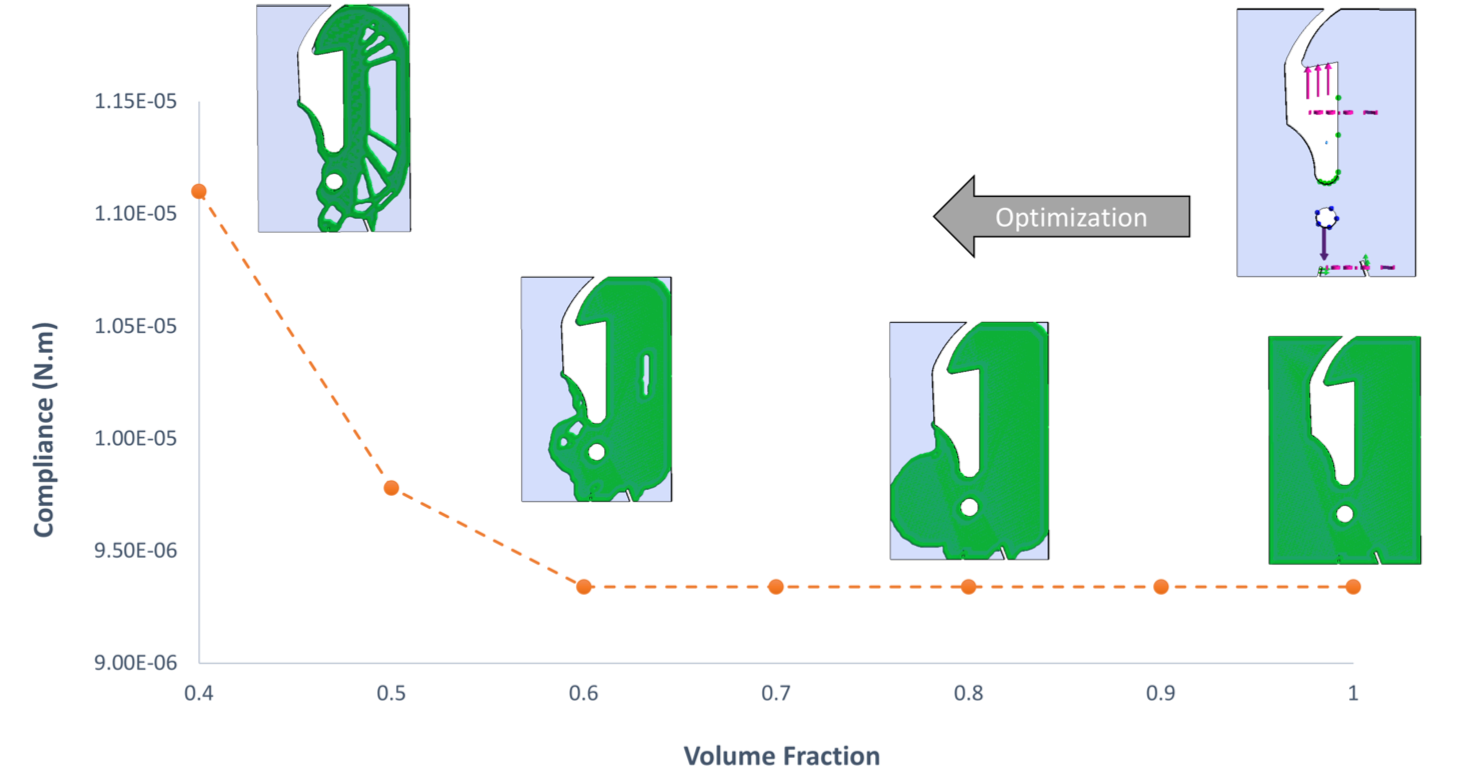}
	\caption{None of the topologically optimized design alternatives obtained from
	applying \textsf{PareTO} to the initial design will satisfy the containment
	constraint. Generate-and-test does not always work.
	}
	\label{fig_latchPareTO}
\end{figure}

Figure \ref{fig_latchPareTO} shows a one-parametric family of solutions on the
Pareto front obtained when using mass and compliance as competing objectives.
\textsf{PareTO} provides a clear advantage over classical TO with a single
solution, by producing many alternatives {\it some} of which might satisfy
additional constraints that were not accounted for in TO. This {\it
	generate-and-test} approach is plausible, and in fact, is a common strategy for
dealing with heterogeneous constraints. However, it turns out that none of the
Pareto-optimized designs in this case will not remain within the envelope after
a clockwise rotation of 21 degrees.

In general, a good rule of thumb to organize solvers in a workflow is to call
the ones that produce the broadest families of designs earlier. The upstream
IP-solvers should generate a large number of designs, as opposed to fixing one
or few designs, to provide more flexibility for downstream solvers. The
downstream solvers can be FP-solvers, testing for new constraints, or
IP-solvers, applying further optimization. However, each solver may prematurely
optimize designs that may fail evaluation criteria considered in downstream
solvers. The ``blind'' process of generating and testing designs without
carefully considering properties of the workflow -- and the associated feasible
design space for each solver in the workflow -- will scale poorly with increasing
number of constraints/solvers and their complexity. A systematic approach to
arranging solvers into workflows that guarantees satisfying new constraints
without violating the already satisfied constraints is much demanded, and is the
subject of this paper.

%% file: contributions.tex
\subsection{Contributions \& Outline}

The main contributions of this paper are:
\begin{enumerate}
	\item classification of IP-solvers based on the properties of design
	constraints (e.g., locality and continuity) that each solver can handle;
	\item organization of computational design workflows with heterogeneous
	constraints into two stages of design space `pruning' and `exploration', where
	pruning provides the feasible design space for exploration;
	\item reconciling set constraints with inequality constraints and deriving
	conditions under which they can be expressed in a strictly local (i.e.,
	`pointwise') fashion;
	\item pruning {\it design spaces}, treated as {\it first-class entities},%
	\footnote{In the lexicon of programming languages, a construct is said to have
	first-class status if it can be passed as an argument, returned from a
	subroutine, or assigned into a variable \cite{Scott2000programming}.}
	by intersecting their representative maximal elements defined implicitly by
	pointwise constraints;
	\item strategies to compose FP-solvers into a generative and iterative design
	process by combining kinematic, manufacturing, and physical constraints; and
	\item illustrating real-world design problems solved using multidisciplinary
	solvers, which are rarely addressed in the (otherwise siloed) areas of related
	research.
\end{enumerate}

In Section \ref{sec_lit}, we review some of the recent works in related areas of
computational design and manufacturing.

In Section \ref{sec_classify}, we present basic definitions (FP/IP-solvers,
global vs. local constraints, and design/performance space terminology). We
present a classification of solvers based on the types of constraints they can
handle, and propose a systematic approach for organizing them into workflows.

In Section \ref{sec_prune}, we show that an important class of constraints can
be satisfied upfront by pruning the design space in the early stages without
premature optimization, allowing flexibility for downstream design decisions.
The feasible design subspaces are described uniquely and completely by their
{\it maximal elements}. These representative elements are, in turn, implicitly
defined by a point membership classification (PMC) test implemented in terms of
strictly local (i.e., `pointwise') constraints. We demonstrate examples of
kinematics-based constraints that appear in assembly, packaging, and
manufacturing.

In Section \ref{sec_explore} we deal with more general global and local
constraints, including (in)equality constraints that specify physics-driven
requirements or manufacturability.

We use a Pareto-tracing levelset topology optimization (PareTO)
\cite{Suresh2013efficient} which uses fixed-point iteration to satisfy multiple
design criteria guided by augmented TSFs. We demonstrate that accessibility
constraints can be incorporated in TO by filtering the augmented TSF using the
overlap measure of cutting tool and optimized part.

We conclude in Section \ref{sec_conclusion} by a discussion of the limitations
of our approach and proposed future directions.

%% file: relatedWork.tex
\section{Related Work} \label{sec_lit}

Real-world design problems often involve solving multi-objective optimization
problems where the goal is to find the best trade-off between multiple competing
objective functions. Classical methods such as linear programming (e.g., the
`simplex' algorithm), nonlinear programming (e.g., the steepest descent and
conjugate gradients), or Newton-Raphson \cite{Nocedal2006numerical} are limited
to single-objective optimization problems. Even for single-objective
optimization, finding the global optimum is NP-hard \cite{Garey2002computers}.
Numerous approaches have been developed to converge to locally optimal solutions
in reasonable computation time to multi-objective problems across different
disciplines.

Unlike single-objective optimization, a total ordering for feasible solutions
may not be possible in multi-objective optimization, i.e., there may not exist a
single ``best'' solution due to competing objectives. However, feasible
solutions may be partially ordered according to {\it Pareto efficiency}---also
known as Pareto--Koopmans efficiency or dominance
\cite{Koopmans1951analysis,Charnes1985foundations}. Pareto-optimal solutions are
locally optimal (according to Pareto-efficiency) where improving one objective
comes at the expense of at least one other objective
\cite{Augusto2012multiobjective}. The collection of all Pareto-optimal solutions
is referred to as a \textit{Pareto front}, which represents a curve, surface, or
higher-dimensional manifold in the design space for two, three, or higher number
of competing objectives, respectively. Tracing a Pareto front is a key challenge
in multi-objective and multi-disciplinary design optimization. We will not
attempt an exhaustive review of all approaches such as gradient-free methods
including rule-based techniques
\cite{Sra2012optimization,Lienhard2017design,Burke2003hyper,Burke2013hyper} and
evolutionary algorithms \cite{Vikhar2016evolutionary,Michalewicz1995genetic,
	Kennedy2006swarm,Shi2004particle,Storn1997differential}. Rather, we focus on a
special class of algorithms that automatically generate designs on the Pareto
front of multiple objectives given an initial design.

TO 
\cite{Eschenauer2001topology,Rozvany2009critical,Deaton2013survey} has emerged
as a practical class of computational methods for designing high-performance
light-weight structures and has been applied in numerous areas such as designing
automobiles components \cite{Wang2004automobile},  aircraft components
\cite{Kesseler2006multidisciplinary,Alonso2009aircraft}, spacecraft modules
\cite{Coverstone2000optimal}, cast parts \cite{Harzheim2005review}, compliant
mechanisms \cite{Sigmund1997design}, and many other products. Numerous
approaches such as density-based
\cite{Bendsoe1999material,Sigmund2001design,Bendsoe2004topology}, levelset-based
\cite{Allaire2002levelset,Allaire2005structural}, and evolutionary
\cite{Chu1996evolutionary,huang2008bidirectional,Liu2008genetic} methods for TO
have been developed.

\subsection{Optimal Design for Manufacturing}

TO typically focuses on optimizing designs for performance (e.g., physical
response to loads during operation) but less on other important factors such as
{\it manufacturability}. Apart from traditional processes such as machining and
molding, more recent technologies such as AM have introduced the ability to
fabricate complex topologically optimized designs while presenting new
manufacturing challenges \cite{Thompson2016design}. Process limitations must be
considered {\it during} the design/optimization stage as much as possible to
avoid repeated prototyping and iterations until the optimized designs are
manufacturable. Specifically, applying corrections to the geometry
\cite{Nelaturi2015manufacturability} or topology
\cite{Behandish2018characterizing} of a solution {\it after} TO to make it
manufacturable may sacrifice the achieved optimality.

\begin{figure} [ht!]
	\centering
	\includegraphics[width=0.48\textwidth]{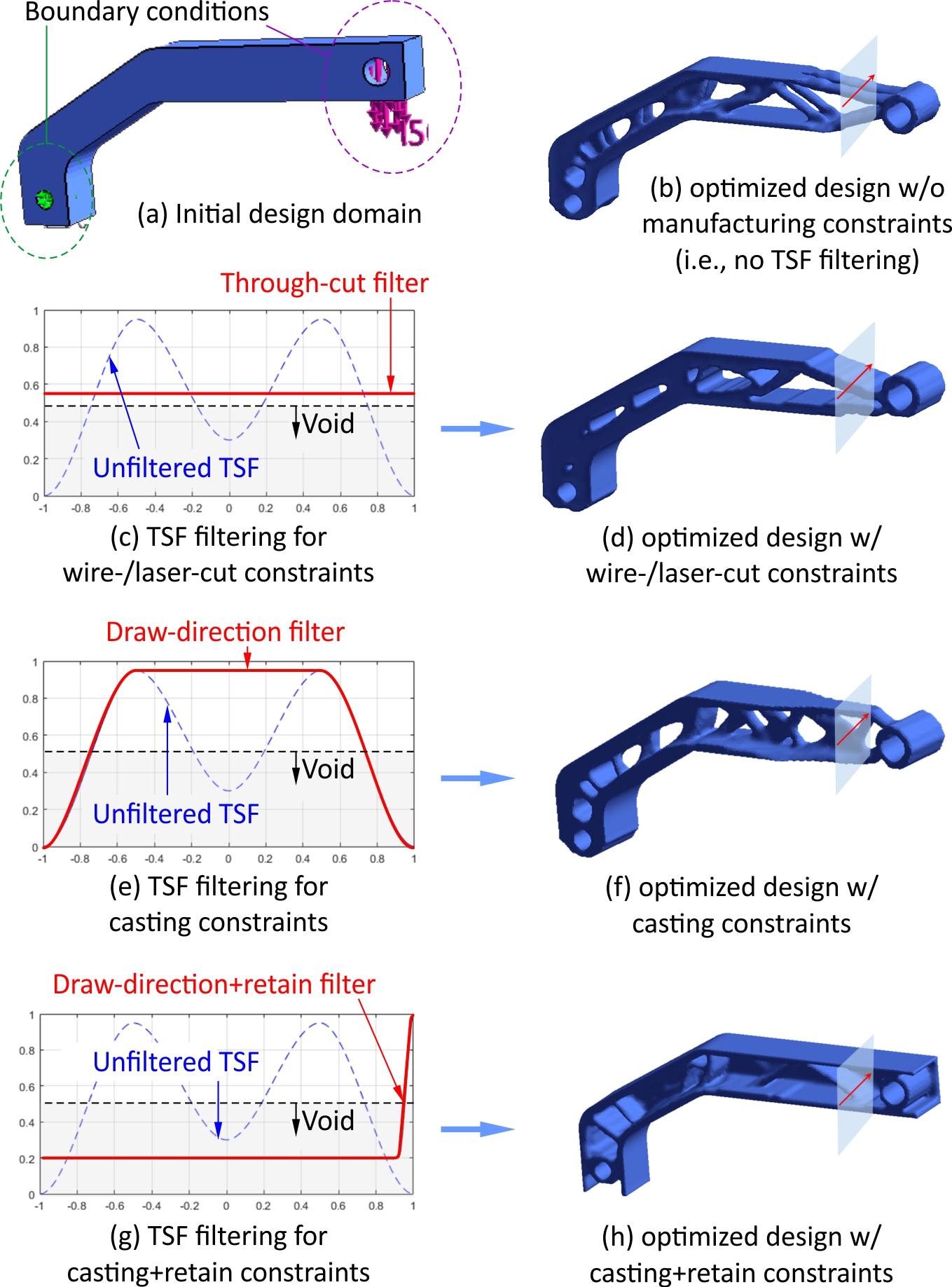}
	\caption{Examples of enforcing manufacturability constraints by TSF filtering.
	Different filters (c, e, g) -- plotted along the cross-sections shown on the
	right -- produce different solutions (d, f, h).} \label{fig_swingarm}
\end{figure}

One solution is to impose {\it design rules} obtained from domain expertise and
experience. These rules relate specific combinations of shape, materials, and
process to impose simplified constraints that can be built into the TO framework
to restrict the feasible design space. For example, when designing for AM via
fused deposition modeling using polymers, one should require that all facets
oriented at an angle greater than 45 degrees with respect to the build direction
be supported with additional scaffolding material. When designing for casting
and injection molding processes, one should ensure that the part has features of
almost uniform thickness and no entrapped holes are present, so that the mold
can be removed and the molten material cools down uniformly throughout the part
\cite{Pandelidis1990optimization,Lee1995optimization}. When designing for wire-
or laser-cutting, one should ensure that the final design has a uniform
cross-section, i.e., is 2.5D along the cutting direction. These constraints can
be imposed during TO through filtering of the sensitivity field as illustrated
in Fig. \ref{fig_swingarm}.

\begin{figure}[ht!]
	\centering
	\includegraphics[width=0.48\textwidth]{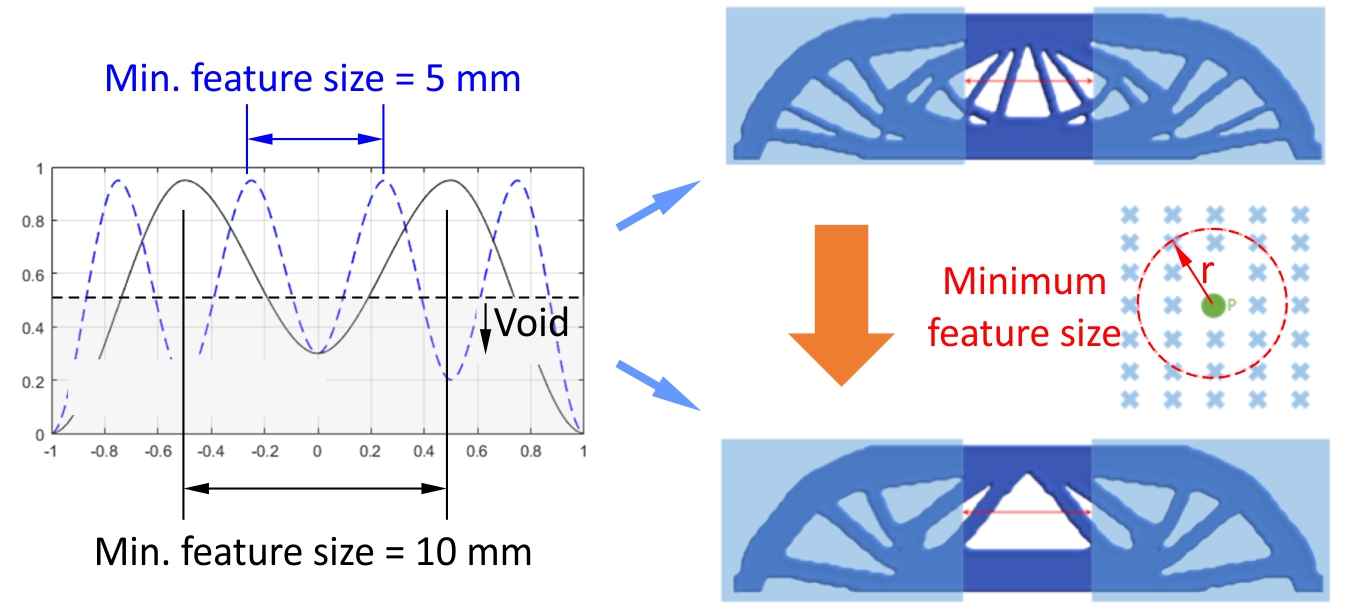}
	\caption{Constraining minimum feature size via TSF filtering
	\cite{Mirzendehdel2017handson}.} \label{fig_minFeatureSize}
\end{figure}

Another important AM consideration is the manufacturing resolution, which can be
directly incorporated into the TO algorithm as a minimum feature size constraint
through either local gradient constraints \cite{Petersson1998slope} or TSF
filtering \cite{Sigmund1997design} (Fig. \ref{fig_minFeatureSize}). It is also
possible to reduce the amount of support structure needed in an AM by either
finding a good build orientation or TSF filtering. Build orientation
optimization often involves solving a multi-objective problem taking into
account other factors such as surface quality
\cite{Pandey2004optimal,Nezhad2010pareto}, build time \cite{Pandey2004optimal},
or manufacturing error \cite{Paul2015optimization}. TSF filtering, on the other
hand, can be achieved by penalizing overhang surfaces
\cite{Gaynor2016topology,Langelaar2016topology}, penalizing undercut surfaces
\cite{Qian2017undercut}, or augmenting new TSFs \cite{Mirzendehdel2016support}.

Other AM constraints may not be as straightforward. For instance, optimizing
designs with respect to shrinkage and warpage during material phase changes
within the AM process may require solving a multi-physics problem at every
iteration \cite{Michopoulos2018multiphysics,Schmutzler2016pre}. Characterizing
material properties of AM parts is also challenging due to process-induced
anisotropy. Lack of inter-layer adhesion and varying thermal history at every
point can introduce unintended porosities and consequently affect material
behavior. Although there are initial results for considering anisotropy in TO
under certain assumptions \cite{Mirzendehdel2018strength}, solving the problem
by simultaneously optimizing the geometry, topology, and process parameters
involves costly in-situ manufacturing simulation or process planning
\cite{Nelaturi2015automatic,Behandish2018automated}.

\begin{figure*}[ht!]
	\centering \includegraphics[width=0.99\textwidth]{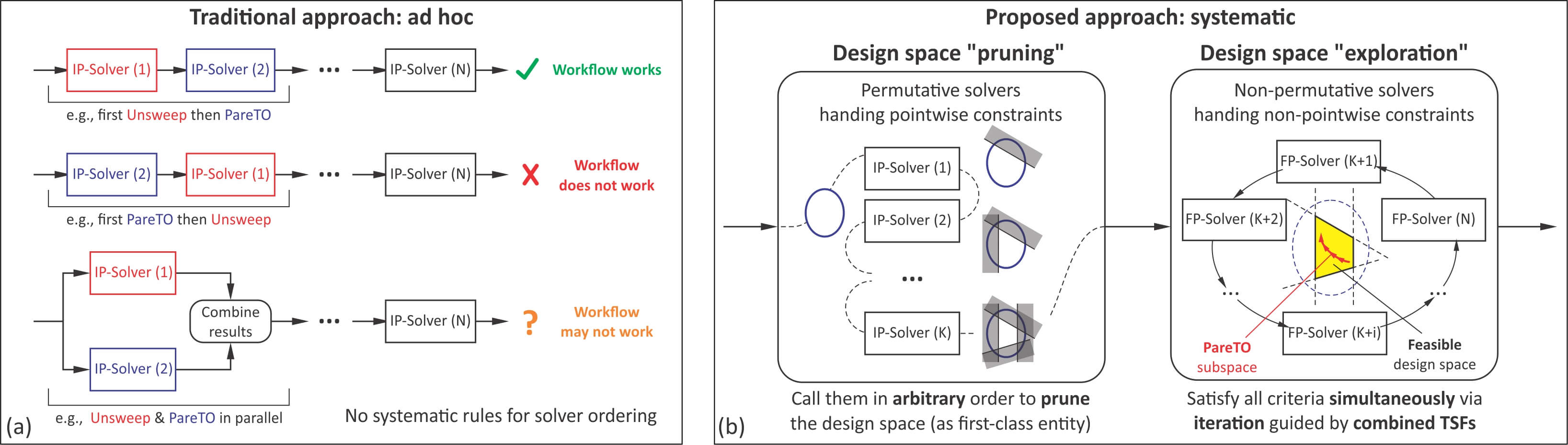}
	\caption{Solving heterogeneous (e.g., kinematic, physical, and manufacturing)
	constraints with multidisciplinary solvers is difficult. Each solver might make
	decisions with care for its target subset of constraints while potentially
	violating the rest of the constraints. (a) Arranging multiple solvers
	sequentially may work in some orders and fail in others, depending on what
	properties they preserve. Parallel composition requires combining solutions in
	ways that do not always preserve the properties either. (b) We propose a
	systematic two-phase approach (Section \ref{sec_phases}) to design space
	pruning (Section \ref{sec_prune}) and design space exploration (Section
	\ref{sec_explore}). The former invokes IP-solvers in an arbitrary order to cut
	out the infeasible design subspace without premature optimization. The latter
	navigates the pruned design space by fixed-point iterations over FP-solvers.
	Section \ref{sec_classify} addresses the question of how to divide a given
	collection of solvers into these two groups.}
	\label{fig_workflows}
\end{figure*}

\subsection{Design for Motion-Related Constraints}

In addition to the above examples of performance and manufacturing requirements,
there are other important design criteria that involve spatial reasoning about
the interactions of moving (translating and rotating) shapes such as collision
avoidance, packaging, robot motion planning, and accessibility analysis. These
requirements cannot be easily enforced by design rules, TSF filtering, or other
techniques commonly used in TO. Rather, they are often expressed as {\it set
	constraints}, i.e., statements in the language of sets (e.g., in terms of affine
transformations, Boolean operations, and containment) rather than the language
of real-valued functions used for (in)equality constraints in TO. A broad class
of inverse problems in practical design and manufacturing reduce to solving set
constraints formulated in the configuration space of rigid motions
\cite{Nelaturi2012rapid,Lysenko2010group,Behandish2015analytic,Wise2000survey}.

Although the problems with set constraints are common and of significant
importance, they are not mainstream in design/optimization workflows due to
non-smoothness and computational intensity \cite{Papalambros2002optimization}.
There are instances of TO frameworks that deal with motion-related problems in
an ad hoc manner; for instance, in modeling collision and contact when designing
compliant mechanisms \cite{Mankame2004topology} or parts made of hyperelastic
materials that undergo large deformations part
\cite{Kumar2019computational,Luo2016topology}. However, it is not immediately
obvious how set constraints can be incorporated in a {\it systematic} fashion
into the design/optimization process without incurring prohibitive computation
costs of spatial analysis at every iteration. The next section presents a
different classification of constraints that enables design space pruning and
exploration, in which set constraints are also restated in terms of
(in)equalities of functions.

%% file: definitions.tex
\section{Classifying Solvers and Constraints} \label{sec_classify}

In this paper we restrict our attention on solvers that reduce material from a
bounded domain in 2D and 3D space to generate designs. Our focus will be on the
properties of {\it constraints} embedded within computational solvers to reason
about their ordering in the workflows.

When all objectives and constraints cannot be handled by a single solver,
current practice relies on a case-by-case domain-specific analysis to properly
construct workflows. Figure \ref{fig_workflows} (a) depicts examples of combining
\textsf{Unsweep} and \textsf{PareTO} illustrated in Section \ref{sec_introExam}.
An arrangement that works in one case may not work in another.

In contrast, we present a systematic approach for organizing solvers
into workflows by first classifying solvers into two fundamentally different types:
\begin{enumerate}
	\item {\it Design space pruning} solvers restrict the feasible design space by
	pruning the subspaces that violate one or more design criteria. They are
	permutative, meaning that they can be called at the beginning of the design
	workflow in an {\it arbitrary order}. By directly operating on the design
	subspaces as first-class entities, they postpone optimization to downstream
	solvers.
	\item {\it Design space exploration} solvers {\it simultaneously} explore the
	(pruned) design subspaces for optimized solutions. In this paper, we choose a
	well-studied Pareto front tracing approach (PareTO) \cite{Suresh2010199} to
	efficiently navigate the trade space of multiple objectives.
\end{enumerate}
Figure \ref{fig_workflows} (b) illustrates the framework. In this section we
present some terminology and a classification of design criteria based on which
the solvers can be grouped into the above two classes. We formalize this
two-phase approach in Section \ref{sec_phases} and discuss how to implement each
phase in depth with examples in Sections \ref{sec_prune} and \ref{sec_explore},
respectively.

\subsection{Design \& Performance Spaces} \label{sec_prelimdefs}

\begin{figure*}[ht!]
	\centering
	\includegraphics[width=0.74\textwidth]{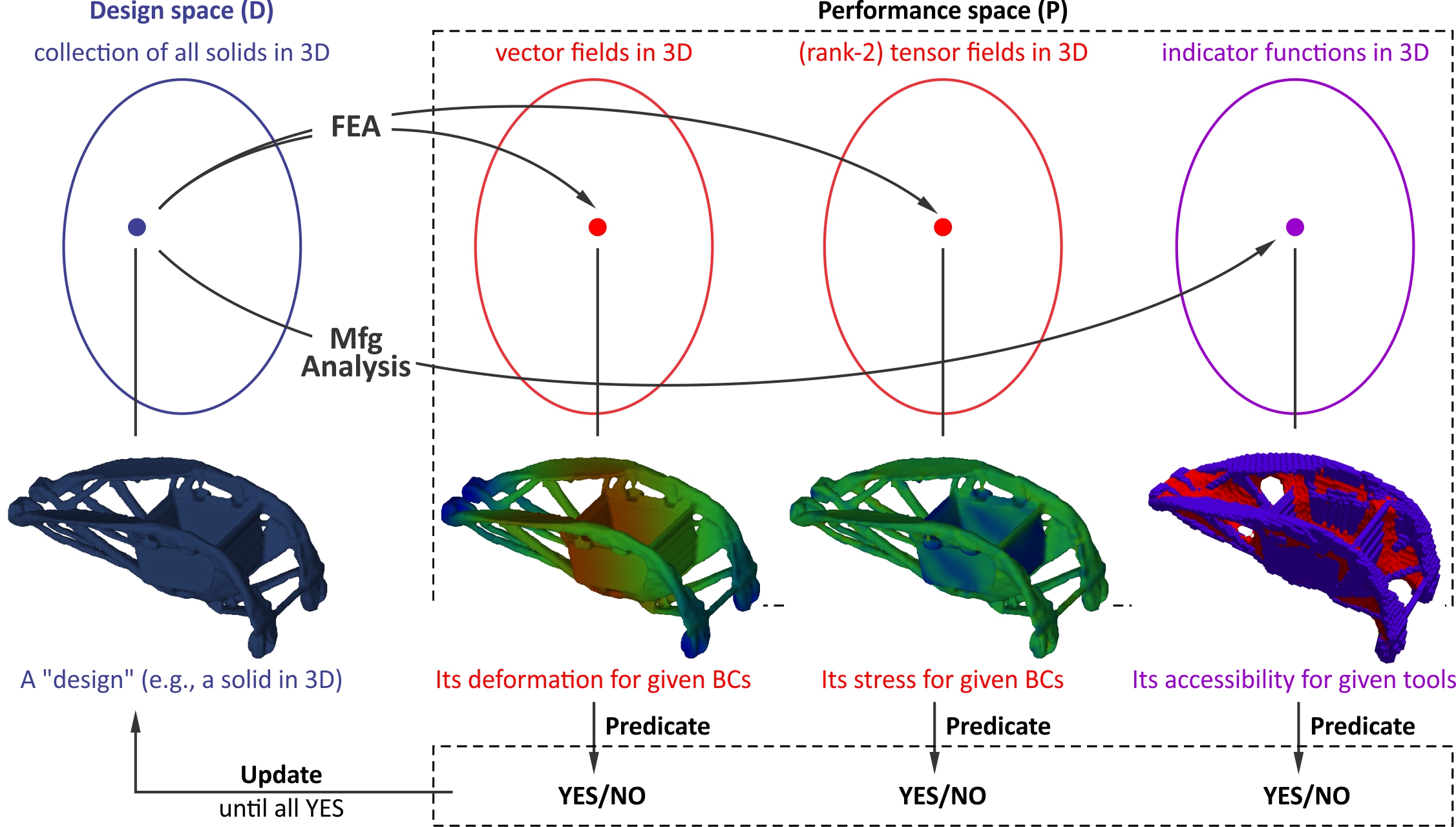}
	\caption{(a) FP-solvers (e.g., FEA and manufacturability analysis) map an
	instance of the design space (i.e., a ``design'') to an instance of the
	performance space (i.e., a ``field-tuple''). Predicates are defined to decide
	whether the design is satisfactory with respect to {\it constraints} in terms
	of performance variables. An IP-solver modifies the design until all
	constraints are satisfied.} \label{fig_multimaps}
\end{figure*}

Throughout this paper, a `design' $\Omega$ refers to a computational model of a
{\it single} designed artifact. We begin the process by specifying a bounded
{\it design domain} $\Omega_0 \subset \R^\dimm$ ($\dimm = 2$ or  $3$) whose
corresponding {\it design space} $\dspace$ is the collection of all `solids',
i.e., closed-regular semianalytic pointsets in $\dimm-$space,%
\footnote{A closed-regular set is defined as a set that equals the closure of
	its interior (i.e., is homogeneously 3D) \com{Semianalyticity rules out other
		forms of pathologies. Both are necessary to ensure that every design is a
		physically realizable 3D object } \cite{Requicha1980representations}.}
contained within the design domain. We use the notation $\dspace =
\powerset^\ast(\Omega_0)$ (read: ``solid powerset'' of $\Omega_0$).
This definition is sufficient for geometric modeling of parts with homogeneous
and isotropic material properties. We initially restrict our attention to this
subclass of artifacts, as considered in classical solid modeling
\cite{Requicha1980representations}. Our goal is to first illustrate nontrivial
challenges in designing with multiple solvers before introducing the additional
complexity of multiple materials, heterogeneity, or anisotropy. However, the
concepts introduced hereafter can  (in principle) be generalized beyond solids.

The `performance' of a given design $\Omega \in \dspace$ is an $n-$tuple
$\analysis(\Omega) := (\analysis_1(\Omega), \ldots, \analysis_n(\Omega))$. Think
of the {\it performance space} as a product space $\pspace := (\fspace_1 \times
\ldots \times \fspace_n)$, where each $\fspace_i$ is a class of fields, i.e.,
each $\analysis_i \in \fspace_1$ is an integrable field $\analysis_i(\Omega):
\Omega_0 \to \range_i$ over the design domain $\Omega_0$ whose value at a given
``query point'' $\bx \in \Omega_0$ is denoted by $\analysis_i(\bx; \Omega) :=
(\analysis_i(\Omega))(\bx) \in \range_i$. Examples of such fields are:
\begin{itemize}
	\item binary-valued fields ($\range_i := \{0, 1\}$), used to describe indicator
	functions of regions of interest within the design domain such as
	non-manufacturable features \cite{Nelaturi2015automatic,Behandish2018turning}
	or regions requiring design correction \cite{Nelaturi2015manufacturability}.
	\item integer-valued scalar fields ($\range_i := \Z$), used to characterize
	local topological properties of 3D printed parts
	\cite{Behandish2018characterizing,Behandish2019detection} or to classify atomic
	units of manufacturing in hybrid (combined AM and SM) manufacturing
	\cite{Behandish2018automated}.
	\item real-valued scalar fields ($\range_i := \R$), 3D vector fields ($\range_i
	:= \R^3$), and higher-rank tensor fields used to represent distributed physical
	quantities such as displacement, velocity, stress, strain, and so on
	\cite{Lai2009introduction}, or manufacturability measures
	\cite{Nelaturi2015representation}.
\end{itemize}

\subsubsection{Forward Problem in Physics \& Kinematics} \label{sec_forward}

Forward problem solvers (FP-solvers) map a given design instance to one or more
performance fields, hence can be viewed as implementations of one or more maps
$\analysis_i: \dspace \to \pspace_i$. The entire forward problem, solved by one
or more FP-solvers, can be viewed as a single map from design space to
performance space $\analysis: \dspace \to \pspace$, which has a unique outcome
for a given design.

For example, consider a finite elements analysis (FEA) FP-solver that computes
(discretized forms of) a displacement field $\bu_\Omega := \analysis_1(\Omega)$
for small deformations \cite{Lai2009introduction} of a given design $\Omega \in
\dspace$ due to boundary conditions such as restraints and external forces (Fig.
\ref{fig_multimaps}). It is also important to compute the stress field
$\sigma_\Omega := \analysis_2(\Omega)$, which depends locally on displacement
and material properties (e.g., the linear elasticity law). The vector/tensor
values of solution fields probed at a query point $\bx \in \Omega_0$ are denoted
by $\bu_\Omega(\bx) = \analysis_1(\bx; \Omega)$ and $\sigma_\Omega(\bx) =
\analysis_2(\bx; \Omega)$. In this case, both functions are zero outside the
design, i.e., $\analysis_{1,2}(\bx; \Omega) = 0$ if $\bx \in (\Omega_0 -
\Omega)$. FEA solves the weak form of the governing differential equation,
discretized into a linear system (e.g., using hat functions) $[K_\Omega]
[\bu_\Omega] = [\bff]$, where the stiffness matrix $[K_\Omega]$ and external
load vector $[\bff]$ depend on the design shape and material properties, as well
as boundary conditions. The equations are solved to obtain the discrete form of
the displacement field $[\bu_\Omega]$ from which the discrete form of the stress
field $[\sigma_\Omega]$ is computed by linear operations.

Another example is accessibility analysis for machining (e.g., milling or
turning) \cite{Nelaturi2015automatic,Behandish2018automated}. For instance,
consider an FP-solver for $3-$axis milling simulation, which computes
(discretized forms of) a volumetric measure of inaccessibility as a field
$\mu_\Omega(\bx) := \analysis_3(\Omega)$ for a given design $\Omega \in \dspace$
and machine tool parameters. This measure at a query point $\bx \in \Omega_0$,
denoted by $\mu_\Omega(\bx) = \analysis_3(\bx; \Omega)$, returns the penetration
volume of the moving tool assembly $T = (H \cup C)$, including the holder $H$
and cutter $C$, into the stationary obstacles $O_\Omega = (\Omega \cup F)$,
including target form $\Omega$ and fixtures $F$. The solver computes the
discrete form of this field (e.g., sampled at point clouds or voxels)
$[\mu_\Omega]$ as a convolution between the discrete forms of the indicator
functions of stationary solids $[\indic_{O_\Omega}]$ and moving solids
$[\indic_T]$ using a fast Fourier transform (FFT) \cite{Kavraki1995computation}.
The maximal set of accessible configurations -- in this case, pure translations
in 3D for a fixed orientation -- is then obtained as the null set $M_\Omega =
\mu_\Omega^{-1}(0)$, i.e., the translations that do not lead to undesirable
collisions. The maximal removable volume is obtained by sweeping the cutter with
the maximal motion, i.e., $R_\Omega := \sweep(M_\Omega, C)$. Its indicator
function $\indic_{R_\Omega}(\bx) = \analysis_4(\bx; \Omega)$ can be viewed as a
predicate for accessibility, i.e., it returns 1 (resp. 0) if the query point
$\bx \in \Omega_0$ is (resp. is not) accessible. The discrete form of this
binary field $[\indic_{R_\Omega}]$ can also be obtained by thresholding
FFT-based convolution of discrete forms $[\indic_{M_\Omega}]$ and $[\indic_C]$.

The computations performed by the above two FP-solvers (FEA and accessibility
analysis) are abstracted by $\analysis := (\analysis_1, \analysis_2,
\analysis_3, \analysis_4)$ that maps a given design to a ``field tuple'' that
represents analysis results. Figure \ref{fig_multimaps} illustrates one instance
of each such field for a topologically optimized bracket---solution of a DARPA
design challenge problem for TRAnsformative DESign (TRADES) program.

\subsubsection{Feasible Design Subspaces \& Predicates}

Inverse problem solvers (IP-solvers), on the other hand, find one or more
designs that satisfy a given collection of functional requirements. Most
IP-solvers employ an iterative process to:
\begin{enumerate}
	\item generate one or more valid candidate design(s);
	\item perform analysis on the candidate design(s) to compute the performance(s)
	of interest (using one or more FP-solver(s));
	\item evaluate the performance(s) against given functional requirements; and,
	\item if the requirements are not met, decide on the next generation of
	candidate design(s) based on the current evaluation and update rules.
\end{enumerate}
The process is repeated until the requirements are met. The evaluation process
(item 3) can be conceptualized as finite number of predicates defined over the
performance space as $\test_i: \dspace \to \{0, 1\}$ for $i = 1, 2, \ldots, n$.
Each predicate's outcome $\test_i (\Omega) \in \{0, 1\}$ is determined by means
of a {\it constraint} imposed on the performance field $\analysis_i(\Omega) \in
\fspace_i$ simulated by an FP-solver:%
\footnote{Without loss of generality, we assume that every requirement depends
	on one performance field only. If more than one field is used in calculating a
	predicate, we tuple those fields into another field. If more than one
	requirement is computed on one field, we think of it as two copies of the same
	field.}
\begin{equation}
	\test_i(\Omega) :=
	\begin{dcases}
	1 &  \text{if the $i$\th constraint is satisfied}, \\
	0 &  \text{otherwise}.
	\end{dcases} \label{eq_g}
\end{equation}
These can be (in)equality constraints (Sections \ref{sec_ineqcons}), which are
common in physics-based design formulations as in TO, and set constraints
(Section \ref{sec_setcons}), which are ubiquitous in design under
kinematics-based constraints such as packaging, assembly, and accessibility for
manufacturing.

We can think of the performance criteria evaluation as a map $\test := (\test_1,
\test_2, \ldots, \test_n) : \dspace \to \{0, 1\}^n$ i.e., $\test(\Omega)$ is a
binary string whose bits indicate whether a given design satisfies each of the
criteria (Fig. \ref{fig_multimaps}). A design $\Omega \in \dspace$ is called
`feasible' if it simultaneously satisfies all criteria, i.e., $\test(\Omega) =
(1, 1, \ldots, 1)$. The {\it feasible design subspace} $\feaspace \subseteq
\dspace$ is the subset of all feasible designs, defined as:
\begin{equation}
	\feaspace := \big\{ \Omega \in \dspace ~|~ \test(\Omega) = 1^n \big\} =:
	\test^{-1} (1). \label{eq_fspace}
\end{equation}
Here, $(\cdot)^{-1}$ denotes inversion of a mathematical relation.%
\footnote{Given $f: X \to Y$, we define $f^{-1}(y) = \{x \in X ~|~ f(x) = y\}$.
	Note that in general, $f^{-1}(y) \subseteq X$ is a set, i.e., $f^{-1}: Y \to
	\powerset(X)$. It is a singleton set (i.e., has one element) if the function is
	bijective. \label{refnote}}

Unlike forward problems, inverse problems have non-unique, often infinitely
many, solutions (i.e., $|\feaspace|  > 1$). The feasible design space can also
be defined as the intersection of the {\it feasibility halfspaces} $\halfspace_i
:= \test_i^{-1} (1)$, each implicitly describing one of the design subspaces
that satisfy one criterion at-a-time:
\begin{equation}
	\feaspace = \bigcap_{1 \leq i \leq n} \test_i^{-1} (1) = \dspace - \bigcup_{1
	\leq i \leq n} \test_i^{-1} (0). \label{eq_fspace_alt}
\end{equation}
The idea of design space pruning (Section \ref{sec_prune}) is to progressively
cut out portions of the design space that violate any one of the criteria.
Theoretically, pruning can be done in an arbitrary order---noting that
intersections or unions of the halfspace in \eq{eq_fspace_alt} are permutative.
Computationally, however, it is only possible if the design subspaces
$\halfspace_i = \test_i^{-1} (1)$ can be manipulated by algorithms as
first-class entities. Our goal is to understand under what conditions such
manipulations are possible and how an entire design subspace can be represented.
The existence (Proposition \ref{prop_exists}) and completeness (Proposition
\ref{prop_complete}) of such representations are discussed in Section
\ref{sec_prune}.

\subsection{Global, Local, \& Strictly Local Inequality Constraints} \label{sec_ineqcons}

The predicates introduced earlier are implemented in practice by testing whether
a candidate design's performance satisfies an (in)equality constraint.%
\footnote{Note that every equality constraint $\gglob(\cdot) = 0$ can be
	represented by two inequality constraints $\gglob(\cdot) \leq 0$ and
	$-\gglob(\cdot) \leq 0$.}
We classify such constraints into three types; namely, global, local, and
strictly local (in)equality constraints.

\subsubsection{Global Inequality Constraints} \label{sec_glob_ineq}

It is common to have design criteria specified in terms of global constraints
$\gglob_i(\Omega) \leq 0$, i.e., by defining a predicate of the general form:
\begin{equation}
	\test_i(\Omega) :=
	\begin{dcases}
		1 &  \text{if}~ \gglob_i (\Omega) \leq 0, \\
		0 &  \text{otherwise}.
	\end{dcases} \label{eq_glob}
\end{equation}
in which $\gglob_i: \dspace \to \R$ is a function of the {\it entire} shape of
the design $\Omega \in \dspace$, potentially in addition to fixed external
factors such as boundary conditions, manufacturing process parameters, packaging
envelope, operating conditions (e.g., motion in assembly), etc. The constraint
is often evaluated in terms of a global property of an entire performance field
$\analysis_i(\Omega) \in \fspace_i$, e.g., as an upper/lower-bound on its
maximum/minimum or its integral properties such as $p-$norms. This is denoted by
$\gglob_i(\Omega) = \bar{\gglob}_i(\analysis_i(\Omega))$ where $\bar{\gglob}_i:
\dspace \to \R$. For example, one can constrain the maximal displacement or
maximal stress of a solid under external loads (Fig. \ref{fig_multimaps}) by
using the following constraints in \eq{eq_glob}:
\begin{align}
	\gglob_1(\Omega) = \bar{\gglob}_1(\bu_\Omega) &:= \max_{\bx \in \Omega}
	\|\bu_\Omega(\bx)\| - \updelta_\mathsf{UB}, \label{eq_max_delta} \\
	\gglob_2(\Omega) = \bar{\gglob}_2(\sigma_\Omega) &:= \max_{\bx \in \Omega} \|
	\sigma_\Omega(\bx) \| - \sigma_\mathsf{UB}, \label{eq_max_sigma}
\end{align}
where $\updelta_\mathsf{UB}, \sigma_\mathsf{UB} > 0$ are constant upper-bounds
on the magnitude of the displacement vector and stress tensor, captured by the
constraints $\gglob_1(\Omega) \leq 0$ and $\gglob_2(\Omega) \leq 0$,
respectively. One can in general use the $p-$norm of the fields for finite (but
large) $p \geq 1$, noting that maximum is a special case as $p \to \infty$. This
is especially useful to smooth out the possible singularities (e.g., infinite
stress, due to stress concentrations).

Here is another example to constrain a design to be manufacturable via
machining, using the accessibility analysis mentioned earlier:
\begin{equation}
	\gglob_3(\Omega) = \bar{\gglob}_3(\indic_{R_\Omega}) := \int_{\Omega_0 -
	\Omega} \!\!\!\!\!\!\!\!\! \neg \indic_{R_\Omega} (\bx) ~dv[\bx] -
	V_\mathsf{UB}, \label{eq_max_vol}
\end{equation}
where $\neg \indic_{R_\Omega}(\bx) = 1 - \indic_{R_\Omega}(\bx)$ is a negation.
This constraint restricts the total volume of inaccessible regions $R_\Omega
\subseteq \Omega^c$, obtained as the $1-$norm of their indicator function, by an
upper-bound $V_\mathsf{UB} > 0$.

\subsubsection{Local Inequality Constraints} \label{sec_gloc_ineq}

It is sometimes possible to define a predicate in terms of local constraints
evaluated at a specific point in the design domain; for instance, using one of
the following forms:
\begin{align}
	\test_i(\Omega) &:=
	\begin{dcases}
		1 &  \text{if}~ \forall \bx \in \Omega_0 : \gloc_i (\bx; \Omega) \leq 0,\\
		0 &  \text{otherwise}.
	\end{dcases} \label{eq_gloc_forall} \\
	\test_i(\Omega) &:=
	\begin{dcases}
	1 &  \text{if}~ \exists \bx \in \Omega_0 : \gloc_i (\bx; \Omega) \leq 0,\\
	0 &  \text{otherwise}.
	\end{dcases} \label{eq_gloc_exists}
\end{align}
Note that the two alternative forms are different by the ``for all'' and ``there
exists'' quantifiers, which lead to different global implications. Unlike the
case with \eq{eq_glob}, here $\gloc_i: (\Omega_0 \times \dspace) \to \R$ is
field for a fixed design $\Omega \in \dspace$, i.e., is also a function of the
query point $\bx \in \Omega_0$. In turn, the constraint is evaluated based on
the probed {\it value} of the performance field $\analysis_i(\bx; \Omega) \in
\range_i$ (generally, a tensor) at the query point. We denote this dependency by
$\gloc_i(\bx; \Omega) = \bar{\gloc}_i(\analysis_i(\bx; \Omega))$ where
$\bar{\gloc}_i: \range_i \to \R$. For example, the global displacement and
stress bounds mentioned earlier can be imposed locally (Fig.
\ref{fig_multimaps}):
\begin{align}
	\gloc_1(\bx; \Omega) = \bar{\gloc}_1(\bu_\Omega(\bx)) &:= \|\bu_\Omega(\bx)\| -
	\updelta_\mathsf{UB}, \label{eq_max_delta_local} \\
	\gloc_2(\bx; \Omega) = \bar{\gloc}_2(\sigma_\Omega(\bx)) &:=
	\|\sigma_\Omega(\bx)\| - \sigma_\mathsf{UB}, \label{eq_max_sigma_local}
\end{align}
It is easy to verify that using \eq{eq_max_delta_local} and
\eq{eq_max_sigma_local} with \eq{eq_gloc_forall} is equivalent to using
\eq{eq_max_delta} and \eq{eq_max_sigma} with \eq{eq_glob} in this example. In
general, local constraints $\gloc_i(\bx; \Omega) \leq 0$ used with ``for all''
or ``there exists'' quantifiers in \eq{eq_gloc_forall} and \eq{eq_gloc_exists}
can be equivalently expressed as global constraints (stated independently of
$\bx \in \Omega_0$) via min/max, respectively:
\begin{align}
	\Big[ \forall \bx \in \Omega_0 : \gloc_i(\bx; \Omega) \leq 0 \Big] &~\rightleftharpoons~  \max_{\bx \in \Omega_0} \gloc_i(\bx; \Omega) \leq 0, \label{eq_equiv_max} \\
	\Big[ \exists \bx \in \Omega_0 : \gloc_i(\bx; \Omega) \leq 0 \Big] &~\rightleftharpoons~  \min_{\bx \in \Omega_0} \gloc_i(\bx; \Omega) \leq 0. \label{eq_equiv_min}
\end{align}

As another example, consider the accessibility analysis discussed earlier.
Instead of constraining the total volume of inaccessible regions via the global
constraint of \eq{eq_max_vol}, we can locally constrain the inaccessibility
measure as:
\begin{equation}
	\gloc_4(\bx; \Omega) = \bar{\gloc}_4(\mu_\Omega(\bx)) := (\indic_{O_\Omega} \ast {\indic}_{-T})(\bx) - \mu_0, \label{eq_cons_mu}
\end{equation}
where $O_\Omega = (\Omega \cup F)$ and $T = (H \cup C)$ are the stationary and
moving solids, respectively. $\mu_0 > 0$ is a small constant to provide
allowance for numerical errors. The convolution operator $\ast$ is defined as:
\begin{equation}
	(\indic_{O_\Omega} \ast {\indic}_{-T})(\bx) = \int_{\Omega_0} \indic_\Omega(\bx') {\indic}_{-T}(\bx - \bx') ~d v[\bx'], \label{eq_access_conv}
\end{equation}
where $\indic_{-T}(\bx) = \indic_T(-\bx)$ is a reflection with respect to the
origin, hence ${\indic}_{-T}(\bx - \bx') = \indic_{T}(\bx' - \bx)$ is the
indicator function of the moving object (i.e., tool assembly), translated to the
query point $\bx \in \Omega_0$. The integral is nonzero at integration points
$\bx' \in \Omega_0$ that belongs to the interference of the translated object
with the stationary obstacles.

It is not always possible to convert global constraints to local constraints or
vice versa, without defining new performance variables, e.g., in terms of the
norms of existing performance fields. We show in the next section that when
$\analysis_i(\bx; \Omega)$ is decidable independently of $\Omega \in \dspace$,
the above two constraints lead to maximal/minimal feasible designs (in
set-theoretic terms).

\subsubsection{Strictly Local Inequality Constraints}

An important special case of \eq{eq_gloc_forall} occurs if the predicate is {\it
	decidable without a priori knowledge of the design itself}. In other words, the
constraints can be evaluated purely from a knowledge of the query point's
position $\bx \in \Omega$ and external factors, if any (e.g., a known rigid body
motion applied to the entire design). The predicate's result can be obtained
without knowing the overall shape of the design. This is the case if
$\analysis_i(\bx; \Omega) = \analysis_i^\ast(\bx)$, i.e., the forward problem's
solution can be evaluated pointwise---emphasized by the overline notation. The
corresponding constraint $\gsloc_i(\bx) := \bar{\gloc}_i(\analysis_i^\ast(\bx))
\leq 0$ is hereafter called a {\it strictly local} (i.e., pointwise) constraint.
The predicates in \eq{eq_gloc_forall} or \eq{eq_gloc_exists} in this case are
dependent on $\Omega$ only due to the logical quantifiers for the pointwise
testing:
\begin{align}
	\test_i(\Omega) &:=
	\begin{dcases}
		1 &  \text{if}~ \forall \bx \in \Omega : \gsloc_i (\bx) \leq 0,\\
		0 &  \text{otherwise}.
	\end{dcases} \label{eq_gsloc_forall} 
	\\
	\test_i(\Omega) &:=
	\begin{dcases}
		1 &  \text{if}~ \exists \bx \in \Omega : \gsloc_i (\bx) \leq 0,\\
		0 &  \text{otherwise}.
	\end{dcases} \label{eq_gsloc_exists}
\end{align}
Hence, one can define {\it pointwise predicates} in this case:
\begin{equation}
	\test^\ast_i(\bx) :=
	\begin{dcases}
	1 &  \text{if}~ \gsloc_i(\bx) \leq 0,\\
	0 &  \text{otherwise}.
	\end{dcases} \label{eq_gsloc_pred} 
\end{equation}
We show in Section \ref{sec_prune} that the pointwise predicate defines a point
membership classification (PMC) that implicitly determines the entire
feasibility halfspace $\halfspace_i = \test_i^{-1}(1)$ using its
``representative'' maximal/minimal feasible design $\Omega_i^\ast :=
{\test^\ast_i}^{-1}(1)$ using \eq{eq_gsloc_forall} or \eq{eq_gsloc_exists},
respectively. In this paper, we restrict our attention to the former (``for
all'' quantifier and maximal designs), since it lends itself to material
reducing downstream solvers such as TO.

The physics-based constraints exemplified earlier might reduce to pointwise
constraints in rare examples---e.g., the stress tensor $\sigma^\ast(\bx)$ for
hydrostatic pressure in a liquid container at rest depends on the query point's
depth from the surface, but not on the container's designed shape. Nevertheless,
pointwise constraints are ubiquitous in kinematics-based constraints that are
central to applications ranging from assembly and packaging to manufacturing. In
the next section, we present an important subclass of pointwise constraints that
manifest as {\it set constraints}.

\subsection{Converting Set Constraints to Inequality Constraints} \label{sec_setcons}

Many kinematics-based design criteria lead to constraints expressed in the
algebra of sets. A common form of set constraints is in terms of {\it
	containment}: $\Gamma(\Omega) \subseteq E$ (for a fixed envelope $E \subseteq
\R^\dimm$). The exact same constraint can be written in terms of {\it
	non-interference}: $(\Gamma(\Omega) \cap O) = \emptyset$ (for a fixed obstacle
$O \subseteq \R^\dimm$), where $\Gamma: \dspace \to \powerset(\R^\dimm)$ is a
{\it set transformation} and $E := O^c$ (i.e., complement of $O$).

At a first glance, these constraints appear to have a completely different form
than the inequality constraints described in Section \ref{sec_ineqcons} in the
algebra of fields. Here, we show that set constraints can {\it always} be
reformulated as (global or local) inequality constraints by virtue of describing
sets with their indicator functions. However, converting them to strictly local
(i.e., pointwise) constraints is possible under certain conditions.

\subsubsection{Global or Local Set Constraints}

For every solid $\Omega \in \dspace$, its indicator (i.e., characteristic)
function $\indic_\Omega: \Omega_0 \to \{0, 1\}$ is defined as:
\begin{equation}
	\indic_\Omega (\bx) :=
	\begin{dcases}
		1 &  \text{if}~ \bx \in \Omega, \\
		0 &  \text{otherwise}.
	\end{dcases}, ~\text{i.e.,}~ \Omega = \indic_\Omega^{-1}(1).
\end{equation}
Hence, every containment constraint is restated as an inequality constraint of
the form used in \eq{eq_gloc_forall}:
\begin{equation}
	\Gamma(\Omega) \subseteq E ~\rightleftharpoons~ \Big[ \forall \bx \in \R^\dimm
	: \indic_{\Gamma(\Omega)} (\bx) \leq \indic_E(\bx) \Big], \label{eq_indic_cons}
\end{equation}
i.e., using the standard form $\indic_{\Gamma(\Omega)} (\bx) - \indic_E(\bx)
\leq 0$.

Recall from Section \ref{sec_ineqcons} that the above inequality constraint can
also be rewritten as a global constraint of the form used in \eq{eq_glob} by
upper-bounding the maximum:
\begin{equation}
	\Gamma(\Omega) \subseteq E ~\rightleftharpoons~ \max_{\bx \in \R^\dimm} \big[
	\indic_{\Gamma(\Omega)} (\bx) - \indic_E(\bx) \big] \leq 0,
	\label{eq_indic_cons_max}
\end{equation}

Notice that we have not made any assumption on the properties of the pointsets
$\Gamma(\Omega), E \subseteq \R^\dimm$. In most practical scenarios, both are
solids within a bounded domain, taken as the design domain $\Omega_0 \in
\dspace$. In such cases, the properties of the mapping and envelope can be
exploited to compute the maximum in \eq{eq_indic_cons_max} efficiently.

Moreover, if the set constraint $(\Gamma(\Omega) \cap E^c) = \emptyset$ is in
terms of regularized intersection, it can be rewritten as a global (in)equality
constraint in terms of the volume $\vol(\Gamma(\Omega) \cap E^c] = 0$ (or $\leq
0$, for consistency), which, in turn, is an inner product of indicator
functions:
\begin{align}
	\vol[\Gamma(\Omega) \cap E^c] &= \langle \indic_{\Gamma(\Omega)}, \neg \indic_E
	\rangle \\ &= \int_{\R^\dimm} \neg \indic_{\Gamma(\Omega)}(\bx) \indic_E
	~dv[\bx]. \label{eq_inner}
\end{align}
Further, if $\Gamma(\Omega)$ is a rigid transformation of $\Omega$, the inner
product turns into a {\it convolution} of $\indic_{\Omega}$ and $\neg \indic_E$
over the configuration space of motions \cite{Lysenko2010group}.

Let us consider the inaccessibility analysis one more time. For manufacturing
with precision requirements (e.g., for assembly/fit), we can restrict the
inaccessible regions $\Gamma(\Omega) := (\Omega_0 - \Omega) - R_\Omega$ to be
completely contained within a tolerance zone $E \subseteq \R^\dimm$, hence
formulating the problem as a set constraint $\Gamma(\Omega) \subseteq E$, where:
\begin{equation}
	\Gamma(\Omega) = (\Omega_0 - \Omega) - \sweep(M_\Omega, C).
	\label{eq_mfg_Mink_1}
\end{equation}
For translational motions $M_\Omega \subseteq \R^\dimm$, the sweep in
\eq{eq_mfg_Mink_1} can be simplified into a Minkowski sum:
\begin{equation}
	\Gamma(\Omega) = (\Omega_0 - \Omega) - (M_\Omega \oplus C).
	\label{eq_mfg_Mink_2}
\end{equation}
The maximal collision-free motion \cite{Nelaturi2015automatic} is obtained as
the complement of the $\conf-$space obstacle, which is in turn computed by
Minkowski operations (for translational full-dimensional motions):%
\footnote{The operators $(\cdot)^c, \cup, \cap, -, \oplus, \ominus$ are all
	regularized to ensure their algebraic closure within the design space $\dspace$
	\cite{Tilove1980closure}. The caveat is that collision-free lower-dimensional
	motions (e.g., of a peg inside a hole with no clearance) would be pruned from
	collision-free motions.}
\begin{align}
	\Gamma(\Omega) &= (\Omega_0 - \Omega) - \big( (\Omega~ \oplus (-T))^c \oplus C
	\big) \label{eq_mfg_Mink_3} \\
	&= (\Omega_0 - \Omega) - \big( (\Omega^c \ominus (-T))~ \oplus C \big),
	\label{eq_mfg_Mink_4}
\end{align}
where $T = (H \cup C)$ represents the tool assembly, including the holder $H$
and the cutter $C$. It is understood that Minkowski sums in \eq{eq_mfg_Mink_3}
can be obtained from the 0-superlevel set of the convolution fields of the
participating sets \cite{Lysenko2010group}.%
\footnote{In general, $Y = (X_1 \oplus X_2) ~\rightleftharpoons~ \indic_{Y} =
	\sign(\indic_{X_1} \ast \indic_{X_2})$ \cite{Lysenko2010group}.}
More precisely:
\begin{equation}
	\indic_{\Gamma(\Omega)} = \neg \indic_{\Omega} - \sign \left( \neg \sign
	(\indic_\Omega \ast {\indic}_{-T}) \ast \indic_C \right), \label{eq_print_conv}
\end{equation}
where $\ast$ is the convolution operator in $\R^\dimm$ and $\sign (x) = 1$
(resp. 0) if $x > 0$ (resp. $x \leq 0$) is the sign function. The latter is
needed to convert the real-valued convolutions to binary-valued indicator
functions. See \cite{Lysenko2010group} for a proof.

In summary, set constraints, as defined in this section in terms of containment
or non-interference, can {\it always} be restated as global or local inequality
constraints defined in Section \ref{sec_ineqcons}. In the next section, we show
the conditions under which set constraints can be converted to strictly local
(i.e., pointwise) constraints, to enable design space pruning in Section
\ref{sec_prune}.

\subsubsection{Set Constraints as Pointwise Constraints} \label{sec_gloc_set}

Depending on the properties of $\Gamma: \dspace \to \powerset(\R^\dimm)$, the
inequality constraint $\indic_{\Gamma(\Omega)}(\bx) - \indic_E(\bx) \leq 0$,
used in global or local forms of \eq{eq_indic_cons} and \eq{eq_indic_cons_max},
respectively, may or may not be restated as a strictly local (i.e., pointwise)
constraint. Our goal in this section is to articulate the conditions under which
this is possible.

To enable pointwise formulation, we need to eliminate the dependency of the PMC
for $\Gamma(\Omega)$ on $\Omega$ so that $\indic_{\Gamma(\Omega)}(\bx)$ on the
left-hand side of the inequality constraint in \eq{eq_indic_cons} depends only
on the query point $\bx \in \Omega_0$ and the fixed envelope $E \subseteq
\R^\dimm$. This can be done if the set transformation $\Gamma$ is itself a
pointwise transformation, meaning that it can be defined by {\it extending} a
transformation of 3D points $\gamma: (\R^\dimm ~\text{or}~ \Omega_0) \to
\powerset(\R^\dimm)$ to a transformation of 3D pointsets $\Gamma: \dspace \to
\powerset(\R^\dimm)$ by simply applying the former to every point of the
pointset and unifying the results:
\begin{equation}
	\Gamma(\Omega) := \bigcup_{\bx \in \Omega} \gamma(\bx) = \big\{ \bx' ~|~ \bx'
	\in \gamma(\bx), \bx \in \Omega \big\}. \label{eq_pointwise_trans}
\end{equation}
Note that $\gamma(\bx)$ is itself a pointset, not a point, to capture the most
general case. For example, it can be a curve segment or surface patch
representing the 1D or 2D trajectory of a point under a given one- or
two-parametric motion, respectively.

The above refactoring is possible for many important applications. For example,
if the design has to move (when deployed in assembly) according to a known
motion set $M \subseteq \SE{3}$ without exiting a safe region of space $E
\subseteq \R^\dimm$ that contains no obstacles \cite{Ilies2000shaping}, the
above constraint can be used with $\Gamma(\Omega) := \sweep(M, \Omega)$, where
\begin{equation}
	\sweep(M, \Omega) = \bigcup_{\tau \in M} \tau\Omega = \big\{ \tau\bx ~|~ \tau
	\in M, \bx \in \Omega \big\}, \label{eq_def_sweep}
\end{equation}
is the sweep of the designed part as it travels by the given motion (known a
priori). In this case the sweep indicator function $\indic_{\Gamma(\Omega)}$ can
be directly obtained as follows:
\begin{equation}
	\indic_{\Gamma(\Omega)}(\bx) = 1 ~\text{iff}~ \exists \tau \in M : \bx \in \tau
	\Omega, ~\text{i.e.,}~ \tau^{-1}\bx \in \Omega,
\end{equation}
In other words, a PMC test for $\Gamma(\Omega)$ can be obtained by applying the
inverse motion $M^{-1} = \{ \tau^{-1} ~|~ \tau \in M \}$ to the query point and
checking if it intersects the design. The inequality constraint in
\eq{eq_indic_cons_max} can thus be computed rapidly by sampling query points in
the design domain and testing intersections of their inverse trajectories with
the design \cite{Ilies2000shaping}. Importantly, the computation for one point
does {\it not} need one to have explicit knowledge of the results for other
points. Other than perfect parallelization (e.g., on GPU), this property enables
pruning the design space -- leading to development of the \textsf{Unsweep} solver
discussed earlier -- before optimizing the design for other criteria.

In general, for pointwise transformations as in \eq{eq_pointwise_trans}, the
global constraint $\Gamma(\Omega) \subseteq E$ can be restated as:
\begin{equation}
	\Gamma(\Omega) \subseteq E ~\rightleftharpoons~ \Big[ \forall \bx \in \Omega :
	\gamma(\bx) \subseteq E \Big], \label{eq_sloc_gamma_1}
\end{equation}
i.e., $\Omega$ remains within $E$ after a $\Gamma-$transform iff all points
inside it remain within $E$ after a $\gamma-$transform. Note also that
inequality constraint in \eq{eq_indic_cons} can now be rewritten in a pointwise
fashion every $\bx \in \Omega$:
\begin{equation}
	\forall \bx \in \Omega: \forall \bx' \in \R^\dimm : \indic_{\gamma(\bx)} (\bx')
	\leq \indic_E(\bx'), \label{eq_indic_cons_sloc}
\end{equation}

As with other inequality constraints, not every global or local set constraint
can be converted to a pointwise set constraint. For example, the toleranced
accessibility constraints $\Gamma(\Omega) \subseteq E$ for $\Gamma(\Omega) :=
(\Omega_0 - \Omega) - \sweep(M_\Omega, C)$ in \eq{eq_mfg_Mink_1} cannot be
evaluated pointwise, because the maximal collision-free motion $M_\Omega =
(\Omega~ \oplus (-T))^c$ depends on the global shape of $\Omega$, unlike the
case with the fixed motion in the earlier example with \textsf{Unsweep}.

\subsection{Design for Heterogeneous Constraints} \label{sec_phases}

We conclude this section by a fairly general formulation of a design problem
subject to $n \geq 1$ heterogeneous (e.g., kinematics- and physics-based)
constraints.

Without loss of generality, let $n_\mathrm{C} = n_\mathrm{G} + n_\mathrm{L} +
n_\mathrm{P}$ where $0 \leq n_\mathrm{G}, n_\mathrm{L}, n_\mathrm{P} \leq
n_\mathrm{C}$ are the number of global, local, and strictly local (i.e.,
pointwise) constraints, respectively. All constraints, including set
constraints, are expressed as inequality constraints for uniformity. The {\it
	design problem} is to identify the feasible design space $\dspace^\ast =
\test^{-1}(1)$, stated as a constraint satisfaction problem:
\begin{align}
	&\text{Find}~ \dspace^\ast \subseteq \dspace, ~\text{such that for all}~ \bx
\in \Omega \in \dspace^\ast: \nonumber \\
	& 
	\begin{array}{ll}
		\gsloc_i(\bx) \quad \leq 0, &\text{for}~ 0 < i \leq n_\mathrm{P}, \\
		\gglob_i(\Omega) \quad \leq 0, &\text{for}~ n_\mathrm{P} < i \leq n_\mathrm{P}+n_\mathrm{G}, \\
		\gloc_i(\bx; \Omega) \leq 0, &\text{for}~ n_\mathrm{P}+n_\mathrm{G} < i \leq n_\mathrm{C}, \\
	\end{array}
\end{align}
We assume that none of the $\gloc_i(\bx; \Omega) \leq 0$ can be simplified into
one of $\gsloc_i(\bx) \leq 0$ or $\gglob_i(\Omega) \leq 0$ forms. Hereafter, we
use the P-, G-, and L-subscripts for various notions related to pointwise,
global, and local constraints, respectively; for instance, $\dspace^\ast =
\dspace^\ast_\mathrm{P} \cap \dspace^\ast_\mathrm{G} \cap
\dspace^\ast_\mathrm{L}$ where $\dspace^\ast_\mathrm{P} =
\test_\mathrm{P}^{-1}(1)$ is the design subspace that is feasible with respect
to pointwise constraint alone, and so on. We solve the design problem in two
phases, depicted in Fig. \ref{fig_workflows} (b):
	
\paragraph{\bf Phase 1}
Prune the design space from $\dspace$ to $\dspace^\ast_\mathrm{P} =
\test_\mathrm{P}^{-1}(1)$, i.e., solve the following (simpler) problem:
\begin{align}
	&\text{Find}~ \dspace^\ast_\mathrm{P} \subseteq \dspace, ~\text{such that for
	all}~ \bx \in \Omega \in \dspace^\ast_\mathrm{P}: \nonumber \\
	& \gsloc_i(\bx) \leq 0, \quad\text{for}~ 0 < i \leq n_\mathrm{P}.
	\label{eq_stage1}
\end{align}
In Section \ref{sec_prune}, we solve the above problem by computing a maximal
set $\Omega^\ast_\mathrm{P} := \max \dspace^\ast$ in the partial ordering of
designs via set containment.

Unfortunately, this is not possible for $\dspace^\ast_\mathrm{G}$ and
$\dspace^\ast_\mathrm{L}$. In most cases, one can at best generate a finite
sample of feasible designs that are superior in some way.

\paragraph{\bf Phase 2}
Explore the pruned design space $\dspace^\ast_\mathrm{P}$ to find a sample
$\dspace^\ast_\mathrm{dom} \subset \dspace^\ast$ of (locally)
``Pareto-dominant'' \cite{Koopmans1951analysis,Charnes1985foundations} designs
that satisfy the remaining constraints:
\begin{align}
	&\text{Find}~ \dspace^\ast_\mathrm{dom} \subset \dspace^\ast_\mathrm{P},
	~\text{such that for all}~ \bx \in \Omega \in \dspace^\ast_\mathrm{opt}:
	\nonumber \\
	& 
	\begin{array}{ll}
		\gglob_i(\Omega) \quad \leq 0, &\text{for}~ n_\mathrm{P} < i \leq
		n_\mathrm{P}+n_\mathrm{G}, \\
		\gloc_i(\bx; \Omega) \leq 0, &\text{for}~ n_\mathrm{P}+n_\mathrm{G} < i \leq
		n_\mathrm{C}, \\
	\end{array} \label{eq_stage2}
\end{align}
Pareto-dominance of $\Omega \in \dspace^\ast_\mathrm{dom}$ means that for some
neighborhood $\mathsf{N}(\Omega) \subseteq \dspace^\ast_\mathrm{P}$ and $\Omega
\in \mathsf{N}(\Omega)$ in the pruned design space,%
\footnote{Technically, $\mathsf{N}(\Omega)$ is an open set in the induced
	Hausdorff topology of $\dspace^\ast_\mathrm{P} = \powerset^\ast(\Omega^\ast)$
	(all solid subsets of the maximal element).}
within which no other design is superior to $\Omega$ with respect to all
objective functions $f_1, f_2, \ldots, f_{n_\mathrm{O}}: \dspace \to \R$. We can
pose this as a minimization problem:
\begin{equation}
	\text{Find}~ \Omega \in \dspace^\ast_\mathrm{P} ~\text{to}~
	\begin{dcases} 
		\text{minimize}~ f_j(\Omega), ~\text{for}~ 0 < j \leq n_\mathrm{O}, \\
		\text{subject to constraints in \eq{eq_stage2}}.
	\end{dcases} \label{eq:ParetoProblem}
\end{equation} 
The objective functions define another partial ordering over the pruned design
space, whose maximal elements we seek, i.e., $\dspace^\ast_\mathrm{dom} := \max
\dspace$.

In Section \ref{sec_explore}, we solve the above problem by iterative
optimization guided by TSF. The TSF is defined with respect to global objective
functions $f_j(\Omega)$ and global constraints $\gglob_i(\Omega) \leq 0$ and is
penalized/filtered using local constraints $\gloc_i(\bx; \Omega) \leq 0$. Since
global optimization ($\mathsf{N}(\Omega) := \dspace^\ast_\mathrm{P}$) is NP-hard
\cite{Garey2002computers}, we settle for local optimality.

%% file: pruning.tex
\section{Design Space Pruning} \label{sec_prune}

In this section, we present a methodology to solve the  {\bf phase 1} problem
formulated in \eq{eq_stage1} of Section \ref{sec_phases}. Using the terminology
and results of Section \ref{sec_classify}, we show how the design space can be
pruned with respect to pointwise constraints (including set constraints) without
premature optimization. We illustrate the process using examples from
kinematics-based constraints that are common in assembly, packaging, and
manufacturing.

\subsection{Existence \& Completeness of Maximal Designs}

The following results on the existence and uniqueness of maximal pointsets and
their informational completeness as a representation for entire feasible design
spaces are central to design space pruning.

\begin{prop} (Existence and Uniqueness)
	For every strictly local (i.e., pointwise) constraint $\gsloc_i(\bx) \leq 0$,
its feasibility halfspace $\halfspace_i$ has a maximal element $\Omega_i^\ast =
\max \halfspace_i$, defined implicitly by the following PMC test:
	\begin{align}
		\indic_{\Omega_i^\ast}(\bx) &:= 
		\begin{dcases}
		1 \quad &  \text{if}~ \gsloc_i (\bx) \leq 0,\\
		0 &  \text{otherwise},
		\end{dcases} \label{eq_pmc_max_1} \\
		\text{i.e.,}~ \Omega_i^\ast &:= \big\{\bx \in \Omega_0 ~|~ \gsloc_i (\bx) \leq
		0 \big\}. \label{eq_pmc_max_2}
	\end{align}
	The maximality is in terms of set containment, i.e., every satisfactory design
	is contained in the maximal element: $\Omega \in \halfspace_i \Rightarrow
	\Omega \subseteq \max \halfspace_i$. \label{prop_exists}
\end{prop} 

\begin{prop} (Completeness)
	For every strictly local (i.e., pointwise) constraint
	$\gsloc_i(\bar{\analysis}_i(\bx)) \leq 0$, its feasibility halfspace
	$\halfspace_i$ contains every solid $\Omega \subseteq \Omega_i^\ast = \max
	\halfspace_i$, i.e., every solid subset of the maximal element is also
	feasible: $\Omega \subseteq \max \halfspace_i \Rightarrow \Omega \in
	\halfspace_i$. \label{prop_complete}
\end{prop} 

In terms of predicates, the design subspace $\halfspace_i = \test_i^{-1}(1)$
(which satisfies \eq{eq_gsloc_forall}) can now be represented by a single design
$\Omega_i^\ast = \bar{\test}^{-1}_i(1)$ (whose {\it all points} satisfy
\eq{eq_gsloc_pred}). The maximal solid is thus a {\it complete} representation
of the feasibility halfspace as the collection of all of its solid subsets
denoted by $\halfspace_i = \powerset^\ast(\Omega^\ast_i)$.

Here is an intuitive but simplified reasoning:
\begin{enumerate}
	\item The set $\Omega^\ast_i$ defined by \eq{eq_pmc_max_1} or \eq{eq_pmc_max_2}
	contains all points that satisfy the constraints.
	\item Every solid subset $\Omega \subseteq \Omega_i^\ast$ of the maximal set
	satisfies the constraint, because all of its points satisfy the constraint
	independently of the global shape.
	\item Conversely, every feasible solid $\Omega \in \halfspace_i$ is the subset
	of $\Omega^\ast_i$, because it only includes points that satisfy this
	constraint independently of the global shape.
\end{enumerate}
Note that the constraint's independence of the shape of $\Omega$ is crucial for
this to hold. For global or local constraints with dependency on $\Omega$
itself, attempting to write a PMC similar to \eq{eq_pmc_max_1} leads to a
circular definition where the right-hand side depends on the set itself. For
example, the (global or local) constraints given in Sections \ref{sec_glob_ineq}
and \ref{sec_gloc_ineq} on FEA and printability analyses do not lead to maximal
elements because their constraints $\gloc_i(\analysis_i(\bx; \Omega)) \leq 0$
depend on particular design instances. It does make sense to define the maximal
set of a feasible space (e.g., using the set-builder definition in
\eq{eq_pmc_max_2}) in a way that it depends on a particular instance of that
space. On the other hand, the set constraints of Section \ref{sec_gloc_set},
such as containment under a prescribed motion, do give rise to maximal elements,
as we elaborate with examples in the following sections.

The above reasoning does not take topological regularization into
account---there is no reason for a maximal set obtained via \eq{eq_pmc_max_1} or
\eq{eq_pmc_max_2} to be a solid, hence it may not itself be a valid design.
However, we show that there always exists a valid maximal element obtained by
regularizing \eq{eq_pmc_max_1} or \eq{eq_pmc_max_2}. The correct definition is:
\begin{equation}
	\Omega_i^\ast := \mathsf{k} \mathsf{i}\big\{\bx \in \Omega_0 ~|~ \gsloc_i (\bx)
	\leq 0 \big\}. \label{eq_pmc_max_3}
\end{equation}
where $\mathsf{k}$ and $\mathsf{i}$ are the topological closure and interior operators, respectively.

\begin{prop} (Design Space Pruning) 
	Given a number of pointwise constraints $\gsloc_i(\bx) \leq 0$ for $i = 1, 2,
	\ldots, n_\mathrm{P}$, the feasible design space $\dspace^\ast$, defined by
	intersecting all feasibility halfspaces $\halfspace_i^\ast$, has a maximal
	element $\Omega^\ast_\mathrm{P} = \max \dspace^\ast_\mathrm{P}$ that satisfies
	the uniqueness and completeness properties, i.e., $\Omega \subseteq
	\dspace^\ast_\mathrm{P} \rightleftharpoons \Omega \in \Omega^\ast_\mathrm{P}$.
	It can be obtained by intersecting all maximal elements $\Omega^\ast_i = \max
	\halfspace_i$:
	\begin{equation}
		\Omega^\ast_\mathrm{P} = \! \bigcap_{1 \leq i \leq n_\mathrm{P}} \!
		\Omega_i^\ast, \quad\text{i.e.,}\quad \indic_{\Omega^\ast_\mathrm{P}}(\bx) =
		\! \bigwedge_{1 \leq i \leq n_\mathrm{P}} \! \indic_{\Omega_i^\ast}(\bx),
	\end{equation}
	in which the intersection/conjunction operators need to be regularized.
	\label{prop_prune}
\end{prop} 
To see why this is true, note that any query point's membership in an (unknown)
feasible design can be tested against all $m$ constraints independently of other
points' membership. If $\gsloc_i(\bx) \leq 0$ for $i = 1, 2, \ldots,
n_\mathrm{P}$, the point can (but does not have to) be included in the design,
for it to be feasible. The feasible design is hence a subset of all points that
satisfy all pointwise constraints. \\

{\it The above result enables computing on design subspaces as first-class
	objects. Computationally, the feasibility halfspaces are represented uniquely by
	their maximal elements. The pruning of halfspaces (abstract operation) is
	implemented by intersecting maximal elements, i.e., conjuncting point membership
	tests defined by pointwise constraints (concrete algorithm) in an arbitrary
	order.}
\begin{equation}
	\begin{tikzcd}[column sep=10em,row sep=3em]
		\dspace; \halfspace_1, \halfspace_2, \ldots, \halfspace_{n_\mathrm{P}}
		\arrow[r, "\cap", dashed] \arrow[r, "\text{(Not directly computable)}"',
		dashed] \arrow[d, "\max", shift left]
		& \dspace^\ast_\mathrm{P} \arrow[d, "\max", shift left] \\
		\Omega_0^{}; \Omega_1^\ast, \Omega_2^\ast, \ldots, \Omega_{n_\mathrm{P}}^\ast
		\arrow[r, "\cap"'] \arrow[u, "\powerset^\ast", shift left] &
		\Omega^\ast_\mathrm{P} \arrow[u, "\powerset^\ast", shift left]
	\end{tikzcd}
\end{equation}

\begin{figure}[h!]
	\centering
	\includegraphics[width=0.46\textwidth]{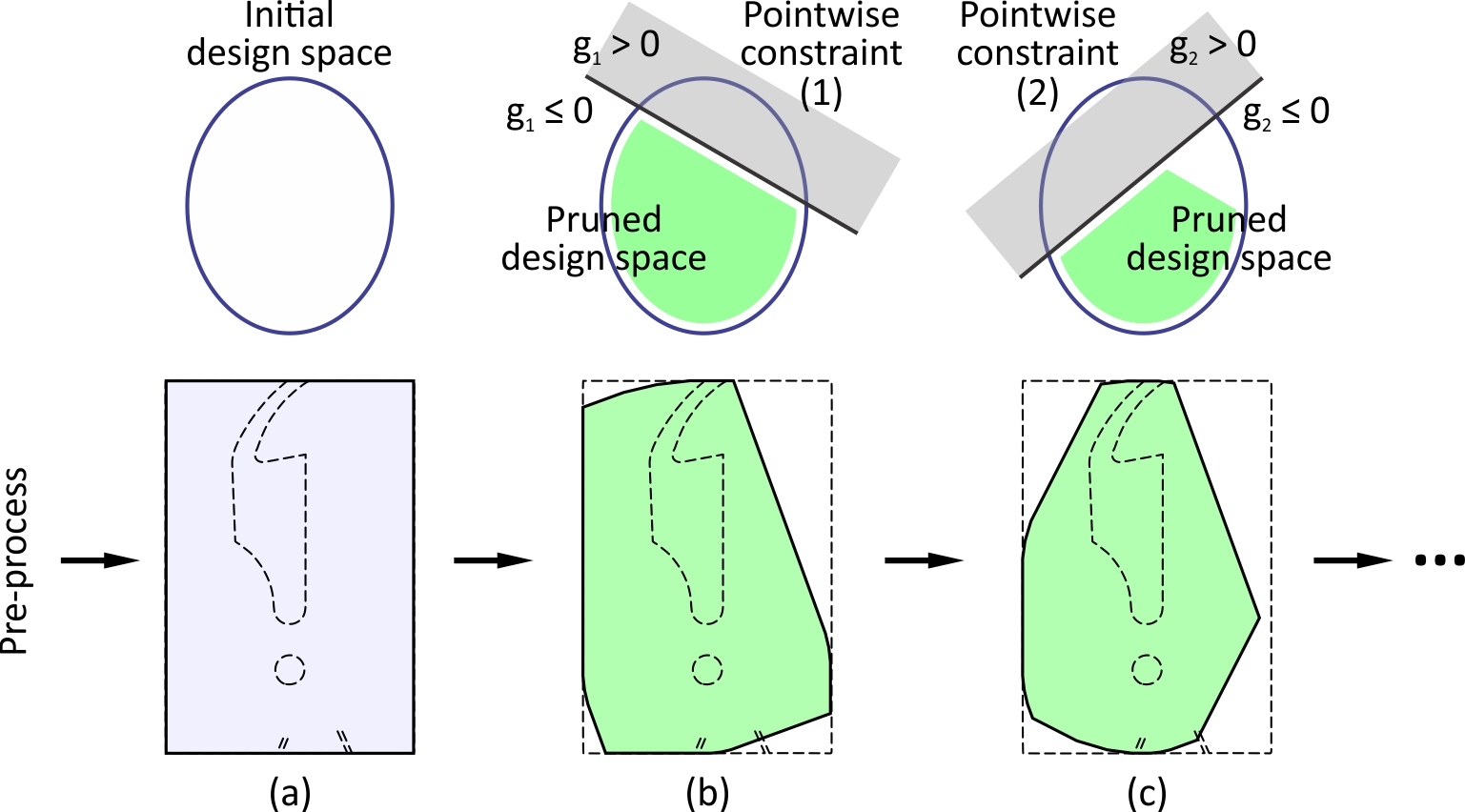}
	\caption{Design space pruning can be abstracted by intersecting the design
	space with feasibility halfspaces. This is not a computable operation in
	general. However, for pointwise constraints, it can be computed by intersecting
	maximal elements in the physical space.} \label{fig_prune}
\end{figure}

Figure \ref{fig_prune} illustrates design space pruning via \eq{eq_fspace_alt}.

Next, we consider maximal elements for set constraints in Section
\ref{sec_prune_set}, and provide examples from real-world engineering problems
in Section \ref{sec_prune_ex1}, \ref{sec_prune_ex2}, and \ref{sec_prune_ex3}.

\subsection{Maximal Pointsets for Set Constraints} \label{sec_prune_set}

Many design requirements, especially those that relate the interaction of moving
shapes, may be expressed as set constraints of the form $\Gamma(\Omega)
\subseteq E$, where $\Gamma: \dspace \to \powerset(\R^3)$ is a set
transformation. For example,
\begin{itemize}
	\item In packaging and assembly problems, the part's shape is often designed so
	that it is restricted to remain within a specified envelope while moving
	according to a prescribed motion.
	\item When designing part to be machined using a given tool that moves in the
	presence of obstacles (e.g., the part itself and fixtures), the surface of the
	part has to be accessible without collisions with the obstacles.
\end{itemize}
We will address examples of these problems in the following sections. The common
theme to many motion-based set transformation $\Gamma: \dspace \to
\powerset(\R^3)$ is that they can re-factored via \eq{eq_pointwise_trans} via a
pointwise transformation $\gamma: \Omega_0 \to \powerset(\R^3)$ {\it that
	depends on the motion}. The maximal pointset that satisfies the containment
$\Gamma(\Omega) \subseteq E$ or non-interference test $(\Gamma(\Omega) \cap E^c)
= \emptyset$ is defined implicitly by its PMC given in terms of $\gamma$:
\begin{align}
	\indic_{\Omega_i^\ast}(\bx) &:= 
	\begin{dcases}
	1 \quad &  \text{if}~ \gamma(\bx) \subseteq E,\\
	0 &  \text{otherwise},
	\end{dcases} \label{eq_pmc_max_set_1} \\
	\text{i.e.,}~ \Omega_i^\ast &:= \big\{\bx \in \Omega_0 ~|~ \gamma(\bx) \in E
	\big\} \label{eq_pmc_max_set_2} \\
	&~= \gamma^{-1}(E) \cap \Omega_0, \label{eq_pmc_max_set_3}
\end{align}
where the set operators are regularized, as before. Note that $\gamma^{-1}$ may
not be a function, because $\gamma$ is not necessarily invertible (see footnote
\footref{refnote}).

Here are a few classical examples from solid modeling:
\begin{itemize}
	\item For one-parametric sweep $\Gamma(\Omega) := \sweep(M, \Omega)$, one has
	$\gamma(\bx) = M\bx$ where $M = M(t) \in \SE{3}$ is a continuous one-parametric
	set of motions for $t \in [t_{\min}, t_{\max}]$. The maximal shape that
	satisfies $\Gamma(\bx) \subseteq E$ is given by an unsweep $\unsweep(M, E)$
	\cite{Ilies1999dual}.
	\item For Minkowski sum $\Gamma(\Omega) := (\Omega \oplus B)$, one has
	$\gamma(\bx) = (\bx + B)$ where $B \subset \R^3$ is typically a solid. The
	maximal shape that satisfies $\Gamma(\bx) \subseteq E$ is given by a Minkowski
	difference with $(E \ominus (-B))$ \cite{Serra1983image}.
	\item For general dilation (which subsumes the above two) with general rigid
	motions, the maximal shape is given by general erosion
	\cite{Nelaturi2011configuration}.
	\item For non-rigid (but pointwise pre-determined) deformations, the maximal
	shape is obtained by its PMC in terms of the pointwise displacement function.
\end{itemize}

\paragraph{\bf Procedure} 
Propositions \ref{prop_exists} through \ref{prop_prune} suggest a systematic
procedure to {\it prune} the design space, i.e., reduce the design space
$\dspace$ to $\dspace^\ast_\mathrm{P} = \dspace \cap (\halfspace_1 \cap
\halfspace_2 \cap \cdots \cap \halfspace_{n_\mathrm{P}})$ for kinematic criteria
expressed in terms of pointwise set constraints:
\begin{itemize}
	\item {\bf Step 0.} Initialize the feasible design space with the design
	domain, i.e., in algorithmic terms, $\Omega^\ast_\mathrm{P} \leftarrow
	\Omega_0$.
	\item {\bf Step 1.} Express the set constraint that outlines one of the
	conditions for a given design $\Omega \in \dspace$ to be feasible in the form
	$\Gamma(\Omega) \subseteq E$. Check if it can be restated as a pointwise
	constraint $\gamma(\bx) \subseteq E$ as in \eq{eq_sloc_gamma_1}.
	\item {\bf Step 2.} Formalize the forward problem in terms of the PMC test for
	the maximal element obtained from the pointwise constraint as prescribed by
	\eq{eq_pmc_max_set_1}.
	\item {\bf Step 3.} Invoke an IP-solver for the inverse problem, which computes
	(an exact or approximate representation of) the maximal element $\Omega^\ast_i$
	in \eq{eq_pmc_max_set_2}.
	\item {\bf Step 4.} Prune the design space by intersecting the maximal element
	$\Omega^\ast_\mathrm{P}$ obtained so far the new $\Omega^\ast_i$, i.e.,
	$\Omega^\ast_\mathrm{P} \leftarrow (\Omega^\ast_\mathrm{P} \cap \Omega^\ast_i)$
	is a smaller maximal element representing a pruned feasible subspace.
	\item Repeat {\bf steps 1--4} for all pointwise set constraints.
\end{itemize}
Notice that the above procedure can be applied to different constraints via
independent invocation of IP-solvers in an arbitrary order.

Importantly, the IP-solver in {\bf step 3} can be implemented in two
fundamentally different ways to obtain either an {\it implicit} representation
(i.e., using PMC test in \eq{eq_pmc_max_set_1}) or an {\it explicit}
representation (i.e., using inversion in \eq{eq_pmc_max_set_3}) of the maximal
element:
\begin{itemize}
	\item If we have access to an IP-solver that computes an explicit
	representation (e.g., B-rep) of the inverse transformation $\gamma^{-1}(E)$, we
	can directly intersect it (using any CAD kernel) with the design domain to
	obtain the maximal element as prescribed by \eq{eq_pmc_max_set_3}.
	\item If we have access to a FP-solver that computes (explicitly or implicitly)
	the forward transformation $\gamma(\bx)$ for a given query point $\bx \in
	\Omega_0$, we can compute an approximate representation (e.g., point cloud or
	voxelization) of the maximal element by:
	\begin{enumerate}
		\item sampling the design domain with a sufficiently dense set of query
		points;
		\item invoking the FP-solver to PMC-test them using \eq{eq_pmc_max_set_3};
		keep the ones that pass the test and discard the ones that do not; and
		\item (optional) use adaptive local re-sampling around the points that passed
		the test to obtain a better approximation.
	\end{enumerate}
	Because of the independence of pointwise test, invocation of the FP-solver for
	different query points can be done with perfect parallelization.
\end{itemize}
Let us apply this procedure to a few examples.

\subsection{Pruning for Containment of Moving Parts} \label{sec_prune_ex1}

Consider the latch design problem introduced earlier in Section
\ref{sec_introExam}, where the goal is to design a car hood latch that remains
within an envelope $E \subseteq \Omega_0$ while moving according to a motion $M
\subseteq \SE{3}$.

\paragraph{\bf Step 1} Every feasible latch design $\Omega \in \dspace$ must
satisfy the set constraint $\Gamma(\Omega) \subseteq E$ where $\Gamma(\Omega) :=
\sweep(M, \Omega)$; i.e., the swept volume by the latch after being transformed
by all configurations $\tau \in M$ (including any combination of translations
and rotations, parameterized or otherwise) remains within the envelope. The sweep
is a pointwise transformation, i.e., can be computed as the union of all
$\gamma(\bx) := M\bx = \bigcup_{\tau \in M} \tau \bx$, which represents the
trajectory traced by the query point $\bx \in \Omega_0$ along the prescribed
motion. Hence, the containment constraint be tested in a pointwise fashion by
$\gamma(\bx) \subseteq E$.

\paragraph{\bf Step 2}
Using the definitions in \eq{eq_pmc_max_set_1}, we construct a PMC test for the
maximal shape in the design space that satisfies this pointwise constraint:
\begin{equation}
	\indic_{\Omega^\ast_1} (\bx) :=
	\begin{dcases}
		1 &  \text{if}~ \forall \tau \in M: \tau \bx \in E,\\
		0 &  \text{otherwise}.
	\end{dcases} \label{eq_pmc_set_sweep}
\end{equation}
The forward problem involves following the trajectory, either exactly or
approximately (e.g., by sampling), and testing whether it remains entirely
within the envelope.

\begin{figure}[h!]
	\centering
	\includegraphics[width=0.46\textwidth]{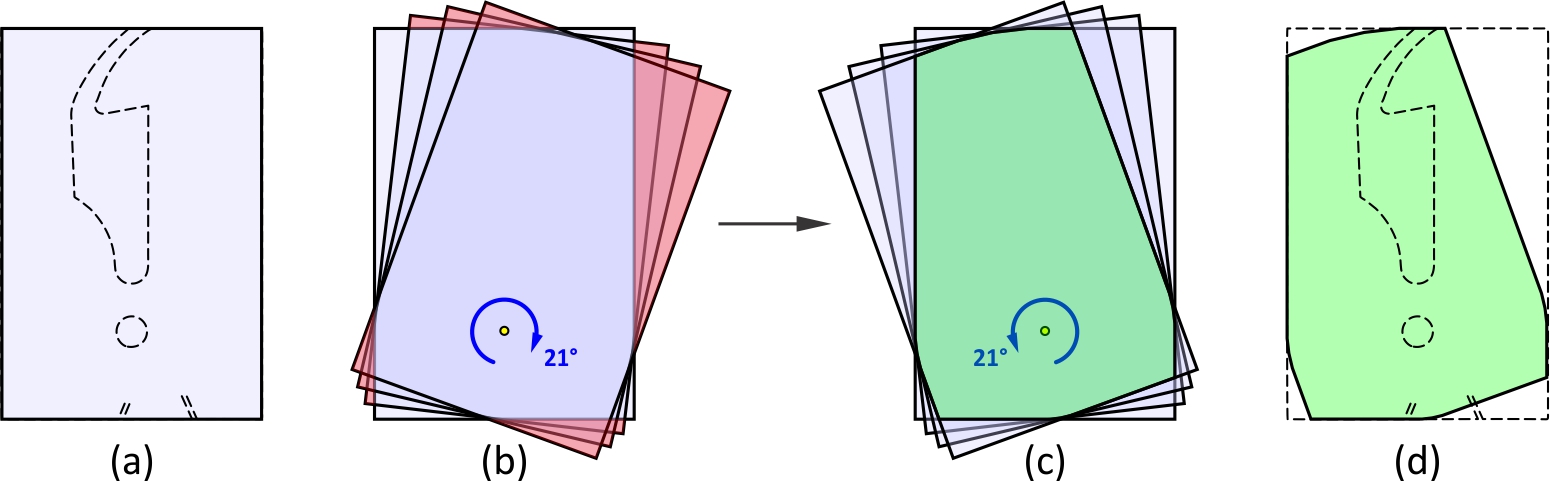}
	\caption{$\unsweep(R, E)$ is the largest set that remains within the square
	envelope $E$ shown in (a) while moving by a $21^\circ$ clockwise rotation $R$
	around a fixed pivot \cite{Ilies2000shaping}, i.e., without violating the
	containment constraint as depicted in red in (b). The unsweep can be computed
	in general by intersecting all moved instances of the envelope with an inverse
	motion \cite{Ilies1999dual}.} \label{fig_unsweep}
\end{figure}

\paragraph{\bf Step 3}
The dual properties of the FP- and IP-solvers (i.e., \textsf{Sweep} and
\textsf{Unsweep}) \cite{Ilies1999dual} can be leveraged to construct an exact or
approximate representation of $\Omega^\ast_1$, as illustrated in Fig.
\ref{fig_unsweep}. If we have access to an \textsf{Unsweep} solver, we can
directly compute $\Omega^\ast_1 = \unsweep(M^{-1}, E) \cap \Omega_0$. However,
if we only have access to an efficient \textsf{Sweep} solver, we can still
compute an approximate representation of $\Omega^\ast_1$ using the PMC test in
\eq{eq_pmc_set_sweep} for a sufficiently dense sample of query points and
retaining the points whose forward trajectory remains within the envelope
\cite{Ilies2000shaping}:
\begin{align}
	\Big[ \forall \tau \in M: \tau\bx \in E \Big] &~\rightleftharpoons~ \bx \in
	\bigcap_{\tau \in M} \tau^{-1}(E), ~\text{i.e.,} \\
	\sweep(M, \Omega) \subseteq E~ &~\rightleftharpoons~ \Omega \subseteq
	\unsweep(M^{-1}, E).
\end{align}
The {\it invertibility} of rigid transformations $M \mapsto M^{-1}$ is key to
efficient direct implementation of \textsf{Unsweep}.

The remaining {\bf steps 4} and {\bf 5} are straightforward.

$\Omega^\ast_1$ can be sent as the initial design to design space exploration
(Section \ref{sec_explore}) via TO or any other downstream material reducing
IP-solver. It is guaranteed that every valid design $\Omega \subseteq
\Omega^\ast_1$ that is the output of the downstream IP-solver, no matter how
complicated, will continue to satisfy the set constraint $\sweep(M, \Omega)
\subseteq E$.

\subsection{Pruning for Accessibility with Obstacles} \label{sec_prune_ex2}

\begin{figure*}
	\centering
	\includegraphics[width=0.96\textwidth]{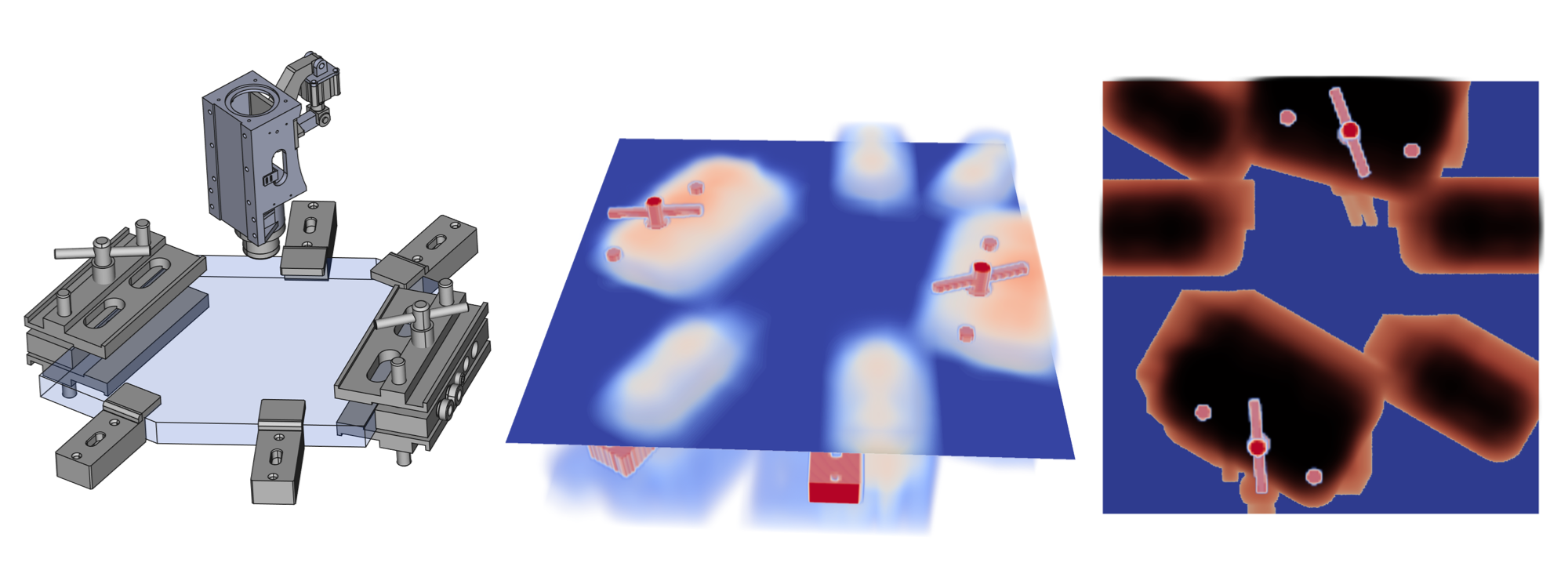}
	 \caption{(a) A machining setup showing six clamps used to locate and hold a
	designed part, and a $2-$axis instrument that can move in the plane. The
	envelope within which the shape must be designed is also highlighted. (b) The
	convolution of the reflected head with the fixtures gives a 3D field whose 2D
	cross-section via the plane of motion captures the collision volume at
	different translations. The position with zero collision volume (dark blue)
	are accessible. (c) The 2D cross-section of the 2.5D maximal manufacturable
	shape is obtained as the zero-set of the convolution field. This pointset
	serves as the initial design for TO (Section \ref{sec_explore}).}
	\label{fig_fixtureDesign}
\end{figure*}

Consider another example in a very different setting. Suppose we are designing a
part that needs to be fixtured in a crowded workholding environment so that a
machining instrument is able to access specific locations without colliding with
surrounding fixtures.

For simplicity, assume that the raw stock $\Omega_0$ is a thick sheet of metal,
fixtured on the machine bench as illustrated in Fig. \ref{fig_fixtureDesign}
(a). The manufacturing process is an EDM wire-cut or CO$_2$ laser-cut in which
the tool assembly $T = (H \cup L)$ moves according to a planar (i.e., 2D) motion
$M \subseteq \R^2$ parallel to the workpiece. We assume that the translation
with a vector $(x, y) \in M$ brings the wire or laser beam, abstracted by a
vertical line of zero thickness $L$, in contact with a line segment with $\bx =
(x, y, z) \in \Omega_0$ for a range of $z-$values along the sheet thickness.

The head $H$ cannot collide with the workpiece because they are located at
different elevations, i.e., $(\sweep(M, H) \cap \Omega) = \emptyset$ is a priori
guaranteed, thus imposes no constraint. Nonetheless, the head $H$ may collide
with the fixtures $F$, which may extend above the workpiece, i.e., $(\sweep(M,
H) \cap F) = \emptyset$ imposes a constraint on the motion. This, in turn,
imposes a constraint for manufacturability, because the motion defines the
boundary of the cut shape, restricted to a curve on the 2D plane: $M =
\partial\Omega \cap \R^2$. In other words, designing the as-manufactured part's
shape amounts to designing the motion, because every translation of the wire or
laser beam is in one-to-one correspondence with a point on the part's boundary
at which there is no collision between the head and the fixtures.

\paragraph{\bf Step 1}
The moving head $H$ may not collide with the fixtures $F$ when swept under the
motion $M$, i.e., $(\sweep(M, H) \cap F) = \emptyset$, i.e., $\sweep(M, H)
\subseteq F^c$. This is written in the standard form $\Gamma(\Omega) \subseteq
E$. Equivalently, for all translations $(x, y) \in M$ we have $((H + (x, y, 0))
\cap F) = \emptyset$, i.e., $(H + (x, y, 0)) \subseteq F^c$ which is in
pointwise form $\gamma(\bx) \subseteq E$ for $\bx = (x, y, 0)$. Notice that
$\gamma(\bx) := (H + \bx)$ does not depend on $\Omega$, as required.

\paragraph{\bf Step 2}
Using the definitions in \eq{eq_pmc_max_set_1}, we construct a PMC test for the
maximal shape in the design space that satisfies this pointwise constraint:
\begin{equation}
	\indic_{\Omega^\ast_2} (x, y, z) =
	\begin{dcases}
		1 &  \text{if}~ (H + (x, y, 0)) \subseteq F^c, \\
		0 &  \text{otherwise}.
	\end{dcases} \label{eq_pmc_set_mfg}
\end{equation}
It is important to note that our success in defining a pointwise constraint
(hence a PMC) depended on the assumption of planar translation at a higher
elevation than the part, which guaranteed $(H + (x, y, 0)) \subseteq \Omega^c$.
Otherwise, the correct constraint in \eq{eq_pmc_set_mfg} would have $(F \cup
\Omega)^c = (F^c \cap \Omega^c)$ instead of $F^c$ on the right-hand side, making
it stricter. But we cannot define the maximal element $\Omega^\ast_2$ and its
corresponding feasibility halfspace $\halfspace_2 =
\powerset^\ast(\Omega^\ast_2)$ in terms of a particular instance $\Omega \in
\halfspace_2$ (circular definition).

\paragraph{\bf Step 3}
Once again, we can leverage the dual properties of the FP- and IP-solvers to
compute an explicit or implicit representation of the 2.5D maximal
manufacturable solid.

The FP-solver can be any collision detection algorithm between arbitrary solids,
taking as input the displaced head $(H + (x, y, 0))$ above a particular query
point $(x, y, z) \in \Omega_0$ and the stationary fixtures $F$. Therefore, one
can sample the design domain (i.e., the raw stock) over a 2D grid $G$ and
construct a bitmap image, representing the 2D section of the 2.5D maximal solid,
by testing \eq{eq_pmc_set_mfg} for all $(x, y) \in G$. Every test requires
invoking the collision detection algorithm, and can be done in parallel.

Alternatively, one can construct an IP-solver to compute the collection of all
collision-free 2D translations (i.e., the configuration space obstacle
\cite{Lozano-Perez1983spatial}). Since we are dealing with solids and
regularized intersections, we can rewrite the set constraint $(H + (x, y, 0))
\cap F = \emptyset$ in terms of measures $\vol[(H + (x, y, 0)) \cap F] = 0$ and
convert it to an (in)equality constraint via \eq{eq_inner}. Hence,
\eq{eq_pmc_set_mfg} becomes:
\begin{align}
	&\indic_{\Omega^\ast_2} (x, y, z) = \neg \sign \circ \vol[(H + (x, y, 0)) \cap
	F] \\ &= \neg \sign \int_{\R^3} \indic_H(\bx' - (x, y, 0)) ~\indic_F(\bx')
	~dv[\bx'], \label{eq_pmc_set_vol}
\end{align}
noting that $\indic_{H + (x, y, 0)}(\bx') = \indic_H(\bx' - (x, y, 0))$ is the
indicator function of the translated head. The integral on the right-hand side
is a convolution $({\indic}_{-H} \ast \indic_F)(\bx)$, evaluated at $\bx := (x,
y, 0)$, after a reflection ${\indic}_{-H}(\bx') = \indic_H(-\bx')$. The
integrand is nonzero only at $\bx' \in \R^3$ where both indicator functions are
nonzero, hence the integral does not vanish within the (measurable) regions of
intersection. Substituting this relation into \eq{eq_pmc_set_vol} yields:
\begin{equation}
	\indic_{\Omega^\ast_2} (x, y, z) =
	\begin{dcases}
		1 &  \text{if}~ (\tilde{\indic}_H \ast \indic_F)(x, y, 0) = 0,\\
		0 &  \text{otherwise}.
	\end{dcases} \label{eq_pmc_set_conv}
\end{equation}
The convolution can be converted to pointwise multiplications in the frequency
domain using (forward and inverse) Fourier transforms:
\begin{equation}
	({\indic}_{-H} \ast \indic_F) = \Fourier^{-1} \big\{ \Fourier\{{\indic}_{-H} \}
\cdot \Fourier\{\indic_F\} \big\},
\end{equation}
which, in turn, can be rapidly computed as fast Fourier transforms (FFT) and
accelerated on GPUs. Figure \ref{fig_fixtureDesign} illustrates the results of
this computation. The convolution computes a 3D image in one shot using three
FFT computations on two 3D bitmaps (voxelized $-H$ and $F$). However, we only
need its 2D cross-section at $z = 0$, whose zero-set gives a 2D bitmap image
representation of the 2.5D maximal solid $\Omega^\ast_2$.

The remaining {\bf steps 4} and {\bf 5} are straightforward.

Note that everything in the above analysis would remain valid if we allowed the
manufacturing instrument to rotate in the plane, except that we would have to
change the constraint on the right-hand side of \eq{eq_pmc_set_mfg} to hold for
at least one planar rotation $R \in \SO{2}$ of the head $H$:
\begin{equation}
	\indic_{\Omega^\ast_2} (\bx) =
	\begin{dcases}
		1 &  \text{if}~ \exists R \in \SO{2}: (RH + (x, y, 0)) \subseteq F^c, \\
		0 &  \text{otherwise}.
	\end{dcases} \label{eq_pmc_set_mfg_rot}
\end{equation}
Accordingly, the convolution in \eq{eq_pmc_set_conv} is adjusted:
\begin{equation}
	\indic_{\Omega^\ast_2} (\bx) =
	\begin{dcases}
		1 &  \text{if}~ \exists R \in \SO{2}: ({\indic}_{-RH} \ast \indic_F)(x, y, 0)
		= 0,\\
		0 &  \text{otherwise}.
	\end{dcases} \label{eq_pmc_set_conv_rot}
\end{equation}
in which the rotation can be parameterized as $R = R(\theta)$ for $\theta \in
[0, 2\pi)$ and ${\indic}_{-RH}(\bx') = \indic_{H}(-R^{-1}\bx')$ where
$R^{-1}(\theta) = R(-\theta)$ is an inverse rotation. To compute the PMC, one
has to sample the rotation angles $\theta \in [0, 2\pi]$ and for each trial
rotation, resample the rotated head's 3D bitmap into the same grid in which the
fixtures' are rasterized to compute the discrete convolution via FFT.

\subsection{When Pruning Fails: A More Complex Example} \label{sec_prune_ex3}

In more general manufacturing scenarios, a number of assumptions that enabled
pointwise formulation in Section \ref{sec_prune_ex2} are invalidated. For
example, in a $5-$axis CNC machine, one deals with 6D rigid motions $(R, \bt)
\in \SE{3}$ composed of 3D rotations $R \in \SO{3}$ and 3D translations $\bt \in
\R^3$. The tool assembly $T = (H \cup C)$ is no longer guaranteed to avoid
collisions with the workpiece, leading to global constraints that depend on the
part's shape as well as the head, cutter, and fixtures. The configuration space
obstacle $M_\Omega := \obs(O_\Omega, T)$ where $O_\Omega = (\Omega \cup F)$ is
stated as a group convolution $\star$ operation \cite{Lysenko2010group}, which,
in turn, can be computed as a Euclidean convolution $\ast$ as before:
\begin{align}
	\indic_{M_\Omega} (R, \bt) &= \sign \circ (\indic_{O_\Omega} \star
	{\indic}_{-T}) (R, \bt) \\
	&= \sign \circ (\indic_{O_\Omega} \ast {\indic}_{-RT}) (\bt),
\end{align}
Attempting to write a PMC similar to \eq{eq_pmc_set_conv_rot} fails for several
reasons; let us give it a try:
\begin{equation}
	\indic_{\Omega^\ast_3} (\bx) \overset{?}{=}
	\begin{dcases}
		1 &  \text{if}~ \exists R \in \SO{3} : (\indic_{O_\Omega} \ast {\indic}_{-RT})
		(\bt) = 0,\\
		0 &  \text{otherwise}.
	\end{dcases} \label{eq_pmc_set_conv_gen}
\end{equation}
The first obvious problem is the dependency of the right-hand side on $\Omega$,
which makes for a circular definition. Moreover, the cutter's shape cannot be
ignored (unlike the case with wire-/laser-cut). Hence, there is no obvious way
to assign a correspondence between the translations $\bt \in \R^3$ and the
points $\bx \in \Omega_0$ within the design domain, unless we consider all
possible contact configurations and treat boundary points differently from
interior points. Last but not least, passing a collision check at the contact
configuration is not sufficient for accessibility, because there may not exist a
connected path from the initial configuration of the tool assembly to the
cutting pose of interest. For example, if a downstream TO creates cavities in
the design in 3D, none of the will be accessible (unlike 2.5D).

In the next section, we deal with constraints that cannot be stated in pointwise
form due to global dependencies. We assume that the design has been pruned
(using methods of this section) for all pointwise constraints to produce an
initial design $\Omega^\ast_\mathrm{P} = (\Omega^\ast_1 \cap \Omega^\ast_2 \cap
\cdots \Omega^\ast_{n_\mathrm{P}}) \subseteq \Omega_0$ for design space
exploration with regard to the remaining $(n - n_\mathrm{P}) = (n_\mathrm{G} +
n_\mathrm{L})$ global and/or local constraints.

%% file: exploration.tex
\section{Design Space Exploration} \label{sec_explore}

In this section, we present a methodology to solve the  {\bf phase 2} problem
formulated in \eq{eq_stage2} of Section \ref{sec_phases}. Our goal is not to
propose new optimization algorithms besides the many existing ones (reviewed in
Section \ref{sec_lit}). Rather, we propose a general strategy to deal with
constraints that {\it cannot} be stated in a pointwise fashion, to guide
gradient-descent optimization.

In Section \ref{sec_multiobj}, we formulate multi-objective and multi-constraint
optimization. We extend existing approaches to deal with heterogeneous
constraints, including not only physics-based constraints obtained from FEA but
also kinematics-based constraints obtained from accessibility analysis for
machining.

In order to move {\it deterministically} in the design space in directions that
consistently reduce the violation of these constraints, we quantify their
sensitivities to hypothetical local changes in the design. Different
gradient-like quantities can be defined for different design representations. In
Section \ref{sec_TSF}, we demonstrate our approach specifically for defining,
augmenting, and filtering topological sensitivity fields (TSF)
\cite{Novotny2007topological} with global and local constraints.

\subsection{Multi-Objective and Multi-Constraint Optimization}
\label{sec_multiobj}

Fixed-point iteration (a.k.a. Picard iteration) \cite{burden1997numerical} is an
effective approach for numerically solving multi-objective optimization
problems, where the problem is iteratively solved through series of outer- and
inner-loops. As the value of each objective function is changed in the
outer-loop, its value is kept fixed in the inner loop. The fixed objective
functions are treated as equality constraints for the single-objective
inner-loop optimization. Among the many popular approaches, we use a Pareto
tracing levelset TO approach (PareTO) \cite{Suresh2013efficient}, because it
produces valid designs (i.e., solids) at all intermediate steps. It was shown in
\cite{Suresh2010199} that Pareto tracing can also be extended to density-based
approaches such as solid isotropic material with penalization (SIMP)
\cite{Sigmund2013topology}.

\begin{figure*}[h!]
	\centering
	\includegraphics[width=0.96\linewidth]{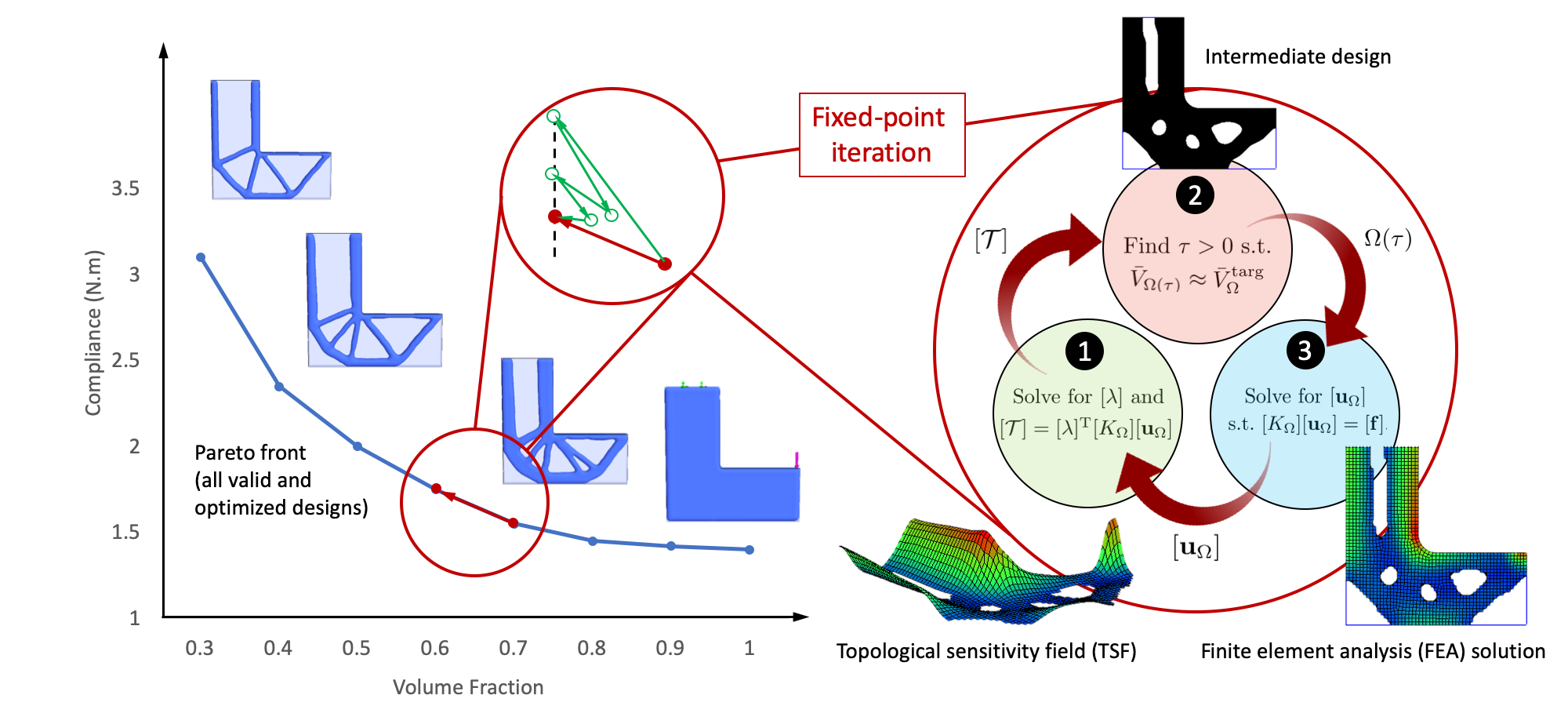}
	\caption{Using fixed-point iteration, the design is iteratively modified for
	every small reduction in the volume such that the optimality conditions and
	performance constraints remain satisfied at every point along the Pareto
	front.} \label{fig_fixedPoint}
\end{figure*}

For example, in classical TO, the goal is to obtain light-weight stiff
structures, leading to two competing objectives (mass and compliance) with a
one-dimensional Pareto frontier, illustrated for different examples in Figs.
\ref{fig_latchPareTO}, \ref{fig_latchPreprocess}, and \ref{fig_fixedPoint} (a). The
problem can be formulated as follows:
\begin{equation}
	\text{Find}~ \Omega \in \dspace^\ast_\mathrm{P}:
	\begin{dcases} 
		\text{minimize}~ \bar{V}_{\Omega} ~\text{and}~ J_\Omega = [\bff]^\mathrm{T}
		[\bu_\Omega], \\
		\text{subject to}~~ [K_\Omega][\bu_\Omega] = [\bff],
	\end{dcases} \label{eq:ParetoProblem_1}
\end{equation} 
where the volume fraction $\bar{V}_{\Omega} :=
\vol[\Omega]/\vol[\Omega^\ast_\mathrm{P}]$ is the ratio of the (unknown)
$\vol[\Omega]$ to the initial design's volume $\vol[\Omega^\ast_\mathrm{P}]$,
where $\Omega^\ast_\mathrm{P} \subseteq \Omega_0$ is the maximal feasible
pointset obtained from pruning in Section \ref{sec_prune}. Classical TO in the
absence of pruning is subsumed as a special case when $\Omega^\ast_\mathrm{P} =
\Omega_0$. The second objective function $J_\Omega  = [\bff]^\mathrm{T}
[\bu_\Omega]$ is the compliance (i.e., strain energy) obtained from FEA, in
which $[\bu_\Omega]$ is the discretized displacement field and $[\bff]$ is the
external load vector given as (Neumann) boundary conditions. The FEA also
appears as an equality constraint $[K_\Omega][\bu_\Omega] = [\bff]$ in which
$[K_\Omega]$ is the stiffness matrix obtained from the design shape, material
properties, and restraints given as (Dirichlet) boundary conditions.

We can reformulate the problem as a single-objective optimization for a fixed
volume fraction:
\begin{equation}
	\text{Find}~ \Omega \in \dspace^\ast_\mathrm{P}:
	\begin{dcases} 
		\text{select target}~ \bar{V}_\Omega^\mathrm{targ} \in (0, 1], \\
		\text{ILI:}
		\begin{dcases}
			\text{minimize}~ J_\Omega = [\bff]^\mathrm{T} [\bu_\Omega], \\
			\text{s.t.}~\bar{V}_\Omega = \bar{V}_\Omega^\mathrm{targ}, \\
			\text{s.t.}~ [K_\Omega][\bu_\Omega] = [\bff],
		\end{dcases}
	\end{dcases} \label{eq:ParetoProblem_2}
\end{equation} 
where ILI stands for inner-loop iteration. Within each ILI, a single-objective
optimization is solved to minimize compliance $J_\Omega$ subject to a fixed
volume fraction constraint $\bar{V}_\Omega = \bar{V}_\Omega^\mathrm{targ}$ for a
fixed $0 < \bar{V}_\Omega^\mathrm{targ} \leq 1$. In PareTO
\cite{Suresh2013efficient}, one starts off on the Pareto frontier at the
right-most extreme with $\Omega := \Omega^\ast_\mathrm{P}$ and
$\bar{V}_\Omega^\mathrm{targ} = 1$, i.e., the best-case scenario for compliance
at the cost of the largest volume. The algorithm incrementally removes material
to decrease $\bar{V}_\Omega^\mathrm{targ}$ by introducing holes in the design,
without deviating too much from the Pareto front. The ILI is a fixed-point
iteration that applies local modifications to the new design to bring it back to
the Pareto front, as shown schematically in Fig. \ref{fig_fixedPoint} (a). See
\cite{Suresh2013efficient,Suresh2013stress,Mirzendehdel2015pareto,Mirzendehdel2016support} for more details on PareTO and its various applications.

The inner-loop optimization can be expressed as local minimization of the
Lagrangian defined as:
\begin{equation}
	\mathcal{L}_\Omega := [\bff]^\mathrm{T} [\bu_\Omega] + \lambda_1
	(\bar{V}_\Omega - \bar{V}_\Omega^\mathrm{targ}) + [\lambda_2]^\mathrm{T} \Big(
	[K_\Omega][\bu_\Omega] - [\bff]\Big). \label{eq_Lag}
\end{equation}
The Karush--Kuhn--Tucker (KKT) conditions \cite{Bendsoe2004topology} for this
problem are given by $\nabla \mathcal{L}_\Omega = 0$ in which the gradient is
defined by partial differentiation with respect to the independent variables;
namely, the design variables used to represent $\Omega$ and the Lagrange
multipliers $\lambda_1$ and $[\lambda_2]$. The latter simply encodes the
constraints into $\nabla \mathcal{L}_\Omega = 0$:
\begin{align}
	\frac{\partial~}{\partial \lambda_1} \mathcal{L}_\Omega &= (\bar{V}_\Omega -
	\bar{V}_\Omega^\mathrm{targ}) := 0, \\
	\frac{\partial~}{\partial \lambda_2} \mathcal{L}_\Omega &=
	[K_\Omega][\bu_\Omega] - [\bff] := [0],
\end{align}
On the other hand, differentiation with respect to $\Omega \in
\dspace^\ast_\mathrm{P}$ {\it depends on the particular parameterization used to
	represent the design by a finite set of decision variables for optimization}.
These variables can be geometric/size variables (e.g., thickness in truss
optimization), density variables (e.g., volume fractions in SIMP), and so on.
Our goal is to present a representation-agnostic form in terms of TSF.

If we use a prime symbol $(\cdot)'$ to represent the generic (linear)
differentiation of a function with respect to $\Omega$, we obtain (via chain
rule):
\begin{align}
	\mathcal{L}_\Omega' &= [\bff]^\mathrm{T} [\bu_\Omega'] + \lambda_1
	\bar{V}_\Omega' + [\lambda_2]^\mathrm{T} \Big([K_\Omega][\bu_\Omega]\Big)',
	\label{eq_chain_rule} \\
	& = \Big( [\bff]^\mathrm{T} + [\lambda_2]^\mathrm{T} [K_\Omega]\Big)
	[\bu_\Omega'] + \lambda_1 \bar{V}_\Omega' + [\lambda_2]^\mathrm{T}
	[K_\Omega'][\bu_\Omega], \nonumber
\end{align}
Computing $[\bu_\Omega']$ is prohibitive, as it requires calling FEA as many
times as the number of independent variables used to represent $\Omega$. The
common solution is to choose $[\lambda_2]$ such that $ [\bff]^\mathrm{T} +
[\lambda_2]^\mathrm{T} [K_\Omega] = [0]$ (adjoint problem)
\cite{Bendsoe2004topology}:
\begin{equation}
	\mathcal{L}_\Omega' = \lambda_1 \bar{V}_\Omega' + [\lambda_2]^\mathrm{T} [K_\Omega'][\bu_\Omega], ~\text{if}~ [\lambda_2] := -[K_\Omega]^{-1}[\bff], \label{eq_Lag_prime}
\end{equation}

In general, if we have $n_\mathrm{O} > 0$ global objective functions
and another
$n_\mathrm{G} \geq 0$ global (in)equality constraints, \eq{eq_chain_rule} can be generalized as:
\begin{equation}
	\mathcal{L}_\Omega' := \!\!\!\!\!\! \sum_{n_\mathrm{C} < i \leq n_\mathrm{C} +
	n_\mathrm{O}} \!\!\!\!\!\! \bar{\lambda}_i f_i' (\Omega) + \!\!\!\!\!\!
	\sum_{n_\mathrm{P} < i \leq n_\mathrm{P}+n_\mathrm{G}} \!\!\!\!\!\!
	\bar{\lambda}_i \gglob_i' (\Omega), \label{eq_grad_generic}
\end{equation}
in which the indexing scheme is to reconcile with that of Section
\ref{sec_phases}. We can simplify the notation by introducing $F_j(\Omega) :=
f_{n_\mathrm{C} + j}(\Omega) - f_{n_\mathrm{C} + j}^\mathrm{targ}$ for $0 < j
\leq n_\mathrm{O}$ and $F_{j}(\Omega) := \gglob_{n_\mathrm{P} - n_\mathrm{O} +
	j}(\Omega)$ for $n_\mathrm{O} < j \leq n_\mathrm{O} + n_\mathrm{G}$, hence:
\begin{equation}
	\mathcal{L}_\Omega' = \!\!\!\!\!\! \sum_{0 < j \leq n_\mathrm{O} +
	n_\mathrm{G}} \!\!\!\!\!\! \lambda_j F_j' (\Omega),
	\label{eq_grad_generic_simple}
\end{equation}

\subsection{Coupling Physical and Manufacturing Constraints} \label{sec_TOforms}

The ILI in \eq{eq:ParetoProblem_2} can be generalized to accommodate other
global objective functions and global constraints. For example, suppose the part
is to be 3D printed at a given build direction. An additional global
(in)equality constraint is imposed in terms of an upper-bound
$\bar{V}_\mathsf{UB} \geq 0$ on the total volume of support material that is
needed based on an overhang angle criterion \cite{Mirzendehdel2016support}:
\begin{equation}
	\text{Find}~ \Omega \in \dspace^\ast_\mathrm{P}:
	\begin{dcases} 
		\text{select target}~ \bar{V}_\Omega^\mathrm{targ} \in (0, 1], \\
		\text{ILI:}
		\begin{dcases}
			\text{minimize}~ J_\Omega = [\bff]^\mathrm{T} [\bu_\Omega], \\
			\text{s.t.}~\bar{V}_\Omega = \bar{V}_\Omega^\mathrm{targ}, \\
			\text{s.t.}~ [K_\Omega][\bu_\Omega] = [\bff], \\
			\text{s.t.}~ \bar{V}_{S_\Omega} \leq \bar{V}_\mathsf{UB},
		\end{dcases}
	\end{dcases} \label{eq:ParetoProblem_3}
\end{equation} 
where $S_\Omega \subseteq \Omega^c$ represents the support structure. Its volume
fraction $\bar{V}_{S_\Omega} = \vol[S_\Omega] / \vol[\Omega^\ast_\mathrm{P}]$
can be computed as a function of the angle between surface normals and the build
direction at every outer-loop iteration \cite{Mirzendehdel2016support}. The
Lagrangian in \eq{eq_Lag} is further augmented by adding another term $\lambda_3
(\bar{V}_{S_\Omega} - \bar{V}_\mathsf{UB})$ and the generic sensitivity in
\eq{eq_Lag_prime} is updated by incorporating $\bar{V}_{S_\Omega}'$ as:
\begin{equation}
	\mathcal{L}_\Omega' = \lambda_1 \bar{V}_\Omega' + [\lambda_2]^\mathrm{T}
	[K_\Omega'][\bu_\Omega] + \lambda_3 \bar{V}_{S_\Omega}'. \label{eq_Lag_prime1}
\end{equation}

\begin{figure}[h!]
	\centering
	\includegraphics[width=0.96\linewidth]{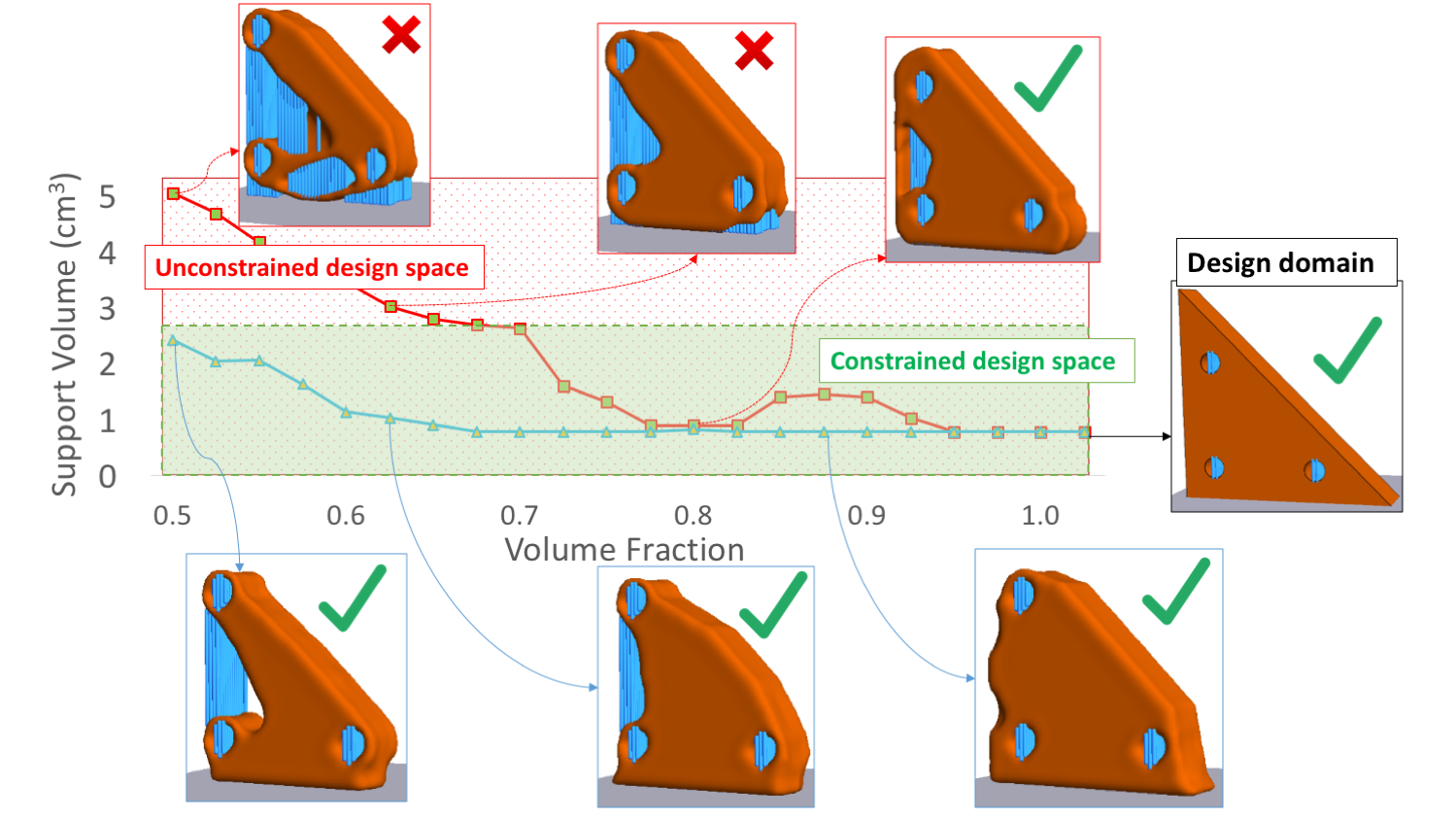}
	\caption{TO with and without augmenting the sensitivity with constraints on the
	support material needed for 3D printing along the vertical build direction
	\cite{Mirzendehdel2016support}. Many of the solutions without considering the
	support constraint will still satisfy that constraint due to the larger volume
	fraction occupied by the design itself. However, as the material is reduced
	from the design below $70\%$, the TO generates designs that require more
	support material.} \label{fig_augmentedTSF}
\end{figure}

Figure \ref{fig_augmentedTSF} compares the solution to a TO problem with and
without constraining the support material volume (adopted from
\cite{Mirzendehdel2016support}). Observe that optimization without the support
constraint exits the feasibility halfspace with respect to this constraint for
design volume fractions less than $70\%$. For lighter designs, the removed
design material comes at the expense of additional support material, hence
costlier manufacturing. The fully constrained optimization with augmented
sensitivity as in \eq{eq_Lag_prime1} dramatically increases the number of
feasible and Pareto-optimal options, even at volume fractions lower than $70\%$.

Another example is TO subject to accessibility constraints for machining. Once
again, we can impose a global (in)equality constraint in \eq{eq_max_vol} to in
terms of an upper-bound $\bar{V}_\mathsf{UB} \geq 0$ on the total inaccessible
volume:
\begin{equation}
	\text{Find}~ \Omega \in \dspace^\ast_\mathrm{P}:
	\begin{dcases} 
		\text{select target}~ \bar{V}_\Omega^\mathrm{targ} \in (0, 1], \\
		\text{ILI:}
		\begin{dcases}
			\text{minimize}~ J_\Omega = [\bff]^\mathrm{T} [\bu_\Omega], \\
			\text{s.t.}~\bar{V}_\Omega = \bar{V}_\Omega^\mathrm{targ}, \\
			\text{s.t.}~ [K_\Omega][\bu_\Omega] = [\bff], \\
			\text{s.t.}~ 1 - \bar{V}_{R_\Omega} - \bar{V}_{\Omega} \leq \bar{V}_\mathsf{UB},
		\end{dcases}
	\end{dcases} \label{eq:ParetoProblem_4}
\end{equation} 
where $R_\Omega \subseteq \Omega^c$ represents the maximal accessible region
outside the design for a combination of tools and approach directions in
$3-$axis milling \cite{Nelaturi2015automatic}. The volume fraction of the
inaccessible regions is $1 - \bar{V}_{R_\Omega} - \bar{V}_{\Omega}$ where
$\bar{V}_{R_\Omega} = \vol[R_\Omega] / \vol[\Omega^\ast_\mathrm{P}]$ and
$\bar{V}_\Omega = \vol[\Omega] / \vol[\Omega^\ast_\mathrm{P}]$, as before. The
Lagrangian in \eq{eq_Lag} is further augmented by adding another term $\lambda_3
(1 - \bar{V}_{R_\Omega} - \bar{V}_{\Omega} - \bar{V}_\mathsf{UB})$, hence:
\begin{align}
	\mathcal{L}_\Omega' &= \lambda_1 \bar{V}_\Omega' + [\lambda_2]^\mathrm{T}
	[K_\Omega'][\bu_\Omega] + \lambda_3 (\bar{V}_{R_\Omega}' - \bar{V}_{\Omega}')
	\\
	&= (\lambda_1 - \lambda_3) \bar{V}_\Omega' + [\lambda_2]^\mathrm{T}
	[K_\Omega'][\bu_\Omega] + \lambda_3 \bar{V}_{R_\Omega}'. \label{eq_Lag_prime2}
\end{align}
One can alternatively formulate the optimization problem for accessibility using
the local constraint in \eq{eq_cons_mu}:
\begin{equation}
	\text{Find}~ \Omega \subseteq \Omega^\ast_\mathrm{P} ~\text{to}~
	\begin{dcases} 
		\text{select target}~ \bar{V}_\Omega^\mathrm{targ} \in (0, 1], \\
		\text{ILI:}
		\begin{dcases}
			\text{minimize}~ J_\Omega = [\bff]^\mathrm{T} [\bu_\Omega], \\
			\text{s.t.}~\bar{V}_\Omega = \bar{V}_\Omega^\mathrm{targ},\\
			\text{s.t.}~ [K_\Omega][\bu_\Omega] = [\bff],\\
			\text{s.t.}~ [\indic_{\Omega}\ast{\indic}_{-T}] = [0],
		\end{dcases}
	\end{dcases} \label{eq:ParetoProblem_5}
\end{equation} 
in which the inaccessibility measure $\mu_\Omega(\bx) = (\indic_{O_\Omega} \ast
\tilde{\indic}_T)$ in \eq{eq_cons_mu}, defined as the convolution in
\eq{eq_access_conv} is discretized to $[\mu_\Omega] = [\indic_{O_\Omega} \ast
\tilde{\indic}_T]$ and further simplified to $[\mu_\Omega] = [\indic_{\Omega}
\ast \tilde{\indic}_T]$ assuming that the stationary obstacle $O_\Omega =
(\Omega \cup F)$ includes only the target design $O_\Omega = \Omega$, ignoring
the fixtures $F := \emptyset$. The tool assembly $T = (H \cup C)$ includes the
holder $H$ and cutter $C$, as before. We use a conservative measure, aiming for
no allowance for inaccessibility (i.e., $\mu_0 := 0$ in \eq{eq_cons_mu}) hence
$[\indic_{\Omega} \ast \tilde{\indic}_T] = [0]$ over all discrete elements
(e.g., voxels) wherever possible in the design domain. The discrete convolution
is computed using two forward FFTs on $[\indic_\Omega]$ and
$[\tilde{\indic}_T]$, a pointwise multiplication of their frequency domain
grids, and an inverse FFT to obtain $[\indic_{\Omega} \ast \tilde{\indic}_T]$ in
the physical domain (as a voxelized field).

Hereon, we assume that reducing the volume is always an objective/cost function,
hence the outer-loop is always set up to incrementally decrease the volume
fraction budget $\bar{V}_\Omega^\mathrm{targ} \in (0, 1]$ starting from the
initial value $\bar{V}_\Omega^\mathrm{targ} := 1$ on the costlier extreme of the
Pareto front. The optimization problem is formulated in general as:
\begin{equation}
	\text{Find}~ \Omega \in \dspace^\ast_\mathrm{P}:
	\begin{dcases} 
		\text{select target}~ \bar{V}_\Omega \in (0, 1], \\
		\text{ILI:}
		\begin{dcases}
			\text{minimize}~ f_i(\Omega), ~n_\mathrm{C} < i \leq n_\mathrm{C} + n_\mathrm{O}, \\
			\text{s.t.}~\bar{V}_\Omega = \bar{V}_\Omega^\mathrm{targ},\\
			\text{s.t.}~ \gglob_i(\Omega) \quad \leq 0, ~n_\mathrm{P} < i \leq
			n_\mathrm{P} + n_\mathrm{G}, \\
			\text{s.t.}~ \gglob_i(\bx; \Omega) \leq 0, ~n_\mathrm{P} + n_\mathrm{G} < i
			\leq n_\mathrm{C},
		\end{dcases}
	\end{dcases} \label{eq:ParetoProblem_gen}
\end{equation} 

\subsection{Sensitivity Fields for Global \& Local Constraints} \label{sec_TSF}

In this section, we use TSF to define $\mathcal{L}_\Omega'$ and $F_i' (\Omega)$
in \eq{eq_grad_generic_simple} in a representation-independent form. Let us
first look at a few examples with manufacturability constraints in addition to
the physical constraints in \eq{eq:ParetoProblem_2}.

The notion of a TSF is widely applied in the TO literature
\cite{Novotny2007topological,Suresh2013efficient} as a means to guide the
optimization process in moving from one candidate solution to another in the
search for local optima. Intuitively, the TSF is a gradient-like operator for
pointsets that quantifies {\it the global effect of local changes} of a given
function (e.g., violation of a global constraint). We couple the TSFs for
various global and local constraints in three distinct steps:

\begin{itemize}
	\item {\it Defining TSFs for Global Constraints}: For global constraints of the
	general form $\gglob_i(\Omega) \leq 0$, we define one TSF per constraint to
	measure how its violation changes after removing a hypothetical small
	neighborhood (called an ``inclusion'') at a given point.
	\item {\it Augmenting TSFs for Global Constraints:} The individual TSFs are
	linearly combined for all global constraints (including the fixed objective
	functions).
	\item {\it Penalizing TSFs via Local Constraints}: For local constraints of the
	general form $\gloc_i(\bx; \Omega) \leq 0$, we penalize the TSF of the global
	constraints by a linear combination of the violation of local constraints.
\end{itemize}
We elaborate in Sections \ref{sec_def_TSF} through \ref{sec_filt_TSF}.

\subsubsection{Defining Sensitivities for Global Constraints} \label{sec_def_TSF}

For every function $F_i: \dspace^\ast_\mathrm{P} \to \R$ that depends globally
on the design (objective function or constraint), we define a field $\tsf_i :
(\Omega^\ast_\mathrm{P} \times \dspace^\ast_\mathrm{P}) \to \R$ as its TSF via:
\begin{equation}
	\tsf_i(\bx; \Omega) := \lim_{\epsilon \to 0^+} \frac{F_i(\Omega -
	B_\epsilon(\bx)) - F_i(\Omega)}{\vol[\Omega \cap B_\epsilon(\bx)]},
\label{eq_TSF}
\end{equation} 
for $0 < i \leq n_\mathrm{O}+n_\mathrm{C}$. $B_\epsilon(\bx) \subset \Omega_0$
is a small 3D ball of radius $\epsilon \to 0^+$ centered at a given query point
$\bx \in \Omega$. The numerator of the limit evaluates the (presumably
infinitesimal) change in $F_i(\Omega)$ when the candidate design is modified as
$\Omega \mapsto (\Omega - B_\epsilon(\bx))$, i.e., by puncturing an
infinitesimal cavity at the query point. The denominator $\vol[\Omega \cap
B_\epsilon(\bx)]  = O(\epsilon^3)$ as $\epsilon \to 0^+$ measures the volume of
the cavity. For internal points $\bx \in \mathsf{i}\Omega$ (i.e., points that
are not exactly on the topological boundary) one has $\vol[\Omega \cap
B_\epsilon(\bx)] = \vol[B_\epsilon(\bx)]$ as $\epsilon \to 0^+$.

\subsubsection{Combining Sensitivities for Global Constraints} \label{sec_aug_TSF}

It was shown in \cite{Deng2015multi} that the method of augmented Lagrangian can
be extended to TSFs, and its effectiveness was demonstrated by TO of multi-load
structures under deformation and stress constraints. We apply the linear
combination of the generic form in \eq{eq_grad_generic_simple} to compute an
``augmented'' TSF to couple the global (in)equality constraints:
\begin{equation}
	\tsf(\bx; \Omega) := \!\!\!\!\! \sum_{0 < j \leq n_\mathrm{O}+n_\mathrm{G}}
	\!\!\!\!\! \lambda_j \tsf_j(\bx; \Omega). \label{eq_tsf_weighted}
\end{equation}
Note that the above sum provides a representation-independent mathematical
definition for the gradient in \eq{eq_grad_generic_simple} with respect to the
(unparameterized) pointset $\Omega \in \Omega^\ast_\mathrm{P}$. Rather than
quantifying a direction of steepest descent for moving in a particular parameter
space, $\tsf(\bx; \Omega)$ identifies the set of points $\bx \in \Omega$ that
are {\it contributing the most to the violation of constraints}. A proper
direction to move in the (unparameterized) design space
$\dspace^\ast_\mathrm{P}$ is to remove the points with maximal TSF.

The coefficients $\lambda_j > 0$ have to be either computed by solving adjoint
problems -- as we showed for the case of strain energy in \eq{eq_Lag_prime}---or
selected using {\it adaptive weighting} schemes that are mainstream in
multi-objective and multi-constraint TO \cite{deng2016multi,Deng2015multi}.

\subsubsection{Penalizing Sensitivities via Local Constraints} \label{sec_filt_TSF}

The TSF operator maps global constraints to fields that vary depending on $\bx
\in \Omega$. The local constraints are already defined as fields that vary in a
similar fashion (i.e., are of the same ``type'' as the TSF). Hence, we can
penalize the TSF in \eq{eq_tsf_weighted} with local constraints as:
\begin{equation}
	\hat{\tsf}(\bx; \Omega) := \tsf(\bx; \Omega) + \!\!\!\!\! \sum_{n_\mathrm{P} +
	n_\mathrm{G} < i \leq n_\mathrm{C}} \!\!\!\!\! \kappa_i \gloc_i(\bx; \Omega).
	\label{eq_tsf_penalized}
\end{equation}
The choice of coefficients $\kappa_i > 0$ might need experimenting with the TO
to adjust the relative importance of different constraints and improve
convergence properties.

\subsection{Pareto Front Tracing by Fixed-Point Iterations}

The TSF orders the points in the design domain according to the potential impact
of removing their local neighborhoods on objective function and constraints. An
incremental improvement to the design is one that eliminates the points with the
lowest TSF (e.g., the bottom $5\%$). We define an `$\tau-$modified' (potentially
infeasible) design $\Omega(\tau) \subset \Omega$ by a PMC in terms of the
current design $\Omega$:
\begin{align}
	\indic_{\Omega(\tau)}(\bx) &:= 
	\begin{dcases}
	1 \quad &  \text{if}~ \hat{\tsf} (\bx; \Omega) \geq \tau,\\
	0 &  \text{otherwise},
	\end{dcases} \label{eq_pmc_TSF_1} \\
	\text{i.e.,}~ \Omega(\tau) &:= \big\{\bx \in \Omega ~|~ \hat{\tsf} (\bx;
	\Omega) \geq \tau \big\}. \label{eq_pmc_TSF_2}
\end{align}
where the isolevel threshold $ \tau > 0$ determines a step size for incremental
change, e.g., $\tau := 0.05$ means we are removing the least sensitive $5\%$. It
is important to select a small value so that only a small subset with
$\hat{\tsf}(\bx; \Omega) \geq \tau$ is removed to obtain a shape that is not too
different. The new design marginally violates the constraints and slightly
deviates from the Pareto front (Fig. \ref{fig_fixedPoint} (a)). However, it is
close enough to the front that it can be brought back by a fixed-point iteration
(Fig. \ref{fig_fixedPoint} (b, c)). The iteration may not converge if the step
size is too large. But if it does, it produces another feasible and (locally)
Pareto-dominant design that is slightly lighter.

\paragraph{\bf Optimization Loops}
Here is a general algorithm:
\begin{enumerate}
	\item Pick a value $\delta > 0$ for the desired change in volume fraction for
	the outer-loop iteration.
	\item Compute $\hat{\tsf}(\bx; \Omega)$ from \eq{eq_tsf_weighted} and
	\eq{eq_tsf_penalized} and normalize it with its maximum value over the current
	design.
	\item Initialize $\Omega(\tau) \subseteq \Omega$ using \eq{eq_pmc_TSF_2} with a
	reasonably small initial $\tau \leftarrow \tau_0$ to start the fixed-point
	iteration:
	\begin{enumerate}
		\item Cycle over the FP-solvers and update the performance fields
		$\analysis_i(\Omega) \rightarrow \analysis_i(\Omega(\tau))$ (e.g., the
		constrained physical or kinematic properties) for the $\tau-$modified design
		obtained from \eq{eq_pmc_TSF_1}.
		\item Re-evaluate the constraints using the updated performance results;
		recompute the TSF using \eq{eq_tsf_weighted} and \eq{eq_tsf_penalized}
		everywhere accordingly.
		\item Find $\tau > 0$ such that the $\tau-$modified design in
		\eq{eq_pmc_TSF_2} with the updated TSF has the desired reduction in volume
		fraction, i.e., $\bar{V}_{\Omega(\tau)} \approx (\bar{V}_{\Omega} - \delta)$.
		\item Repeat (a--c) until the $\tau-$modified design does not change. The
		result is feasible with respect to the  constraints and is Pareto-dominant.
	\end{enumerate}
	\item Repeat (1--3) until the volume fraction reaches the smallest feasible
	value to sustain the requirements.
\end{enumerate}

\paragraph{\bf Procedure} 
Here is a systematic procedure to {\it explore} the pruned design space, i.e.,
trace a locally Pareto-optimal family of alternative design variants
$\dspace^\ast_\mathrm{dom} \subset \dspace^\ast_\mathrm{P}$ by recurrent
incremental thresholding of (augmented and penalized) TSF, defined in terms of
global and local constraints:
\begin{itemize}
	\item {\bf Step 0.} Start at the extreme end of the Pareto front (maximal
	volume) by initializing the design with the maximal pointset obtained from
	pruning $\Omega \leftarrow \Omega^\ast_\mathrm{P}$.
	\item {\bf Step 1.} Express the global objective functions and global and local
	constraints for a given design $\Omega \in \dspace^\ast_\mathrm{P}$ to
	formulate the problem in the general form of \eq{eq:ParetoProblem_gen}.
	\item {\bf Step 2.} Define a subroutine to evaluate TSFs for each global
	objective function and global constraint using \eq{eq_TSF}, combine them using
	\eq{eq_tsf_weighted}, and penalize them with local constraints using
	\eq{eq_tsf_penalized}.
	\item {\bf Step 3.} Invoke the outer-loop optimization algorithm explained
	above to incrementally reduce the material by thresholding the TSF.
	\item {\bf Step 4.} Within the inner-loop (fixed point iteration) cycle over
	FP-solvers to evaluate the objective functions and constraint upon every
	incremental change in the outer-loop. Repeat until the deviated solution
	converges back on the Pareto front.
	\item Repeat {\bf steps 2--4} sequentially until the algorithm cannot find a
	solution for after removing enough material, i.e., arrives at other extreme end
	of the Pareto front (minimal volume).
\end{itemize} 

\subsection{Exploration after Uncoupled Assembly Constraints}
\label{sec:exploreContainment}

\begin{figure}[h!]
	\centering
	\includegraphics[width=0.8\linewidth]{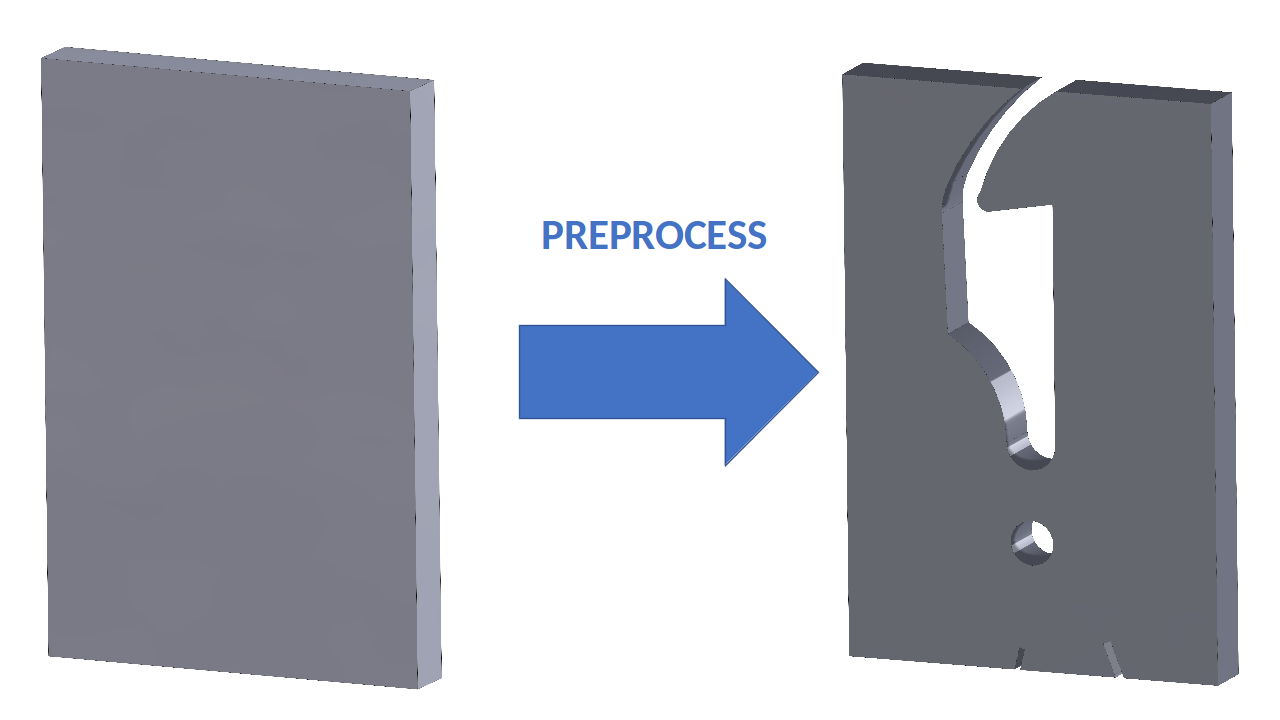}
	\caption{Functional surfaces are specified in pre-processing.}
	\label{fig_latchPreprocess}
\end{figure}

Let us consider the (secondary) car hood latch problem adopted from
\cite{Ilies2000shaping} and presented in Section \ref{sec_intro} with the
following kinematic and physical constraints:
\begin{enumerate}
	\item The latch must retain special features designated by the designer, as
	illustrated in Fig. \ref{fig_latchPreprocess}.
	\begin{itemize}
		\item One feature ensures that its mating pin (moving vertically up and down)
		rotates the latch by $21^\circ$ due to sliding contact maintained through a
		spring (not shown here).
		\item The other feature is for safety considerations; it ensures that if the
		pin moves upwards in a sudden reverse motion -- due to a failure of the
		primary latch -- the secondary latch stops clap it to prevent the car hood
		from opening.
	\end{itemize}
	\item As the latch rotates around its pivot from $0^\circ$ to $21^\circ$, it
	must remain completely within a safe region of space to avoid interference with
	other car parts.
	\item The latch is to be manufactured from stainless steel 304 using a metal AM process.
	\item The latch should not weigh more than $0.30$ pound.
	\item The latch will experience loads at pre-determined points/surfaces,
	including the contact forces with the pin exerted by the spring. Under these
	loads, its maximum deflection must not exceed $0.03$ inches.
\end{enumerate}
Such a diverse set of requirements is quite common, and should be simultaneously
handled by the computational design framework. We note from the first
requirement that modeling design intent and  synthesizing functional features to
satisfy them are difficult without knowing substantial information about the
application. These features are given in a pre-processing step (Fig.
\ref{fig_latchPreprocess}). Nevertheless, a substantial remaining portion of the
geometry is not defined by functional features and can be optimized. The
remaining requirements are systematically solved using the methods presented in
this paper.

\begin{figure}[h!]
	\centering
	\includegraphics[width=0.8\linewidth]{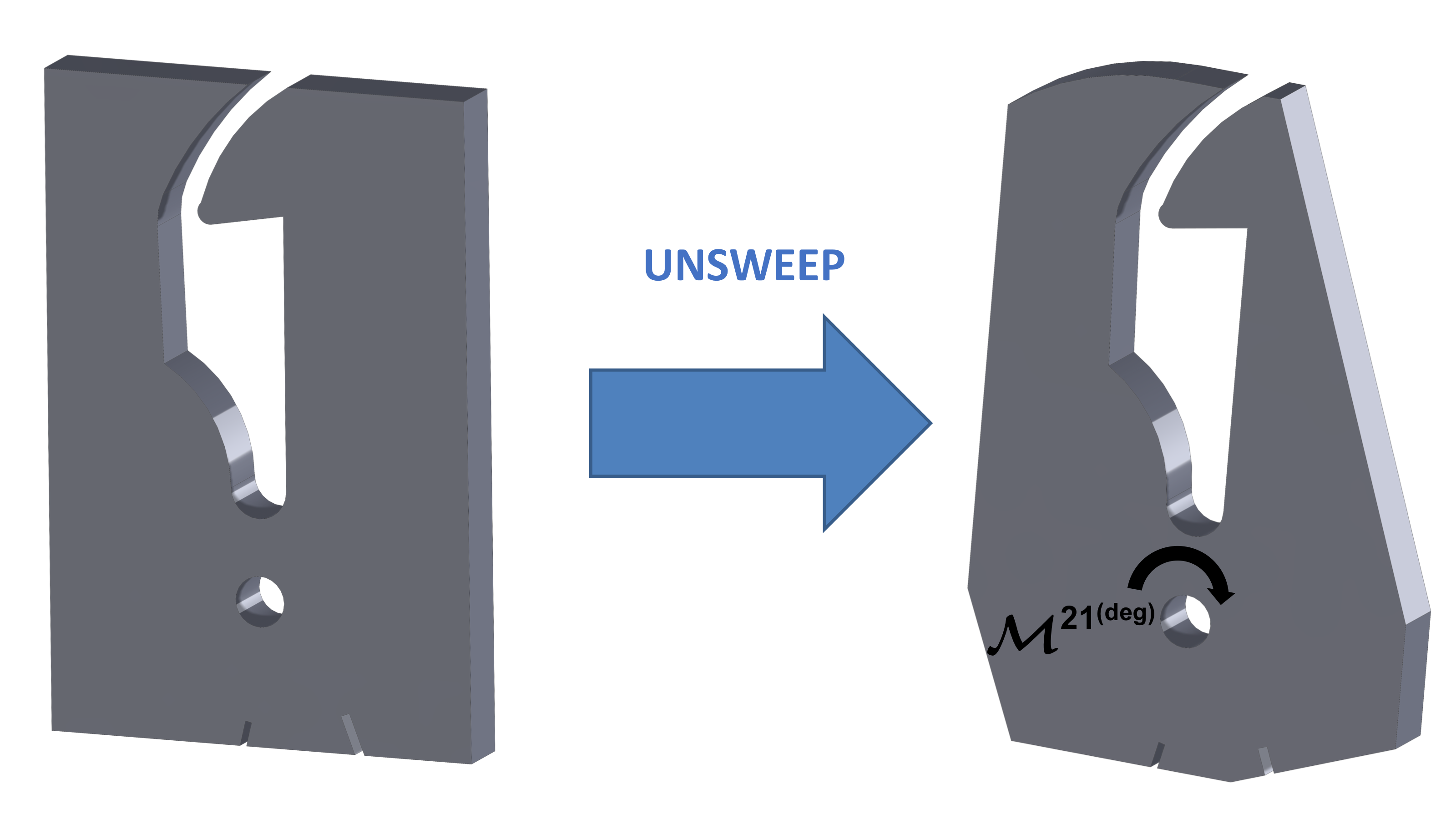}
	\caption{The \textsf{Unsweep} removes parts of the pre-processed initial design that would exit the envelope for any clockwise rotation of $\theta \in [0^\circ, 21^{\circ}]$ around the pivot.}
	\label{fig_latchUnsweep}
\end{figure} 

\paragraph{\bf Step 0}
Recall from Section \ref{sec_prune_ex1} that requirement 2 can be satisfied
upfront (without premature optimization) by pruning the design space via an
IP-solver. We start the TO with the initial design $\Omega^\ast_1 := \unsweep(M,
E)$ where $M = \{ R(\theta) \in \SO{2} ~|~ 0 \leq \theta \leq 21^\circ \}$ is
the collection of all rotations that the latch can go experience, and $E
\subseteq \R^3$ is the containment envelope (Fig. \ref{fig_latchUnsweep}).

Since TO is a material-reducing procedure, the remaining requirements 3--5 can
be satisfied by TO without violating the containment constraint.

\paragraph{\bf Steps 1, 2}
In the absence of manufacturing constraints, the physics-based constraints for
this problem are posed in common form of \eq{eq:ParetoProblem_1}. The
upper-bound on the weight can be converted to an upper-bound on volume fraction
$\bar{V}_\Omega \leq \bar{V}_\Omega^\mathrm{targ}$ where
$\bar{V}_\Omega^\mathrm{targ} = (0.30~\text{lb}/\rho^{}_\mathsf{SS304}) /
\vol[\Omega^\ast_1]$ using the known density of SS304.

The upper-bound on deflection $\updelta_\Omega(\bx) \leq \updelta_\mathsf{UB} :=
0.03''$ need not be stated as a separate constraint, because it implies an
upper-bound on compliance, hence a lower-bound on the volume fraction.

\paragraph{\bf Steps 3, 4}
At every outer-loop iteration, the maximal deflection increases due to removed
material. The algorithm checks if the deflection constraint is violated and
stops at the lightest possible solution.

Within the inner-loop fixed point iteration, the TSF is computed as in
\eq{eq_Lag_prime}, based on which the $\tau-$modified design $\Omega(\tau)$ is
extracted as the $\tau-$superlevel set of the TSF. Subsequently, the FEA solver
is invoked to solve $[K_{\Omega(\tau)}][\bu_{\Omega(\tau)}] = [\bff]$. Based on
the updated stiffness matrix $[K_{\Omega(\tau)}]$ and displacement field
$[\bu_{\Omega(\tau)}]$ in response to the boundary conditions, the Lagrange
multipliers are updated via \eq{eq_Lag_prime} as $[\lambda_2] =
-[K_{\Omega(\tau)}]^{-1}[\bff]$ and the TSF in \eq{eq_tsf_weighted} is
recomputed. The iteration is repeated until the design remains unchanged.

\begin{figure}[h!]
	\centering
	\includegraphics[width=\linewidth]{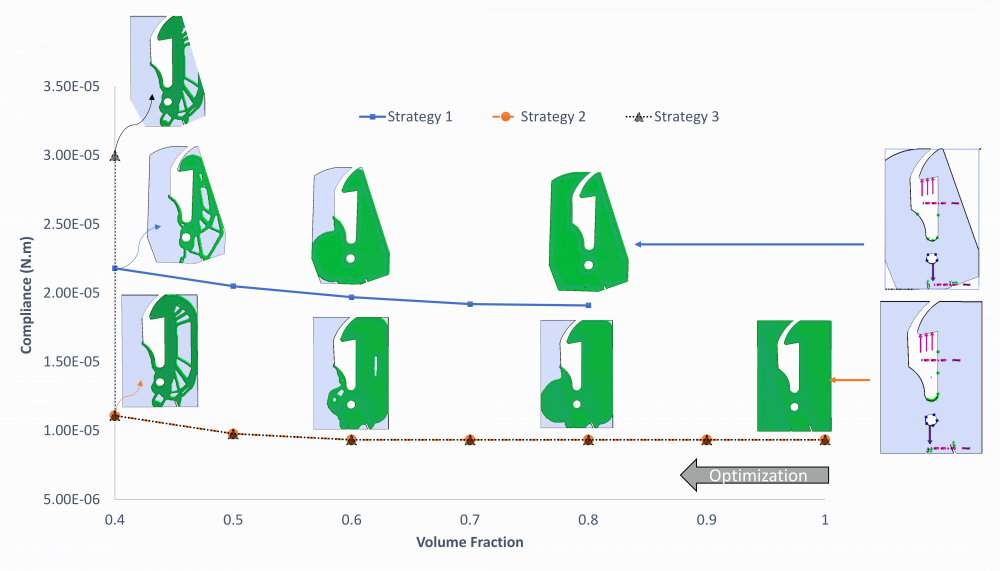}
	\caption{Optimization fronts traced by different strategies. The first strategy
	that applies TO to the pruned feasible design subspace of the containment
	constraint is the most computationally rational, as the entire family of
	solutions satisfy all requirements.} \label{fig_latchPareTOs}
\end{figure}

Figure \ref{fig_latchPareTOs} illustrates the Pareto front for solving the
above problem, starting from the pruned design domain $\Omega^\ast_\mathrm{P}$
as prescribed above (strategy 1). The results are compared against the case when
the algorithm starts from the initial design domain $\Omega_0$ (strategy 2),
ignoring the containment constraint (requirement 2). The latter is an example of
premature optimization, after which there is no guarantee that the design can be
fixed to take the containment constraint into account. The graph also shows an
incorrect attempt to fix the design as in Fig. \ref{fig_latchIntersect} of
Section \ref{sec_intro}, leading to an infeasible and suboptimal design
(strategy 3).

\subsection{Exploration after Uncoupled Accessibility Constraints}
\label{sec:exploreFixed}

\begin{figure}[ht!]
	\centering
	\includegraphics[width=\linewidth]{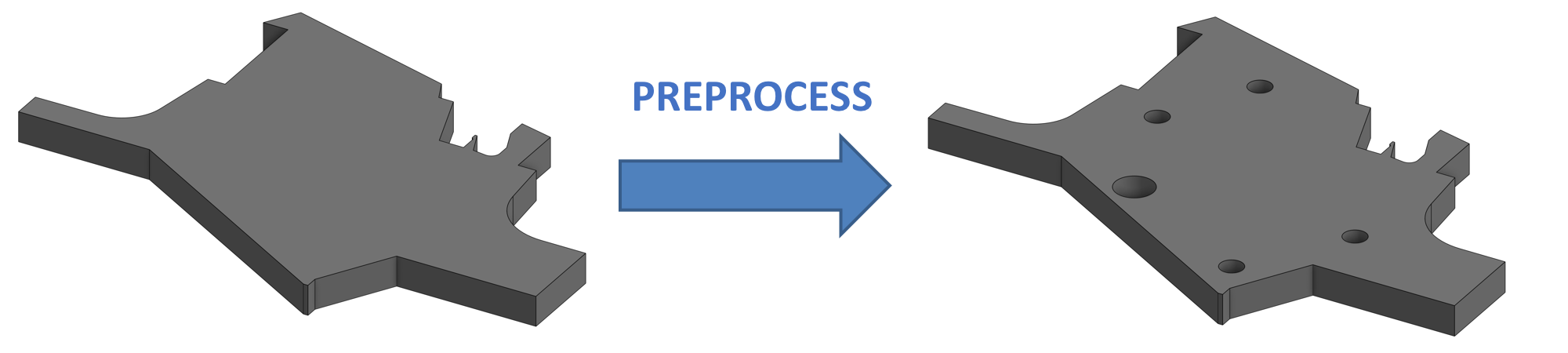}
	\caption{Functional surfaces are specified in pre-processing.}
	\label{fig_fixturePreprocess}
\end{figure}

\begin{figure}[ht!]
	\centering
	\includegraphics[width=\linewidth]{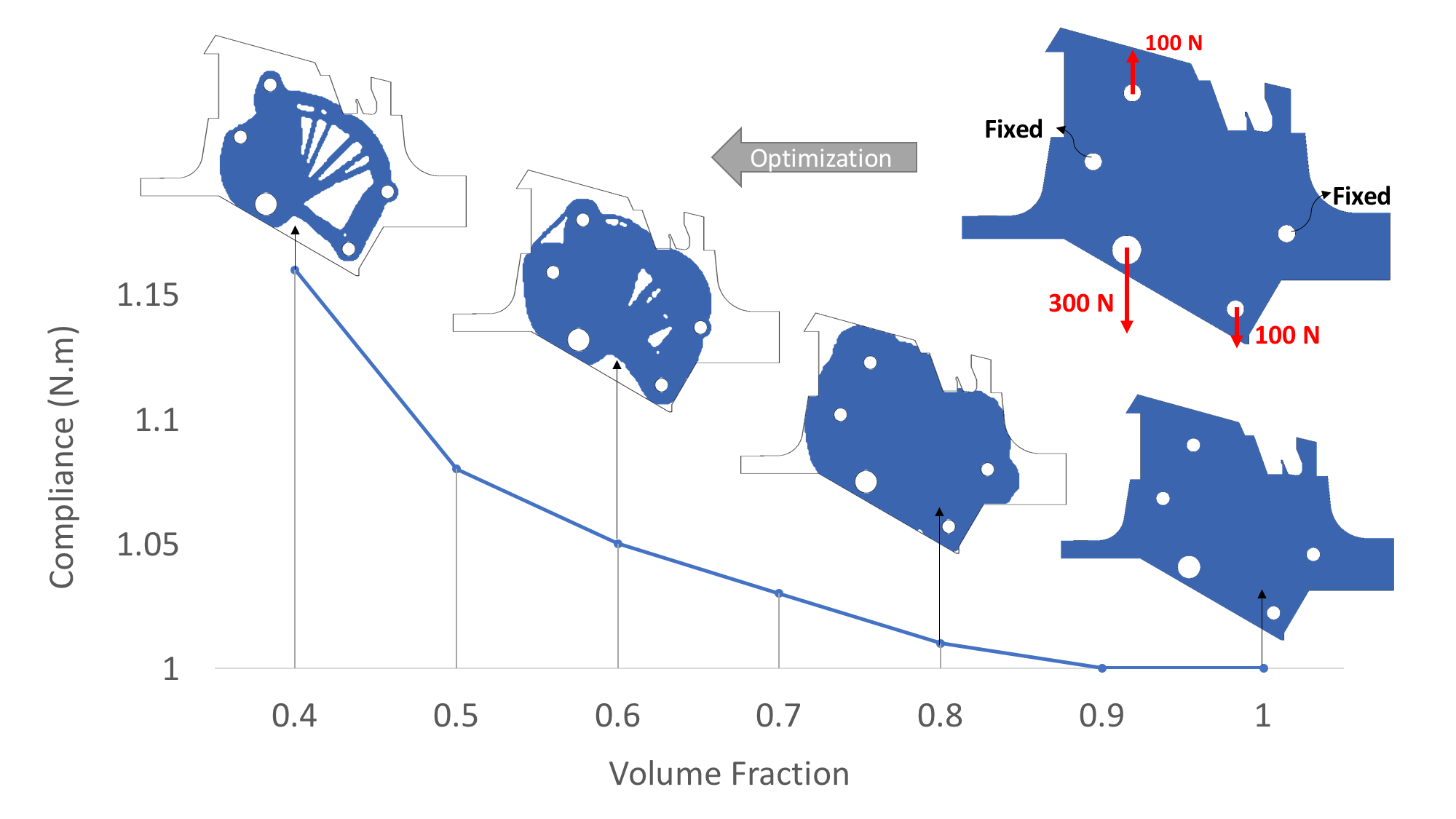}
	\caption{Pareto front of accessible designs optimized under illustrated loading
	conditions. The lightest optimized design at $40\%$ volume fraction is also
	shown in Fig. \ref{fig_fixtureSetup}.} \label{fig_fixturePareTO}
\end{figure}

Let us next consider the problem of design for manufacturability in Section
\ref{sec_prune_ex2} with the setup shown earlier in Fig.
\ref{fig_fixtureDesign}. Once again, functional features are specified in
pre-processing,%
\footnote{We are assuming that these features are also accessible. Otherwise,
	they could be introduced upfront in the raw stock and be modified by designer
	through trial and error.}
as illustrated in Fig. \ref{fig_fixturePreprocess}. The boundary conditions
are shown on the top-right corner of Fig. \ref{fig_fixturePareTO}, including
both forces and restrained surfaces. The underlying material is stainless steel
with Young's modulus $Y = 200$ GPa and Poisson's ratio $\nu = 0.33$.

\paragraph{\bf Step 0}
We showed in Section \ref{sec_prune_ex2} that the design space can be pruned
with respect to the accessibility of a $2-$axis CNC instrument, where the 2D
cross-section of the 2.5D maximal pointset was obtained as the $0-$level set of
a 3D convolution field between the head $H$ and fixtures $F$ (both in 3D), i.e.,
$\Omega^\ast_2 \cong ({\indic}_{-H} \ast \indic_F)^{-1}(0)$.

\paragraph{\bf Steps 1, 2}
We are interested in finding a set of designs with maximal stiffness while
reducing the volume of the pruned design domain by another $60\%$.

All the optimized designs must have uniform cross-sections along
wire-/laser-cutting direction (i.e., are 2.5D). This constraint can be imposed
either by applying a 2D TO to the cross-section of the initial design and
extruding its results, or by using a 3D TO with a through-cut filtering of the
TSF shown in Fig. \ref{fig_swingarm} (c) of Section \ref{sec_intro}. We use the
latter (PareTO in 3D) for this example.

The remaining {\bf steps 3} and {\bf 4} are similar to the previous example. The
only FP-solver in the loop is a standard FEA, in absence of coupled
manufacturing constraints, noting that manufacturability is a priori guaranteed
in the pruning phase. In the next example, we consider a problem in which the
manufacturing constraints cannot be pruned and has to be coupled with the
physical constraints within the inner-loop fixed-point iteration.

\begin{figure}[ht!]
	\centering
	\includegraphics[width=\linewidth]{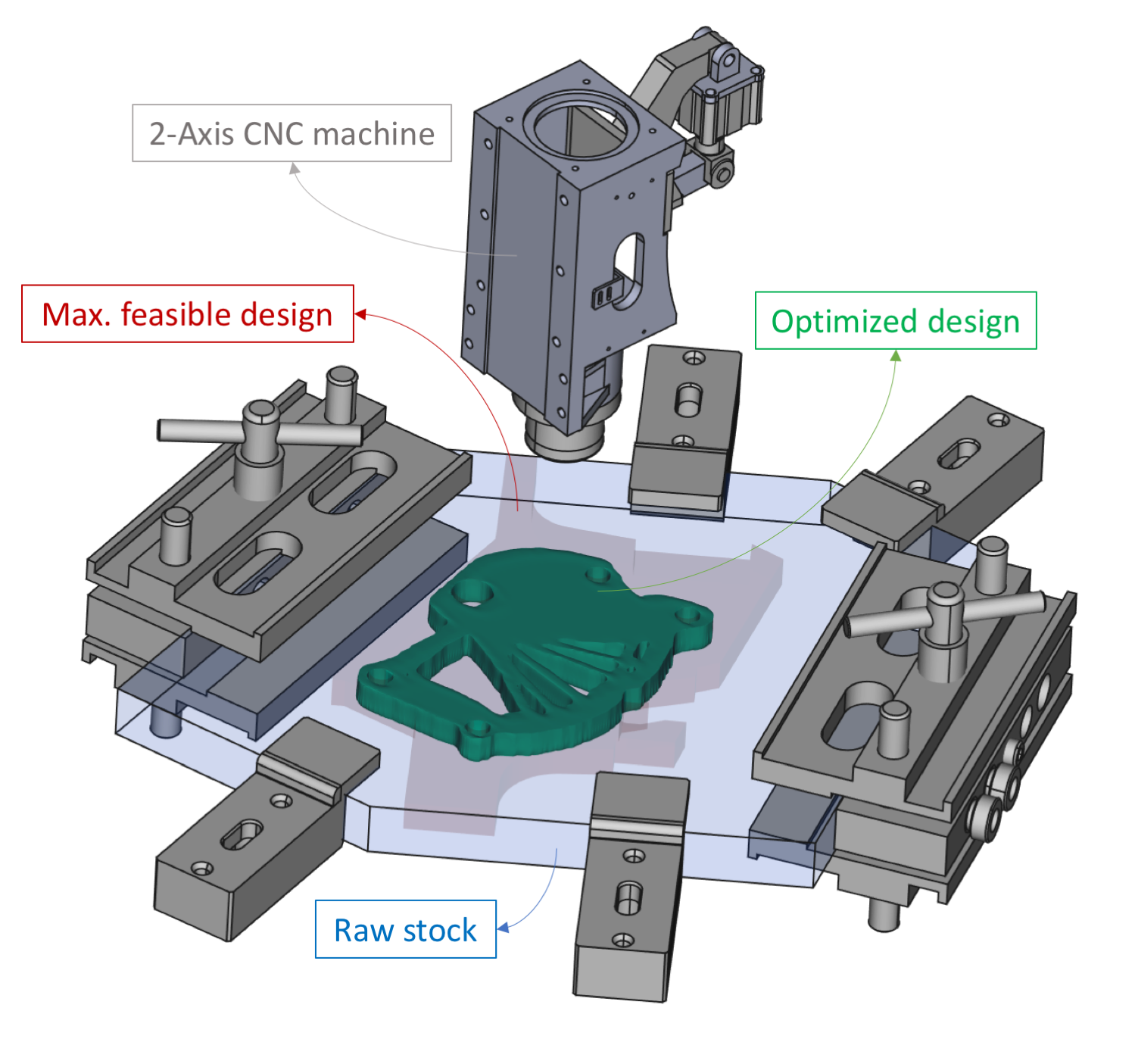}
	\caption{Optimized design at $40\%$ volume fraction shown in the manufacturing
	setup.} \label{fig_fixtureSetup}
\end{figure}

Figure \ref{fig_fixtureSetup} shows the fixturing setup, raw stock, maximal
manufacturable domain (i.e., initial design for TO), and  the optimized design
at $40\%$ volume fraction. Figure \ref{fig_fixturePareTO} shows the Pareto front
as it is traced from $100\%$ to $40\%$ volume fraction. As with the previous
example, the material reducing nature of TO ensures that it does not violate the
manufacturability constraint satisfied in Section \ref{sec_prune_ex2}.

\subsection{Exploration with Coupled Accessibility Constraints}
\label{sec:challenge}

The accessibility constraint of Section \ref{sec:exploreFixed} was solely
dependent on the surrounding fixtures, which enabled us to enforce it by pruning
the design space before TO. We argued in Section \ref{sec_prune_ex3} that this
is not possible for general accessibility problems in which the design's
(unknown) shape affects the evaluation of the constraint. In this section we
focus on accessibility constraints that evolve throughout the design space
exploration via TO. They are coupled with physical constraints (evaluated by
FEA).

In Section \ref{sec_TOforms}, we presented two alternative formulations for
optimization subject to accessibility constraints, one with global formulation
in \eq{eq:ParetoProblem_4} (based on total inaccessible volume), and one with
local formulation in \eq{eq:ParetoProblem_5} (based on inaccessibility measure
as a convolution field). Experimenting with the former fails, as expected,
because the TSF for the global form is discontinuous, as elaborated in Section
\ref{sec_TSF_limits}. The convolution field, on the other hand, is relatively
well-behaved and can be used to penalize the TSF, as confirmed by our numerical
experiments.

\paragraph{\bf Step 0}
In the absence of uncoupled pointwise constraints, start with the initial design
domain $\Omega_0 \in \dspace$.

\paragraph{\bf Step 1}
The objective functions and constraints are given in \eq{eq:ParetoProblem_5}.
The additional constraint $[\indic_{\Omega}\ast\tilde{\indic}_T] = [0]$ requires
that every point in the (voxelized) design $[\indic_{\Omega}]$ be accessible by
the (voxelized) tool assembly $[\indic_T]$, inverted as $[\tilde{\indic}_T]$.
Remember that the convolution's value at a given query point measures the volume
of collision when the tool is displaced in such a way that a representative
point on the tool -- i.e., the origin of its local coordinate system in which
$[\indic_T]$ is represented -- is brought to the query point in the design
domain.%
\footnote{Technically, the convolution field is defined over the configuration
	space of relative motions (translations in this case). The proper selection of
	the local coordinate system is important to ``register'' the convolution field
	with the design domain and other fields defined over it (e.g., TSF). See
	\cite{Lysenko2010group} for more details on the choice of origin.}
We pick the origin of the tool at the tip of the cutter to simplify the
formulation.

\begin{figure}[ht!]
 	\centering
 	\includegraphics[width=0.46\textwidth]{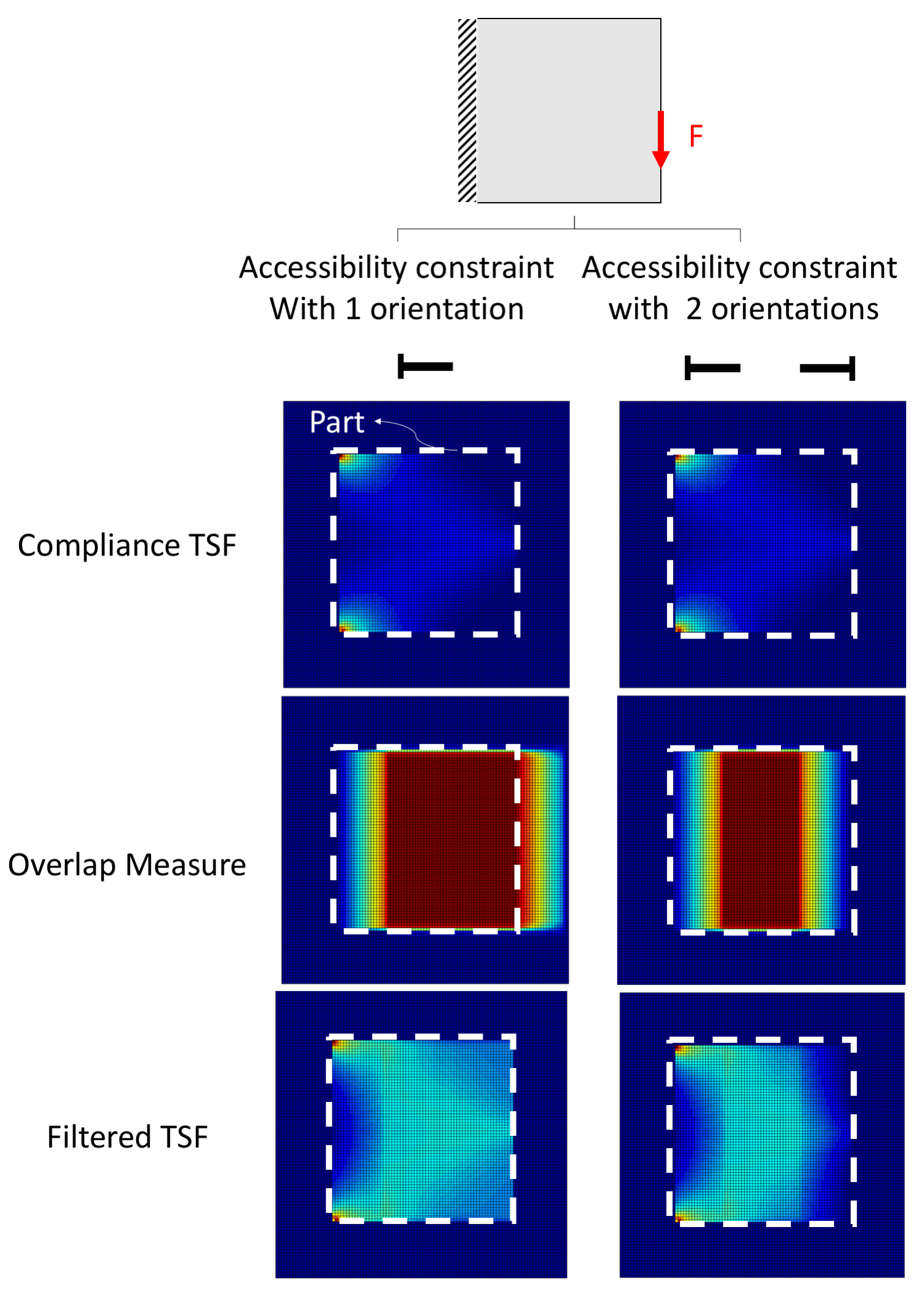}
 	\caption{Penalizing TSF by the inaccessibility measure for T-shaped tool
	approaching from left (a) or from both left and right (b). Note that the field
	is asymmetric for the former.} \label{fig_filteredTSF}
 \end{figure}

\begin{figure*}[ht!]
	\centering
	\includegraphics[width=0.96\textwidth]{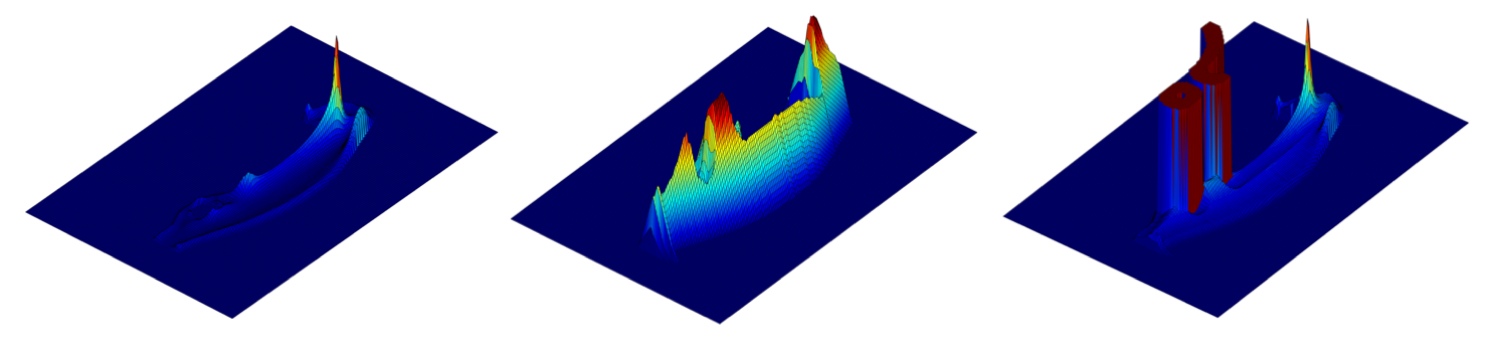}
	\caption{The original TSF for compliance (a), the inaccessibility measure
	obtained from a convolution of the design and tool at $0^\circ$ and $180^\circ$
	(b), and the penalized TSF to incorporate accessibility and retain functional
	surfaces (c) for the final design at a volume fraction of $35\%$.}
\label{fig_examplesTSF}
\end{figure*}

For the interior points, the constraint is obviously violated, because the tool
cannot reach the interior without colliding with the part. The violation is
larger for points that are farther from the boundary, providing a continuous
penalty for TSF. Importantly, not every point in the exterior is accessible
either. Even if the tip of the cutter does not collide with the part, the rest
of the tool assembly might. The penalty is typically smaller for external
points, as illustrated by Figs. \ref{fig_filteredTSF} and \ref{fig_examplesTSF}. when
there are more than one tool or approach orientations, the algorithm picks the
minimum collision measure for penalization.

\paragraph{\bf Step 2}
The TSF for compliance is computed as usual, and is normalized by its maximum.
An independent subroutine computes the convolution via FFTs, as discussed
earlier, and is normalized by the volume of the tool (upper-bound to
convolution). The TSF is penalized via convolution using an adaptive weight (for
instance we start with $\lambda_3 := 0.01$ and increase it to $\lambda_3 := 0.2$
for lower volume fractions). Other design constraints such as minimum feature
size or surface retainment can also be imposed as discussed in Section
\ref{sec_lit}.

\paragraph{\bf Steps 3, 4}
The outer-loop iteration is as before. The inner loop iteration now cycles
through one more FP-solver (the FFT-based convolution routine), as illustrated
earlier in Fig. \ref{fig_fixedPoint} (c).

Let us first consider a simple 2D cantilever beam of Fig. \ref{fig_filteredTSF},
with simple boundary conditions of a downward force $\mathbf{F} = 1$ N, Young's
modulus of $Y = 1$ GPa, and Poisson's ratio of $\nu = 0.3$. Given a T-shaped
tool with cutting part at the thin end, we consider accessibility in two
scenarios with the tool approaching from one orientation (from left) and two
orientations (from left and right). Figure \ref{fig_filteredTSF} illustrates how
the convolution fields differ in the two cases. Subsequently, the compliance
TSF is penalized to capture accessibility under tool orientations.
 
\begin{figure}[ht!]
	\centering
	\includegraphics[width=0.46\textwidth]{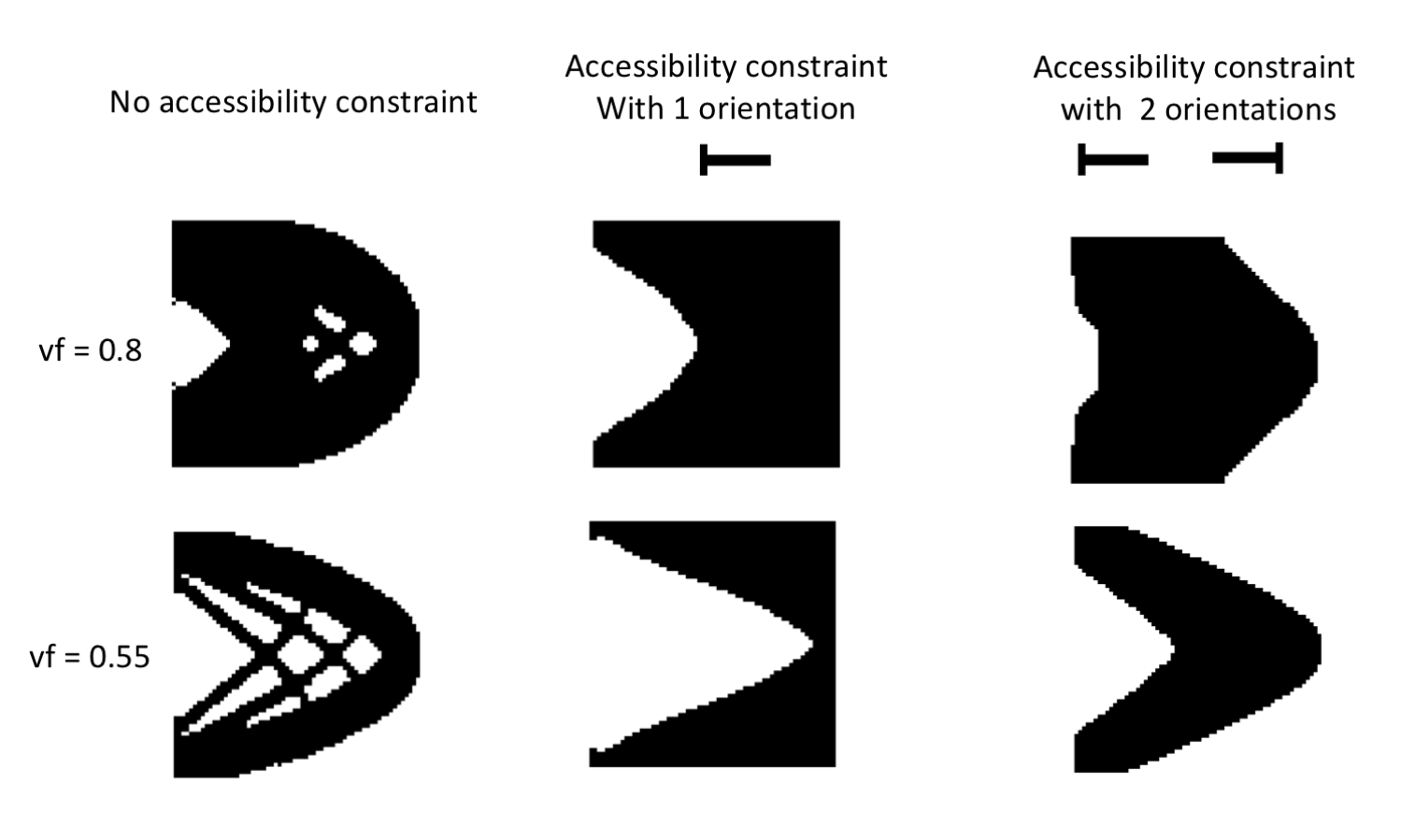}
	\caption{ Optimized topologies at volume fractions 0.55 and 0.80 without
	accessibility constraint (a), with accessibility constraint for tool at
	$0^\circ$ (b), and for tool at $0^\circ$ and $180^\circ$ (c).}
	\label{fig_beams}
\end{figure}

Fig. \ref{fig_beams} shows optimized shapes with and without the accessibility
constraints with one and two tool orientations. Since the tool can only move in
the plane, the TO cannot introduce interior holes without incurring a large
penalty. Moreover, it can only remove material from the boundary in such a way
that the remaining shape is machinable---e.g., no concave features of smaller
feature size than the tool thickness in this case. Note that this is
automatically enforced by penalizing convolution, without appealing explicitly
to any notion of features or feature size.

\begin{figure}[ht!]
	\centering
	\includegraphics[width=0.46\textwidth]{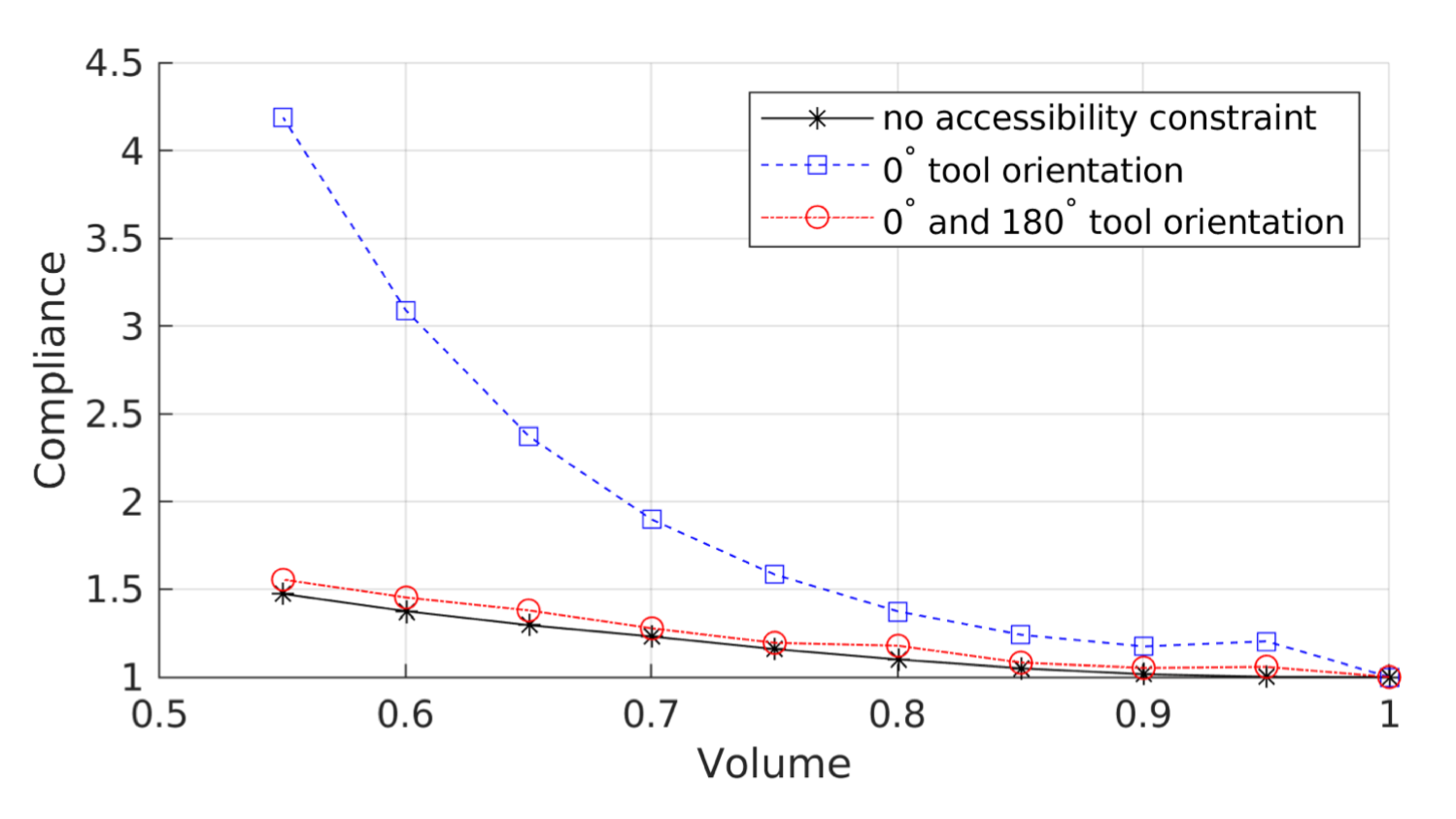}
	\caption{Pareto fronts of cantilever beam optimization without accessibility
	constraint, accessibility with tool only at $0^\circ$ angle, and accessibility
	with tool at $0^\circ$ and $180^\circ$.} \label{fig_beamPareto}
\end{figure}

In the case of the tool approaching at $0^\circ$, material can be accessed and
removed only from the left side. However, with $0^\circ$ and $180^\circ$ angles
for tool orientation, the material can be removed from both sides. Figure
\ref{fig_beamPareto}  shows the Pareto fronts of the three scenarios. As
expected, optimization without accessibility constraint yields the best
performance in terms of compliance while imposing accessibility with one
approach direction significantly increases the compliance. However, when the
tool can approach from both directions comparable performance to the
unconstrained solutions can be achieved.

\begin{figure}[ht!]
	\centering
	\includegraphics[width=0.46\textwidth]{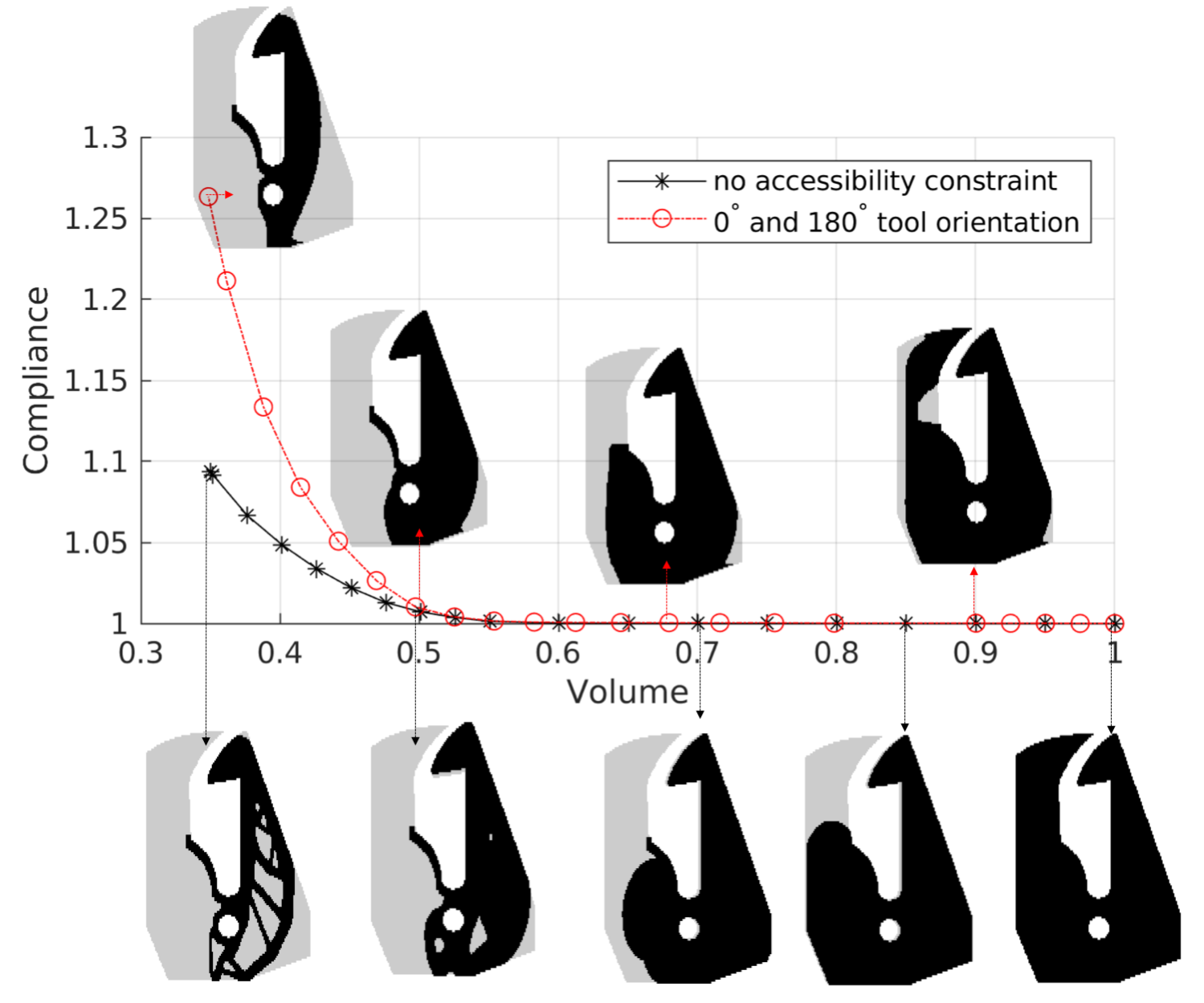}
	\caption{Pareto fronts of optimized designs for the hood latch, starting from
	the maximal pointset pruned via \textsf{Unsweep}, with and without
	accessibility constraint for tool at $0^\circ$ and $180^\circ$.}
\label{fig_latchAccess}
\end{figure}

Let us next consider the car hood latch example of Fig. \ref{fig_latchUnsweep}
one more time (also in 2D). Figure \ref{fig_latchAccess} shows the
optimized latches at $35\%$ volume fraction with and without the accessibility
constraint. The same T-shaped tool is considered  and oriented at both $0^\circ$
and $180^\circ$. Imposing the accessibility constraint increased the relative
compliance from 1.09 to 1.26. Figure \ref{fig_examplesTSF} illustrates the original
TSF for compliance, the inaccessibility measure, and the penalized TSF for the
hood latch at a volume fraction of $35\%$.

It should be noted that in the above examples, we only considered the collision
between the tool and the part (i.e., no fixtures). Moreover, constraining the
convolution field captures only the existence of final collision-free
configurations for the tool to machine the part at different points in the
design domain. It does not guarantee a collision-free tool-path from the initial
tool configuration to the removal site. Path planning is beyond the scope of
this paper.

\subsection{Limitations of Sensitivity-Based Exploration} \label{sec_TSF_limits}

For the definition of the TSF in Section \ref{sec_def_TSF} to be valid, the
limit in \eq{eq_TSF} must exist everywhere in the design domain and for all
intermediate designs. In other words, puncturing the design with infinitesimal
cavities must lead to infinitesimal changes in the violation of objective
functions and constraints. For this to hold, the functions must be sufficiently
smooth.%
\footnote{Technically, we need the function $F_i: \dspace^\ast_\mathrm{P} \to
	\R$ to be differentiable in the Hausdorff topology of $\dspace^\ast_\mathrm{P} =
	\powerset^\ast(\Omega^\ast_\mathrm{P})$, which is relative to the topology of
	$\dspace = \powerset^\ast(\Omega_0)$.}
This is not always the case for general constraints.

For example, if $F_i(\Omega)$ is itself evaluating a topological property,
introducing a puncture (no matter how small) can produce a large change in
$F_i(\Omega)$. For example, topological defects in AM due to resolution limits
can be characterized as integer-valued Euler characteristics 
\cite{Behandish2018characterizing,Behandish2019detection}. The Euler
characteristic will increase by $+1$ after adding a cavity, hence $\tsf_i(\bx;
\Omega) \sim +1/O(\epsilon^3) \to +\infty$ as $\epsilon \to 0^+$. In practice,
this appears as a discontinuity in constraint evaluation which adversely affects
the convergence of the optimization loop (e.g., fixed-point iteration for
PareTO).

As another example, recall the manufacturability constraint of \eq{eq_max_vol}
in which the total volume of inaccessible regions (for machining) was
upper-bounded as a global constraint. Puncturing a hole of volume
$\vol[B_\epsilon(\bx)]$ in the interior of the design adds exactly
$\vol[B_\epsilon(\bx)]$ to the inaccessible volume, hence $\tsf_i(\bx; \Omega) =
1$ for all $\bx \in \mathsf{i}\Omega$. Most of the time, chipping off a visible
(but too small) $\vol[\Omega \cap B_\epsilon(\bx)]$ adds the same exact amount
to the inaccessible volume, because the tool is of finite size and cannot remove
that volume in practice. However, it is possible that a substantial region of
design that was initially inaccessible becomes accessible due to the small
change at the boundary. Hence either $\tsf_i(\bx; \Omega) = 1$ or $\tsf_i(\bx;
\Omega) \to -\infty$ for all $\bx \in \partial\Omega$. This is not a
well-behaved TSF for optimization.

The TSF is well-behaved for most global constraints that are defined as
volumetric integrals of continuous physical fields (e.g., strain energy).%
\footnote{This is the actual motivation behind using a volumetric measure of the
	inclusion in the denominator of \eq{eq_TSF} to normalize volumetric measures in
	the numerator. In principle, one can use a different measure for global
	constraints that vary at a different rate than $O(\epsilon^3)$.}
Small changes in the design often lead to small changes in physical response,
when it is integrated over the entire domain, because integration smooths out
the effects of local singularities. For instance, if we impose a constraint on
the maximal stress as in \eq{eq_max_sigma}, we cannot define a realistic TSF due
to stress concentration. The simplest model of stress concentration yields an
infinite maximal stress $\sigma_\Omega(\bx) \sim O(\epsilon^{-0.5})$ near an
infinitesimal radius of curvature $\epsilon \to 0^+$, hence $\tsf_i(\bx; \Omega)
\sim O(\epsilon^{-3.5}) \to +\infty$ when defined for the maximal stress as in
\eq{eq_max_sigma}. In practice, one can alleviate this issue by using a
volumetric integral (e.g., the $p-$norm) of stress rather than its maximum, in
order to smooth out the effects of local singularities.

The other limitation with the usability of TSF is that the constraint function
should not be locally ``flat''. If the change in the violation of a constraint
is too small, i.e., decays faster than $O(\epsilon^3)$, the limit will vanish
and the TSF will not help the optimization. For example, manufacturability
constraints that have to do with surface properties are insensitive to
volumetric changes in design. We do not consider such cases in this paper.

In summary, the TSF formulation works well for global constraints that change
smoothly with local volumetric material removal. Although this is not true for
all global constraints, the good news is that some of them can be reformulated
as local constraints. Therefore, we can use penalization of the local
constraint (Section \ref{sec_filt_TSF}) instead of defining a TSF for the global
constraint. For example, although the accessibility constraint that we discussed
above does not yield a well-behaved TSF when treated as a global constraint
$\gglob_3(\Omega) \leq 0$ of \eq{eq_max_vol}, we demonstrated in Section
\ref{sec:challenge} that it can be successfully incorporated to constraint
PareTO using penalization of the local form $\gloc_4(\bx; \Omega) \leq 0$ of
\eq{eq_cons_mu} in terms of inaccessibility measure defined as the convolution
in \eq{eq_access_conv}.

%% file: conclusion.tex
\section{Conclusions} \label{sec_conclusion}

Mechanical design requires simultaneous reasoning about multidisciplinary
functional requirements and evaluating their trade-offs. These requirements are
often expressed via heterogeneous types of constraints, including
kinematics-based constraints for assembly and packaging, physics-based
constraints for performance under mechanical or thermal loads, and both for
manufacturability. Automated design optimization algorithms rarely consider all
such requirements, and do not provide mechanisms to explore their trade space.
For example, topology optimization can automatically generate designs with
optimized material layouts for performance criteria such as strength and
stiffness, but often ignores complex motion-based constraints imposed by
collision avoidance in assembly or accessibility in manufacturing.

\subsection{Design Subspaces as First-Class Entities} \label{sec_firstclass}

The challenge in design space pruning and exploration is that IP-solvers are
usually equipped with the tools to satisfy only a subset of the criteria in a
multi-functional design problem. When these solvers are composed sequentially or
in parallel, they can rarely provide guarantees to retain the other criteria
satisfied by the preceding or concurrent solvers in the workflow. Moreover, most
of the existing solvers generate a narrow subset of the design space---most
commonly one or few design(s) that is/are deemed (locally or globally)
``optimal'' within the design subspace that appears feasible to the solver. Such
{\it premature optimization} dramatically limits the subsequent solvers' freedom
to explore (best-case scenario) and might even get deadlocked at infeasible
designs (worst-case scenario).

To solve such challenges, we follow a philosophy of treating {\it design spaces}
(as opposed to individual {\it designs}) as {\it first-class entities}---at
least to the extent that it is possible to do so by proper ordering of solvers
in the workflow. This means that the entity being passed through the design
pipeline---as input/output of consecutive synthesis solvers---is a design
subspace described in its entirety by a representative object. This treatment
allows postponing restrictive decisions and pushing premature optimization
downstream as much as possible. We organize the design workflows by a careful
analysis of the types of constraints and available solvers to address them and
provide a systematic approach to compose FP- and IP-solvers depending on the
type of design constraint(s) they can satisfy.

\subsection{Pointwise Constraints \& Maximal Designs}

An important contribution of this work is a classification of constraints
(namely, global, local, or strictly local) based on which the solvers are
organized systematically in the computational design workflow. In particular,
the strictly local (i.e., pointwise) constraints can be evaluated without a
knowledge of the global shape, hence lead to a point membership classification
(PMC) for a maximal design that satisfies them. The maximal pointset represents
the entire feasible design subspace for a pointwise constraint, in the sense
that containment in the maximal pointset is a necessary and sufficient condition
for feasibility. As such, the design space can be pruned upfront by intersecting
maximal pointsets of pointwise constraints, without premature optimization.

\subsection{Other Constraints \& Sensitivity Analysis}

Most design criteria that depend on physics-based performance do not lead to a
pointwise condition/PMC because the physical response of a design at any given
point is typically dependent on the overall shape, i.e., the membership of one
point is couple with the membership of other points. The dependency may be
long-range, as in the case of static equilibrium throughout a mechanical
structure, or local, as in transient dynamic effects within a bounded
neighborhood over a finite time interval. In either case, further design space
pruning by means of PMC, to postpone decision making on the particular design
layout, is not an option. In such cases, the FP-solvers collaborate in
generating feasible and optimized designs by combining their sensitivity fields
and methods such as fixed-point iteration for tracing the trade space of
multiple objectives.

\subsection{Duality of Forward \& Inverse Problem Solvers}

An important revelation of the classification is that the two different types of
problems (namely, design space `pruning' and `exploration') demonstrate a
different form of {duality} between forward and inverse problem (IP/FP)-solvers
for {\it generative design}:
\begin{itemize}
	\item For pointwise constraints, an IP-solver can be constructed from an
	FP-solver by generating a large sample of points in the design domain, applying
	the FP-solver in a pointwise fashion, evaluating the constraint, and
	retaining/discarding the ones that do/do not satisfy the constraint. In other
	words, the FP-solver provides a PMC test for the IP-solver. The process can be
	perfectly parallelized.
	\item For other constraints, an IP-solver can be constructed from an FP-solver
	by generating a number of candidate designs, evaluating the constraints,
	obtaining a sensitivity field to order the different points in the design
	according to their expected impact on the (dis)satisfaction of the constraint,
	removing the least sensitive points, and try again with the modified design. In
	other words, the FP-solver provides an evaluator for candidate designs to put
	in a feedback loop for the generate-and-test IP-solver. The process is a
	sequential loop that is repeated until convergence.
\end{itemize}

\subsection{Limitations \& Future Work}

An important limitation of the approach is that it does not provide any
guarantees for satisfying constraints that are neither pointwise nor
differentiable, as exemplified in Section \ref{sec_TSF_limits}. For some local
constraints (e.g., accessibility measures for machining), we have shown that
penalizing the sensitivity fields of other global constraints with the local
constraint can be effective. However, it is unclear how to systematically make
such decisions with every new problem and constraint, unlike the case with
pointwise constraints that are pruned upfront, or differentiable non-pointwise
constraints that are filtered via local sensitivity analysis.

\section*{Acknowledgments}

This research was developed with funding from the Defense Advanced Research
Projects Agency (DARPA). The views, opinions and/or findings expressed are those
of the authors and should not be interpreted as representing the official views
or policies of the Department of Defense or U.S. Government.